\documentclass[12pt]{article}
\usepackage{enumerate}
\usepackage{graphicx}
\usepackage{amsmath}
\usepackage{amssymb}
\usepackage{amsthm}
\usepackage{setspace}
\usepackage{hyperref}
\usepackage{caption}
\usepackage{subcaption}
\usepackage{xcolor}
\usepackage{natbib}
\setcitestyle{authoryear,open={(},close={)}} %JT: getting the citation style I want
\begin{document}
\title{\bf Measuring Financial Advice: aligning client elicited and revealed risk} 
\author{{\bf John R.J. Thompson, Longlong Feng, R. Mark Reesor, }\\{\bf Chuck Grace, and Adam Metzler} \\ Department of Mathematics \\ Wilfrid Laurier University \\ Waterloo, Ontario, Canada N2L 3C5}
\date{\today} 
\maketitle

\section*{Abstract}

Financial advisors use questionnaires and discussions with clients to determine a suitable portfolio of assets that will allow clients to reach their investment objectives. Financial institutions assign risk ratings to each security they offer, and those ratings are used to guide clients and advisors to choose an investment portfolio risk that suits their stated risk tolerance. This paper compares client Know Your Client (KYC) profile risk allocations to their investment portfolio risk selections using a value-at-risk discrepancy methodology.  Value-at-risk is used to measure elicited and revealed risk to show whether clients are over-risked or under-risked, changes in KYC risk lead to changes in portfolio configuration, and cash flow affects a client's portfolio risk. We demonstrate the effectiveness of value-at-risk at measuring clients' elicited and revealed risk on a dataset provided by a private Canadian financial dealership of over $50,000$ accounts for over $27,000$ clients and $300$ advisors. By measuring both elicited and revealed risk using the same measure, we can determine how well a client's portfolio aligns with their stated goals. We believe that using value-at-risk to measure client risk provides valuable insight to advisors to ensure that their practice is KYC compliant, to better tailor their client portfolios to stated goals, communicate advice to clients to either align their portfolios to stated goals or refresh their goals, and to monitor changes to the clients' risk positions across their practice.

Keywords: Risk measures, Value-at-risk, Portfolio management, Financial advice, Client-advisor relationship

%Authors
\newpage
\noindent John R.J. Thompson % \emph{(Corresponding Author)}

Department of Mathematics

Wilfrid Laurier University

Waterloo, Ontario N2L 3C5

johnthompson@wlu.ca
\\[3\baselineskip]
Longlong Feng

Department of Mathematics

Wilfrid Laurier University

Waterloo, Ontario N2L 3C5

feng0290@mylaurier.ca
\\[3\baselineskip]
R. Mark Reesor

Department of Mathematics

Wilfrid Laurier University

Waterloo, Ontario N2L 3C5

mreesor@wlu.ca
\\[3\baselineskip]
Chuck Grace

Department of Finance

Ivey Business School

London, Ontario N6G 0N1

cgrace@ivey.ca
\\[3\baselineskip]
Adam Metzler

Department of Mathematics

Wilfrid Laurier University

Waterloo, Ontario N2L 3C5

ametzler@wlu.ca

\newpage

%% To-do

\section{Introduction}
\doublespacing
In previous research, we find that that Know Your Client (KYC) information--such as gender, residence region, and marital status--does not explain client investment behaviours, whereas eight variables for trade and transaction frequency and volume are more informative \citep{thompson20}. With these findings that trading behaviour (recency and volume patterns) is linked to actual client risk preference and capacity, our recommendation to financial regulators and advisors is to use metrics more advanced than those gleaned from KYC information to understand clients better. The dataset used in that paper spanned from January 1st 2019 to August 12th 2019, and consisted of daily KYC information (age, gender, annual income, risk tolerance) and financial information (account type, total value of that account, proportion of total value invested into different securities) for $50,980$ accounts of $27,644$ clients and 311 advisors. 

This paper demonstrates how value-at-risk (VaR) can be used as an advanced metric by financial advisors to manage their clients' portfolio risks. We show that VaR provides unique real-time empirical insights that can help stakeholders visualize gaps between the elicited KYC-derived profile risk and the actual portfolio risk revealed by trading behaviours. We examine how clients' trading behaviour and cash flow impact their portfolios and use VaR to quantify how much of each client's portfolio is genuinely ``at risk" and how far clients are from their stated risk preferences.  While part of our contribution to the literature is a methodology to estimate clients' risk preferences, our primary contribution is a methodology to measure the difference in elicited and revealed risk tolerance.

In Canada, financial advisors and dealers are required by provincial securities commissions and self-regulatory organizations to collect and maintain KYC information for investor accounts. With this information, investors, under their advisor's guidance, decide on investments that are presumed to be beneficial to their investment goals and therefore suitable. Suitability is described by regulators in Canada as a ``meaningful dialogue with the client to obtain a solid understanding of the client's investment needs and objectives $\ldots$" \citep{ont14}, and has been updated to better serve client needs \citep{ont19}. The assumption is that any investment purchases or sales (trading behaviour) will conform to the KYC attributes and therefore be suitable. 

It should be noted that suitability must be informed by both Know Your Product (KYP) and KYC attributes. Part of the KYP information on securities made available by the dealership is a risk rating \citep{ont09}. These risk ratings are used to place products into risk buckets (for example, low, medium, or high risk). The KYC attributes give a portfolio selection across risk buckets, followed by a selection of products within each risk bucket. Our analysis evaluates a client's portfolio based on their KYP attributes--at least in terms of their risk and return characteristics. Our measures do not delve deeper to complete comprehensive due diligence of the investment products as would be prescribed by the regulators. 

Our analysis shows that advisors and clients use the KYC prescribed risk as a "guard rail" against which they define--and do not generally exceed--upper and lower boundaries. However, between the "guard rails", they tailor and customize portfolios in a manner that is consistent with their clients' elicited KYC risks. Do our findings suggest negative consequences for investors? We find that advisors are systemically safe and conservative and thus do not appear to expose clients to undue risk. For clients seeking to preserve capital, that is good news, but clients seeking to maximize growth may not be exposed to enough risk to achieve their expressed goals. Under-risked accounts are just as problematic as over-risked, but easier for advisors to defend to regulators and lawyers given there is no significant loss of funds. An example might be older clients who can preserve their capital but are unable to achieve investment incomes that allow them to maintain their lifestyles. An accurate answer to the question of client impact can only be answered through the lens of the advisor/client interactions--which are by necessity private and confidential. 

Every client has a unique story, and we present methods that can digitize their story to improve personalization, transparency, timeliness and visualization. Our methodology falls into the interest realms of advisors, regulators, and robo-advisors; (1) advisors can implement these mathematical methods into their practice\footnote{We have included an example Excel spreadsheet as part of this article, and this sheet is available upon request from the corresponding author.}, (2) regulators will be able to identify inconsistencies in practitioners' portfolios, and (3) robo-advisors can use these methods to help advisors monitor their clients' portfolios. 

The paper from here unfolds as such: Section \ref{sec:litrev} provides the academic context for our methodology in the broad area of measuring risk tolerance. Section \ref{sec:datadesc} is an overview of the dataset provided by a private Canadian financial dealership. Section \ref{sec:discrepancies} is a study of financial methodologies, including VaR, to measure the discrepancy between profile and portfolio risk. Section \ref{sec:results} shows the results of using the VaR methodology for measuring and comparing risk for clients, advisors, and dealerships. Lastly, Section \ref{sec:discussion} summarizes our results and provides a broader discussion for the impacts of the methodology presented herein.

\section{Context}\label{sec:litrev}

One goal of this paper is to bridge the gap between the ivory tower of academic computational behavioural finance and the real work of the financial agent--a client, advisor, dealership, regulator, or otherwise. This section provides the academic context of how behavioural finance has been applied to understand risk tolerance from questionnaires and trading behaviour, point to real applications of behavioural concepts that have directly benefited practitioners, and elaborate on how our work is helpful to both academics and practitioners. 

\subsection{Behavioural finance: elicited and revealed risk}

In behavioural economics, there is a large body of research dedicated to measuring and understanding the difference between elicited and revealed preferences of consumers, originated by \cite{samuelson48}. In essence, elicited preferences are what people say they prefer, whereas revealed preferences are what they actually prefer, as shown through their behaviours. To apply this concept in finance, elicited and revealed preferences for financial risk were developed, with focuses on estimating the utility of wealth of consumers based on their investing behaviour \citep{dybvig97}, applications of prospect theory \citep{kahneman13}, understanding the perceptions of risk \citep{diacon01}, the effect of cognitive ability on risk preferences \citep{guillemette15}, and--the focus of this paper--measuring risk preferences. 

% Elicited risk
Elicited risk in investment dealerships is typically collected via questionnaires administered in-person yearly by financial advisors to clients. In conjunction with the conversations surrounding the questionnaire, the responses to the questionnaire are used to calculate risk preferences. \cite{finke16} provides a modern review of eliciting risk tolerance from questionnaires; their summative findings include measuring choices surrounding income risk and volatility to assess risk tolerance \citep{guillemette12}, financial literacy effects on risk assessment \citep{linciano12}, and emotional responses to risk and loss aversion \citep{loewenstein01,grable01,michael15}. \cite{wahl20risk} employed a modern questionnaire methodology that measures risk propensity, attitude, capacity, and knowledge to elicit risk tolerance from clients. Our dataset described in Section \ref{sec:datadesc} contains the risk elicited using a questionnaire under the KYC obligation guidelines. A risk score is determined from the questionnaire's responses by calculating a weighted sum based on their answers. That risk score is then transformed by the advisor administering the questionnaire into a profile risk allocation (percentage of assets allocated across risk buckets).  Detailed questionnaire response data is hitherto unavailable, with future plans to exchange and analyze that information with our industry partner to better evaluate a client's elicited risk with guidance from revealed risk.

% Revealed risk:
Revealed risk is typically measured by a client's recovered utility of wealth from their trade and transaction behaviours. Utility of wealth is the idea that income and total wealth affect how a loss is perceived; for example, a \$$10,000$ loss is significant if your total wealth is \$$50,000$ and insignificant if your total wealth is \$$50$ million. This concept is not new; income and total wealth have been shown to affect risk tolerance based on the utility of wealth\citep{samuelson75}. A client's utility of wealth can be recovered from realizations of how they are investing their wealth \citep{dybvig81,dybvig97}.

%VaR
We apply the financial measure of VaR to evaluate both elicited and revealed risk individually. VaR is a quantile of the profit-and-loss distribution--for example, fixing a time horizon of one year, the 99\% VaR is the minimum loss you should expect on your worst day out of one hundred. This concept is well-known in financial management, where it is explained in depth in \cite{jorion07} with many different strategies for implementation \citep{kuester06}.  In this paper, VaR is used to accomplish three tasks: (1) quantify elicited risk, (2) quantify revealed risk, and (3) quantify the discrepancy between elicited and revealed risk. 

% Where does this research fit into literature?
Previous research bridging the gap between elicited and revealed has been conducted to understand risk preferences for crop producers \citep{sharma16}. \cite{corter06} used a weighted sum score methodology to measure revealed risk and compared it with elicited risk measured with scores from a risk tolerance questionnaire. They found that higher risk preferences in the questionnaire were positively correlated with higher risk sum scores. However, our interest is not in reaffirming the correlation between elicited and revealed risk but in quantifying the \textit{difference} between the elicited and revealed preferences. In our approach, VaR is a risk comparison methodology that allows the user to calculate elicited and revealed risk using the same criteria for comparison. While part of our contribution to the literature is using VaR to estimate clients' risk preferences, our main contribution is measuring the difference in elicited and revealed risk tolerance.

\section{Data description}\label{sec:datadesc}

The dataset used in this paper is drawn from the same database provided by a private investment dealership discussed in \cite{thompson20}. The dataset in this paper spans from March 29th 2019 to August 12th 2019, and consists of daily KYC information (age, gender, annual income, risk tolerance) and financial information (account type, total value of that account, proportion of total value invested into different securities) for $50,980$ accounts of $27,644$ clients and 311 advisors. The KYC information is collected via a questionnaire when the account is opened and is updated at least yearly, and the financial information is updated each business day. All identifying data has been anonymized, and individuals will not be identified or referenced in this paper. Also, any subset of the data cannot be publicly shared. Table \ref{tbl:clientDetails} shows the variables of the data from \cite{thompson20} that are considered in this paper. 
\begin{table}[!htbp]
\centering
\caption{Dataset details on clients' KYC information.} 
\label{tbl:clientDetails}
\begin{tabular}{p{2cm}|p{8cm}|p{2.5cm}|p{4cm}}
Variable           & Summary     & Data type & Example values \\ \hline
Account type & Canadian account classifications that have different benefits & Categorical & Cash, LIRA, RDSP, RESP, RIF, RSP, TFSA, Margin\footnotemark \\ \hline
Advisory type & Where the advisor has the discretion to trade for the client & Categorical & Discretionary, non-discretionary, or unknown  \\ \hline
Age                & Ages range from 18 to 98 years old, with average at 57.4 years   & Continuous  & 31 years old \\ \hline
Annual income      & Gross annual income in CAD & Continuous & Multiples of \$$100$ between $\$0$ and $\$15,000,000$ inclusive \\ \hline
Gender             &  $50.5\%$ male and $49.5\%$ female  &  Indicator & M,F               \\ \hline
Investment knowledge & The self-reported investment knowledge of 2.5\% sophisticated, 44.7\% good, 35.2\% fair, or 17.6\% poor &
Ordinal & $1$, $2$, $3$, or $4$ \\ \hline
Marital status     & 67\% married, 18\% single, 11\% unknown and 4\% divorced & Categorical & M,D,S, or \mbox{*}               \\ \hline
Number of accounts & Clients can have more than one account & Ordinal & $1$,$2$,$3$,$\ldots8$                 \\ \hline
Residency          &  Province or Country or Region, with approximately $65\%$ from Ontario & Categorical & ON, MB, AB, $\ldots$ \\ \hline
Retirement indicator & The client's retirement status with 73.9\% not retired, 18.2\% retired, and 7.9\% unknown & Indicator & Yes, No 
\end{tabular}
\end{table}
\footnotetext{Account types are cash, locked-in retirement account (LIRA), registered disability savings plan (RDSP), registered education savings plan (RESP), retirement income fund (RIF), retirement savings plan (RSP), tax-free savings account (TFSA), and margin accounts.}

%Discretionary structure
The new data in this paper that was previously unavailable is the discretionary information of the client-advisor relationship. Advisors associated with the investment dealership have direct control over the securities owned in each account. Clients cannot make direct changes to the accounts without going through their financial advisor. There are two types of advisory relationships: (i) advisors have full discretion on trading for the client's accounts, or (ii) advisors must have trades approved by the client. In our dataset, there are $4,423$ discretionary accounts, $44,712$ non-discretionary accounts, and $1,845$ unknown. Additionally, there exist accounts in the dataset that the advisor personally owns, but that information is currently unavailable due to anonymization. Henceforth, we will treat advisor accounts as client accounts since we are informed that the dealership expects advisors to trade on their accounts, similar to how they would trade for their clients. In fact, internal auditors at the dealership monitor advisor trades to ensure that they are offering trades to or conducting trades for clients before trading the same securities in the advisor's personal accounts, where regulators mandate this behaviour. 

%Cleaning
% Due to a structural aberration in data collection, July 31st 2019 was removed from the data due to large amounts of missing information. Other variables that include missing data are $\ldots$. 

\begin{figure}
    \centering
    \includegraphics[width=10cm]{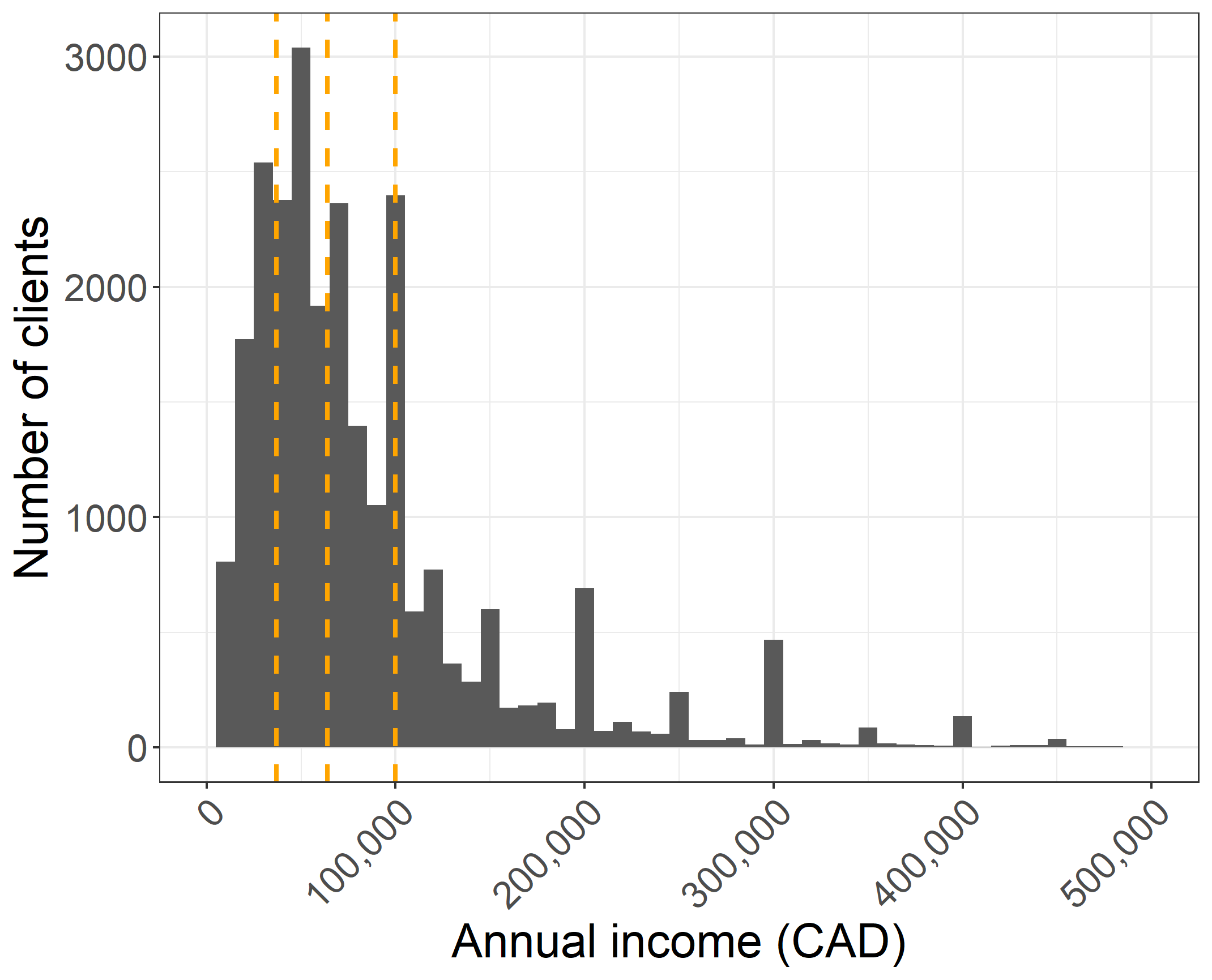} 
    \caption{The distribution of annual incomes for clients on August 12th 2019, with a binwidth of \$$10,000$. The three vertical dashed lines represent the 25th, 50th, and 75th percentiles. There are 286 clients not pictured with annual incomes greater than \$$500,000$, with a maximum annual income of \$$15,000,000$.}
    \label{fig:incomedistribution}
\end{figure}

\begin{figure}
    \centering
    \includegraphics[width=12cm]{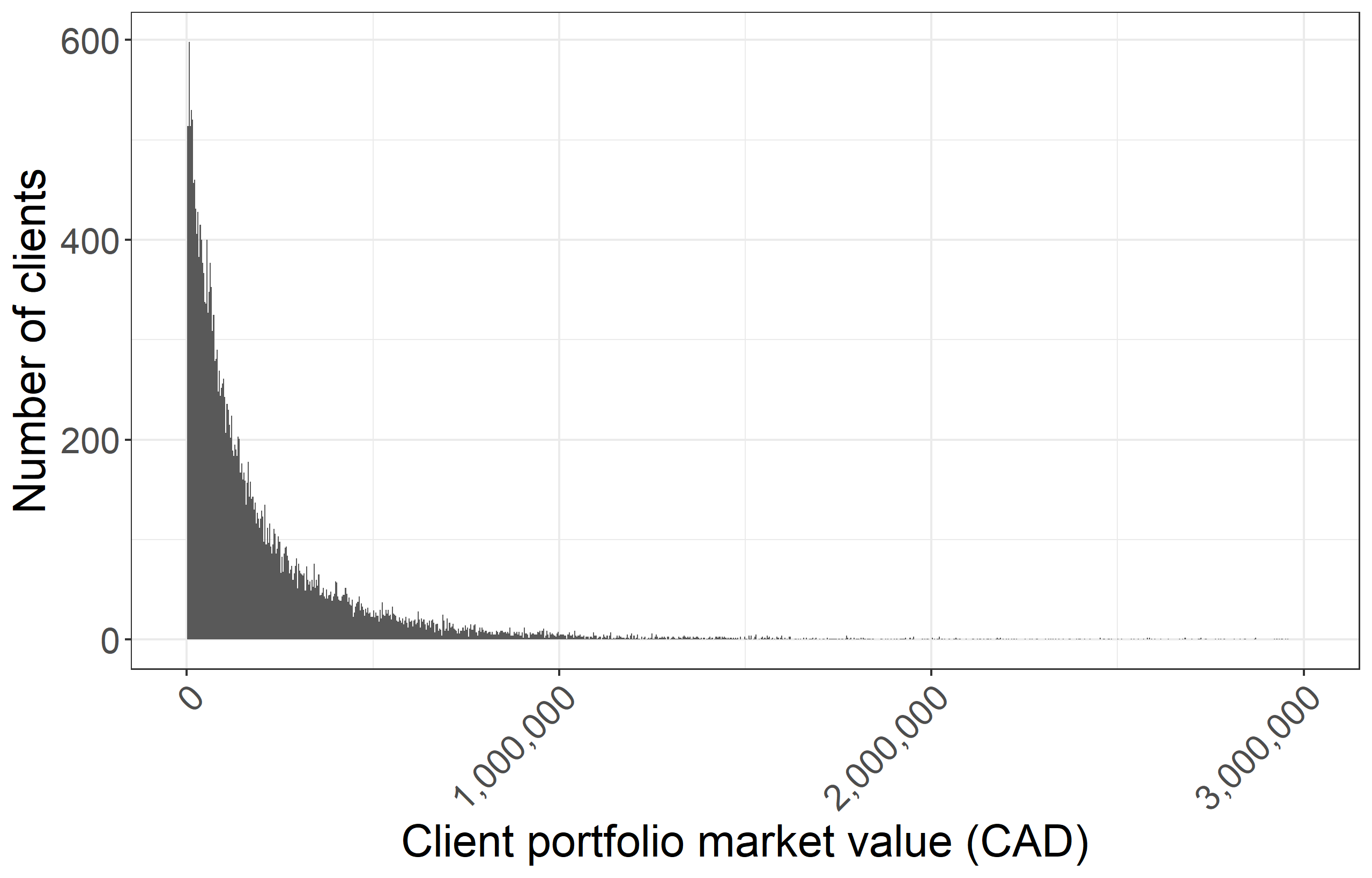}
    \caption{The distribution of the total market value for client portfolios on August 12th 2019, with a binwidth of \$$10,000$. There are 74 clients not pictured with portfolio market values greater than \$$3,000,000$, with a maximum of \$$62,684,999$.}
    \label{fig:marketValue}
\end{figure}

%Plots and graphs of the data
The distribution of account residency is shown in Table \ref{tbl:clientRes}, with the majority of accounts owned by clients in the province of Ontario. Figure \ref{fig:incomedistribution} shows a graph of the annual incomes where we can see that 25\% of all clients earn \$$37,000$ or less, 50\% earn \$$64,000$ or less, and 75\%  earn \$$100,000$ or less. There are income spikes at multiples of \$50k, and there were also 286 clients that earned \$$500,000$ or more. Figure \ref{fig:marketValue} shows the total market values of client portfolios, where 25\% of all clients own assets valued at \$$44,041$ or less, 50\% earn \$$113,147$ or less, and 75\%  earn \$$262,099$ or less.
\begin{table}[!htbp]
\caption{Distribution of residency for client accounts. Locations are Ontario (ON), British Columbia (BC), Alberta (AB), Manitoba (MB) and Nova Scotia (NS).}
\label{tbl:clientRes}
\centering
\begin{tabular}{rrrrrrrrrr}
  \hline
Location & ON & BC & AB & MB & NS & Other \\ 
  \hline
Percentage & 65.36 & 13.85 & 12.80 & 4.07 & 2.17 & 1.75 \\ 
   \hline
\end{tabular}
\end{table}
%market value

\begin{figure}
    \centering
    \includegraphics[width=10cm]{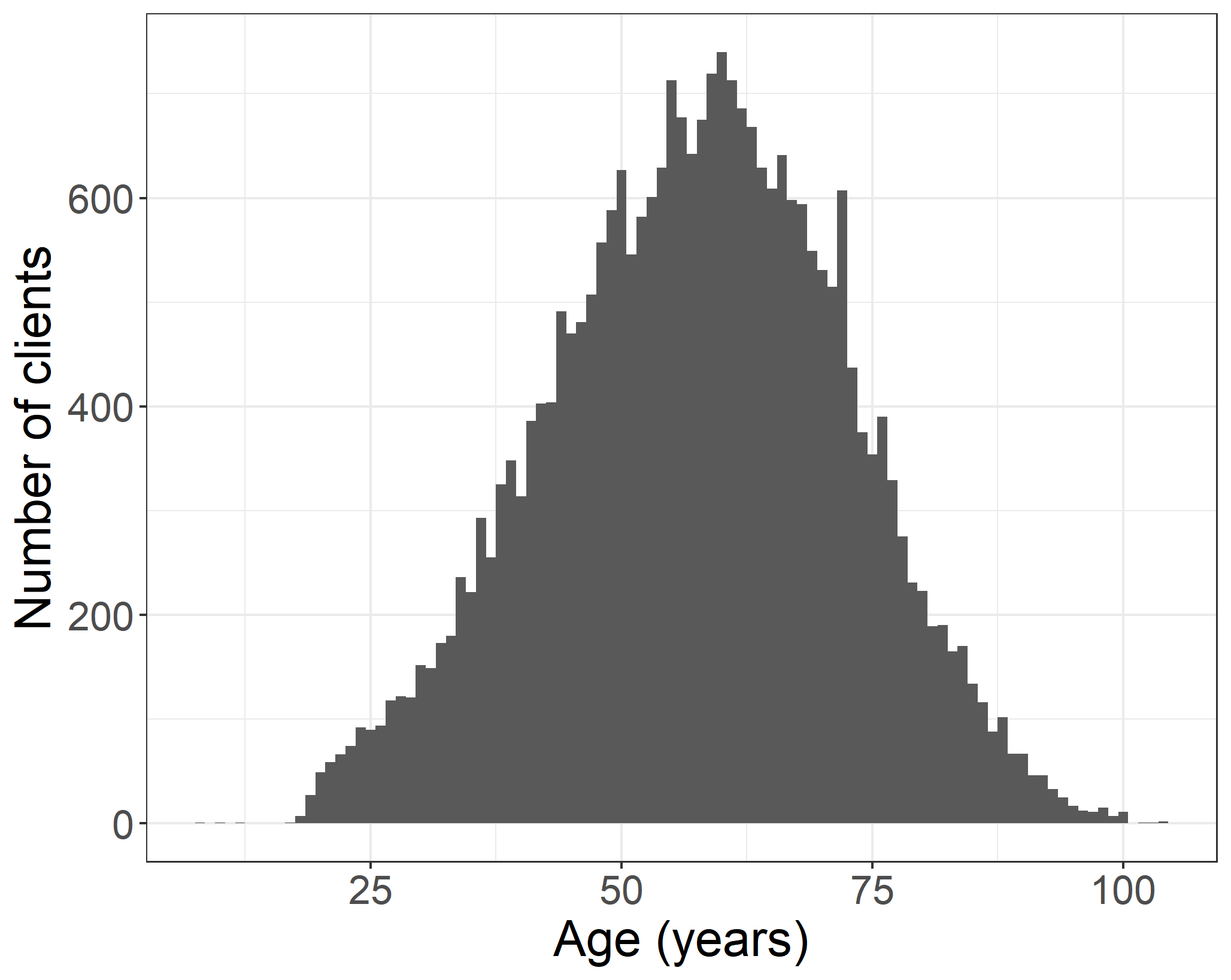}
    \caption{The distribution of client ages, where each bin contains one year.}
    \label{fig:ageDistribution}
\end{figure}

%Graph on age distribution
Figure \ref{fig:ageDistribution} shows clients' ages, where the distribution is unimodal, centred at 57.5 years, has a standard deviation of 14.8 years, and is very slightly left-skewed. The minimum age is 18 years--the legal age to open an account in Canada--and the maximum is 104. Table \ref{tbl:numberOfAccountsUniqueClient} shows the number of accounts for clients, where 77.3\% of clients have two or fewer accounts. Table \ref{tbl:accountTypes} shows the account types, where most own RSP and TFSA accounts.
%Table Number of accounts and account type
\begin{table}[!htbp]
\centering
\caption{The number of accounts owned by clients.}
\label{tbl:numberOfAccountsUniqueClient}
\begin{tabular}{rrrrrrrrr}
  \hline
Unique accounts &   1 &   2 &   3 &   4 &   5 &   6 &   7 &   8 \\ \hline 
Number of clients & 13112 & 8246 & 4380 & 1457 & 330 & 87 &  20 &  12  \\ 
   \hline
\end{tabular}
\end{table}
\begin{table}[!htbp]
\caption{Distribution of account types.}
\label{tbl:accountTypes}
\centering
\begin{tabular}{rrrrrrrrrr}
  \hline
Account type & Cash & LIRA & RDSP & RESP & RIF & RSP & TFSA & Margin \\ 
  \hline
Number of accounts & 7843 & 129 & 204 & 2975 & 6169 & $19,979$ & $11,080$ & 944  \\ 
   \hline
\end{tabular}
\end{table}
In addition, we also have data on the elicited and reveal risk distributions. This data is either collected during the KYC questionnaire (elicited) or observed from trading behaviours (revealed), and is discussed in the next section. 

\subsection{Profile risk allocation and portfolio risk selection}
The dealer's method to satisfy the KYC risk tolerance obligation results in their clients' risk tolerance represented as a set of five percentages, with each percentage associated with the following categories: low, low-medium, medium, medium-high, and high risk. The percentages sum to 100, so all risk tolerance is allocated. From here on, we will refer to these percentages, elicited from a client, as the \textit{profile risk allocation}, or simply profile risk. Each security available for purchase from the dealer is evaluated to have a KYP risk rating classified into one of the five risk categories. An investor's familiarity with a security's brand plays a key role in the optimism of the financial return of an investment in that security \citep{aspara13}, and unfamiliar securities have a higher perception of risk with lower return \citep{ganzach00}. Security risk ratings provide numerical guidance to investors on the realistic risk of an investment.  Our dataset's risk ratings are based on a weighting of the standard deviation of the security price, maximum drawdown, downside deviation, and correlations with other securities. In this paper, the revealed risk distribution on any given day is the percentage of total market value held in each of the five risk categories. Henceforth, it will be referred to as the \textit{portfolio risk selection}, or simply portfolio risk.

\subsection{Clusters} \label{sec:clusters}
Our previous work grouped client behaviours--measured by a recency, frequency, monetary and profile (RFMP) model--though a $k$-prototypes machine learning clustering algorithm \citep{thompson20}. In that work, we found the optimal number of clusters to be five using the silhouette coefficient and the Davies–Bouldin (DB) score. The cluster attributes yield five personas:
\begin{itemize}
\setlength\itemsep{0em}
\item Cluster 1 -- active investors who trade frequently, in large amounts and appear sensitive to market influences,
\item Cluster 2 -- younger savers who make regular, smaller deposits using automated platforms such as preauthorized chequing (PACs) and dollar cost averaging,
\item Cluster 3 -- ``just in time'' traders who make infrequent trades at seemingly random intervals,
\item Cluster 4 -- older investors who make regular withdrawals and cash out dividends and interest payments, and
\item Cluster 5 -- systematic savers who make larger trades but make use of automation for predictable deposits and re-balancing.
\end{itemize}
In this paper, we further analyze those clusters to investigate whether the methodologies presented herein agree with our previous work. We note that risk tolerance was not an input into the clustering algorithm, and previous analysis on the profile risk allocation of the clusters showed little difference between clusters. 

\section{Methodology}\label{sec:discrepancies}
Advisors allocate their clients' wealth to different recommended proportions of risk categories, and they select particular assets within these categories. Once set up, advisors continuously aid clients in modifying their portfolio as markets evolve, based on their stated financial goals. Advisors strive to construct portfolios that are consistent with stated goals, and thus one should be able to quantify how close an advisor is coming to the stated goal. Otherwise, stated goals are inconsequential. Advisors pick assets that are classified into risk categories--mandatory for all financial institutions. In our dataset, each asset is classified into one of five risk categories. It is natural to assume that when an advisor sets up a clients' proportional risk allocation, the advisor will ensure that their assets line up with that risk allocation. 

A main issue is quantifying the difference between (i) a client's prescribed KYC profile risk and (ii) their actual portfolio risk. A risk allocation or selection is represented by an ordered collection of five percentages associated with each risk category (Low, Low-medium, Medium, Medium-high, High), where the percentages sum to one. An example portfolio risk selection is (0.2, 0.1, 0.7, 0, 0) with 20\% of wealth allocated to the low risk bucket, 10\% to low-medium, and 70\% to medium. Readers familiar with linear algebra might (correctly) recognize that we can view an allocation or selection as a point in five-dimensional Euclidean space. In this case, quantifying discrepancy is as simple as computing the Euclidean distance between two points.  Readers familiar with probability theory might (correctly) recognize that we can view an allocation or selection as a probability distribution, in which case quantifying discrepancy would be possible using well-known measures such as Kullback-Liebler Divergence.  Unfortunately, neither of these approaches leads to meaningful comparisons to advisors or their clients. See Appendix A for a detailed investigation of metrics and divergences used for risk category comparisons.  In other words, measuring the discrepancy in a mathematically valid way is straightforward, but doing so in a manner that is both mathematically valid \textit{and} financially meaningful, is surprisingly non-trivial.

The well-known financial risk measure Value-at-Risk (VaR) allows us to evaluate and compare risk allocations and selections in a valid and financially informative way. VaR captures the risk tolerance relative to actual market indices, while also incorporating the relative and correlated risk between risk categories. Typically, VaR is quoted in basis points (bps) so that VaR across clients with different total market values can be compared. VaR can provide an actual dollar value estimate of the minimum a client should expect to lose on their worst day out of one hundred, which is an excellent communication tool between advisors and clients. 

To compare profile and portfolio risk, the first step is to estimate the VaR and the second step is to compute the difference in VaRs to compare them. Estimating VaR requires an understanding of the assumptions on how the investor and advisor interpret each risk category, and the statistical properties of the returns of those risk categories. Our analysis makes specific choices to calculate VaR, but it is essential to highlight that the VaR framework is highly customizable by the user.

\subsection{Value-at-risk} \label{sec:valueAtRisk}
\begin{figure}
    \centering
    \includegraphics[width=18cm]{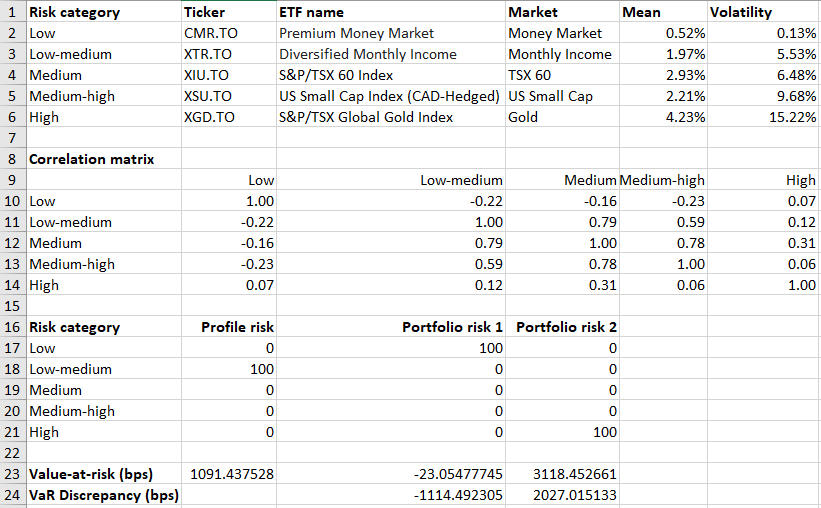} 
    \caption{VaR calculation using Excel spreadsheet software. The spreadsheet is available to readers.}
    \label{fig:excel}
\end{figure}

Using the assets in a portfolio is necessary for the exact calculation of portfolio VaR. However, this approach fails for assigning a profile risk VaR--perceptions and securities held across advisors can be very different. Our solution to calculating profile VaR similarly across clients and advisors is to use an exchange-traded fund (ETF) as a representative investment for each risk category. This methodology is customizable to any set of ETFs or any assets of the user's choosing. The risk representative iShares ETFs\footnote{\url{https://www.blackrock.com/ca/investors/en/products/product-list}} from a given class that had the longest publicly-available data\footnote{\url{https://ca.finance.yahoo.com}} are shown in Figure \ref{fig:excel} with their expected annualized returns, volatility, and correlations\footnote{We have included an Excel spreadsheet \citep{msexcel} that reproduces the VaR calculation, and Appendix B that provides mathematical details of the calculation.}.

The expected annualized return of each ETF and the correlations between them are used to calculate VaR. The VaR calculation gives the minimum expected loss in basis points (bps) on the worst trading day out of one hundred. Figure \ref{fig:excel} shows an example VaR calculation for a single client that is considering two possible portfolio selections. The client has a profile risk of 100\% low-medium, and their risk is a loss of at least 10.91\% of their total market value. They are considering a portfolio VaR of 100\% low, which yields a VaR of -0.23\%--negative means that on your worst day out of one hundred, you expect to gain 0.23\% or less--and a 100\% high account, which yields 31.18\% VaR. The discrepancy between the low and low-medium accounts is -11.14\%, and between low and high accounts is 20.27\%. A negative VaR discrepancy means the portfolio is relatively under-risked and a positive VaR discrepancy means it is relatively over-risked.

\section{Results} \label{sec:results}

In this section, we apply our risk comparison methodology to a real trade and transaction dataset. We show the benefits and drawbacks of using VaR to evaluate and compare profile and portfolio risk.  We start by investigating a single client's performance over time, followed by an advisor's portfolio with a large clientele, and conclude with results at the financial dealership level.

\subsection{Client level}
\begin{figure}
    \centering
    \includegraphics[width=16cm]{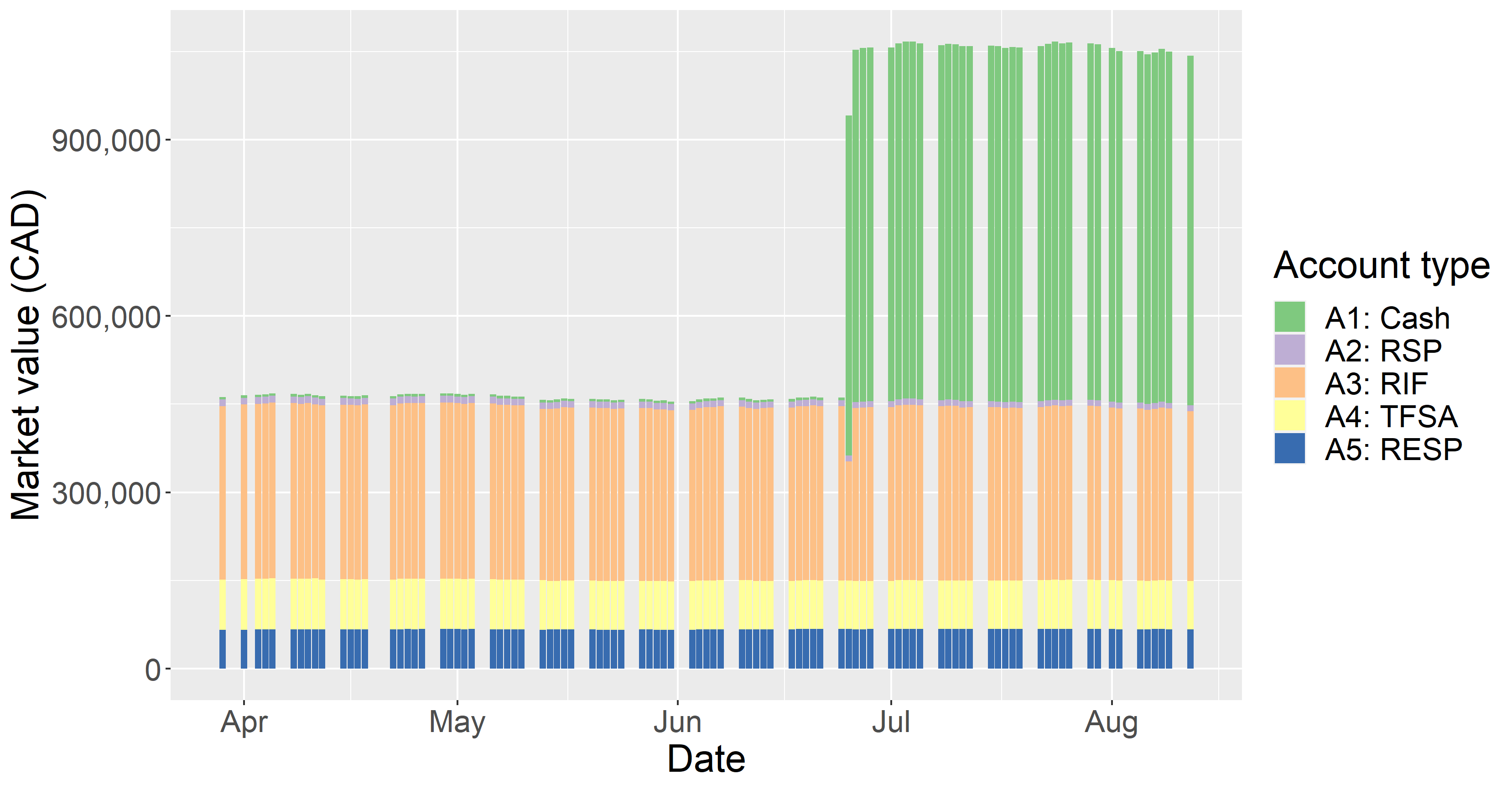} 
    \caption{The portfolio market value of each account of the example client over time.}
    \label{fig:singleClient_accountholding}
\end{figure}

First, we look at how the two methodologies can evaluate the difference between the client's profile and portfolio risk. This investigation will show not only the usefulness of the VaR methodology, but also show possible analyses and visualizations that could appear on a client dashboard. The example client is a 62-year-old investor with five accounts with five types: Cash, retirement savings plan (RSP), retirement income fund (RIF), tax-free savings account (TFSA), registered education savings plan (RESP). The holdings of each account over time is shown in Figure \ref{fig:singleClient_accountholding}, where the account starts with approximately \$$300,000$ in their RIF accounts, \$$75,000$ in each of their TFSA and RESP accounts, and change in each of their RIF, RSP, and Cash accounts. A large addition of approximately \$$600,000$ is made in the fourth week of July to their Cash account. Table \ref{tbl:client} shows the market value, profile VaR, portfolio VaR, and VaR discrepancy for each account on on August 12th 2019, where we can see they have the same KYC profile risk (100\% medium) across accounts. The accounts are all under-risked to their stated goal, shown by the negative differences in VaR.
% latex table generated in R 3.6.3 by xtable 1.8-4 package
% Wed Mar 10 19:39:53 2021
\begin{table}[ht]
\centering
\caption{The example client market values, profile VaR, portfolio VaR, and VaR discrepancy on August 12th 2019 by account type. The averages in the last row are weighted by the market value of each account's holdings.} 
\label{tbl:client}
\begin{tabular}{lrrrr}
  \hline
  Account & Market Value (CAD) & Profile VaR (bps) & Portfolio VaR (bps) & VaR discrepancy (bps) \\
 \hline
A1: Cash & 595152 & 1216 & 971 & -244 \\ 
  A2: RSP & 9883 & 1216 & 1089 & -126 \\ 
  A3: RIF & 288552 & 1216 & 882 & -333 \\ 
  A4: TFSA & 82302 & 1216 & 1089 & -126 \\ 
  A5: RESP & 67028 & 1216 & 1089 & -126 \\ 
   \hline
Average & 208583 & 1216 & 965 & -251 \\ 
   \hline
\end{tabular}
\end{table}

\begin{figure}
    \centering
\includegraphics[width=13cm]{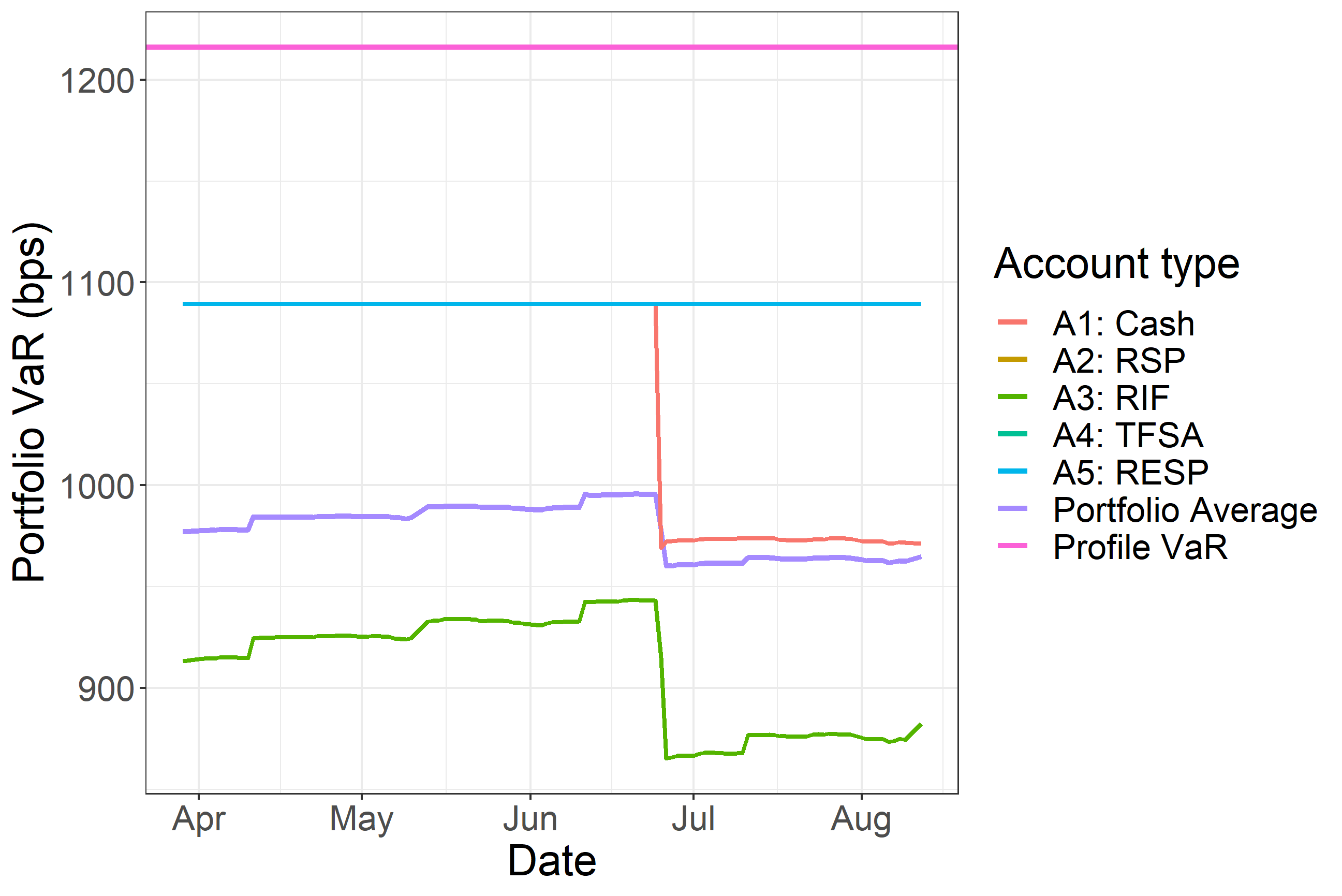} 
\caption{The example client's account portfolio VaRs over time, with constant profile VaR of 1216 bps for all accounts. The unseen TFSA account line is same as the RESP line.}
\label{fig:clientOverTime}
\end{figure}

Figure \ref{fig:clientOverTime} shows the client's profile VaR, individual account portfolio VaRs, and an overall portfolio VaR (weighted by market value) over time. Similar to their account holdings in Figure \ref{fig:singleClient_accountholding}, we see there is a re-balancing of asset risks in their cash and RIF accounts, which affects their overall portfolio average.

\subsection{Advisor level} 

\begin{figure}
    \centering
    \begin{subfigure}[t]{0.49\textwidth}
        \centering
        \includegraphics[width=\linewidth]{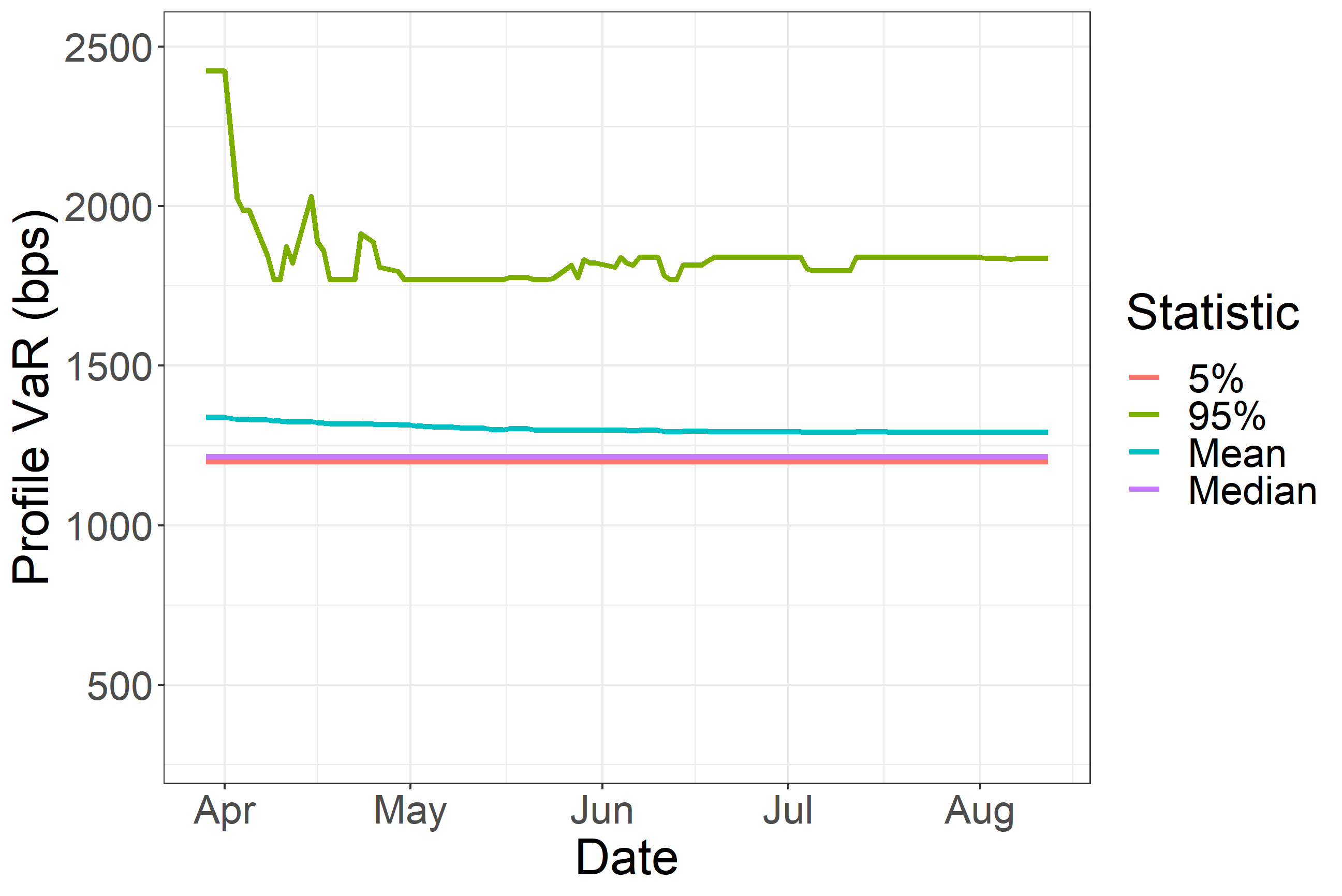}
        \caption{Profile VAR} \label{fig:advisorPrescribedVaR}
    \end{subfigure}
    \hfill
    \begin{subfigure}[t]{0.49\textwidth}
        \centering
        \includegraphics[width=\linewidth]{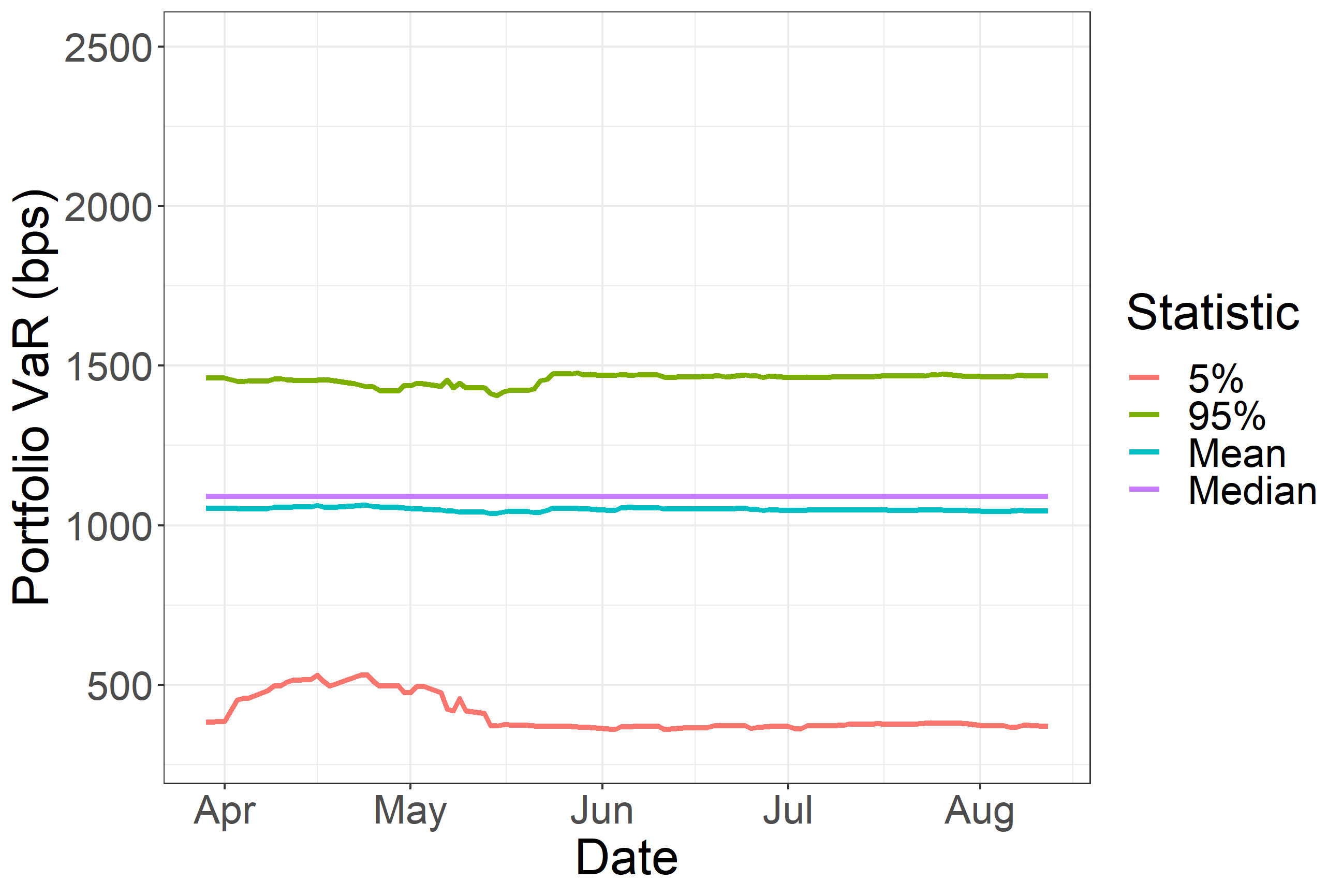} 
        \caption{Portfolio VaR} \label{fig:advisorActualVaR}
    \end{subfigure}
    \vspace{1cm}
    \begin{subfigure}[t]{0.5\textwidth}
        \centering
        \includegraphics[width=\linewidth]{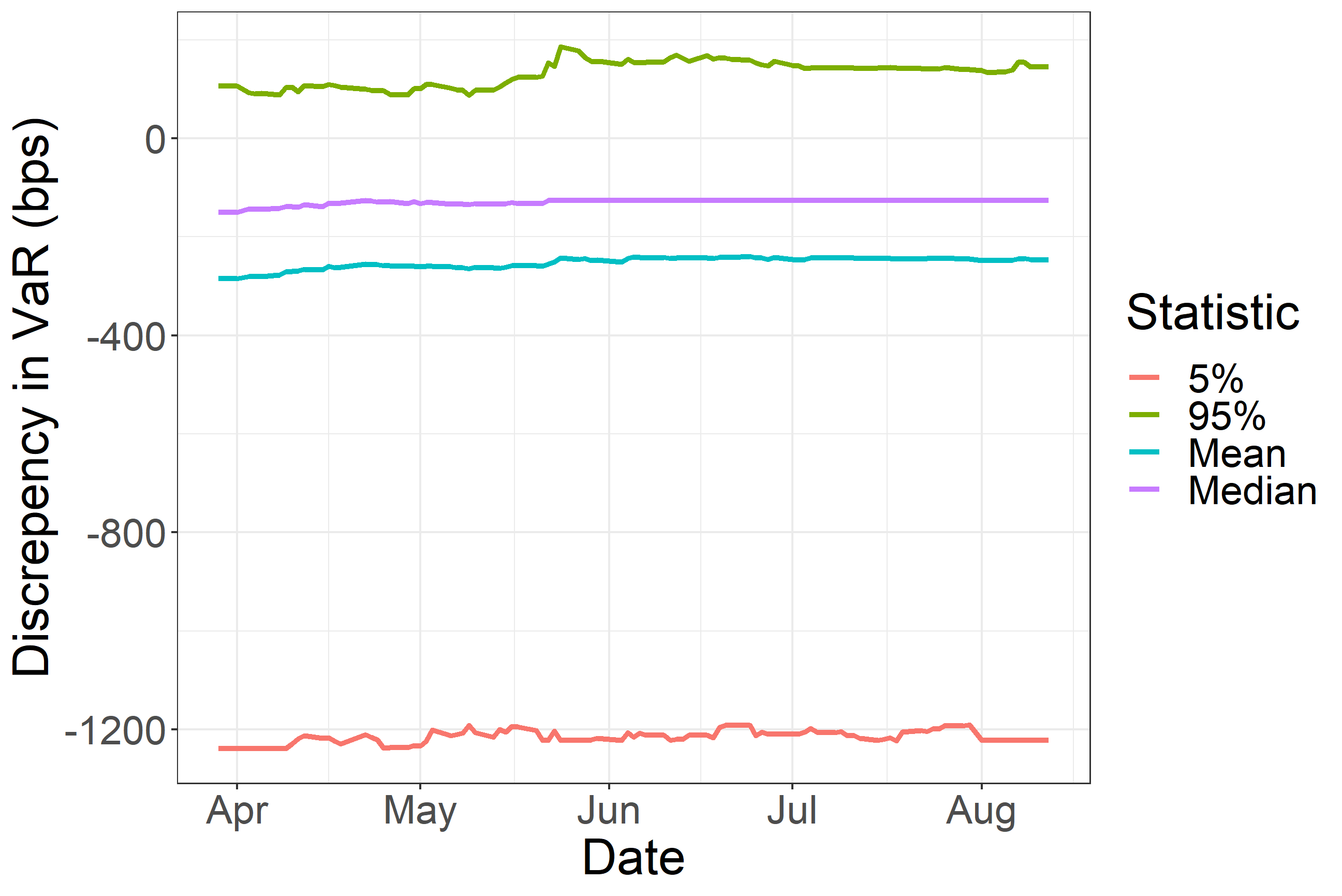} 
        \caption{Discrepancy in VaR} \label{fig:advisorDiscrepancyOverTime}
    \end{subfigure}
     \caption{Profile VaR (upper left panel), portfolio VaR (upper right panel), and the discrepancy between the two VaRs (lower panel) over time for a single advisor . The advisor starts with 283 clients with \$$48,748,731$ CAD in total asset market value on March 29th 2019 and has 662 clients with \$$104,530,928$ in total asset market value on August 12th 2019.}
     \label{fig:singleAdvisor}
\end{figure}

\begin{figure}
    \centering
    \begin{subfigure}[t]{0.49\textwidth}
        \centering
        \includegraphics[width=\linewidth]{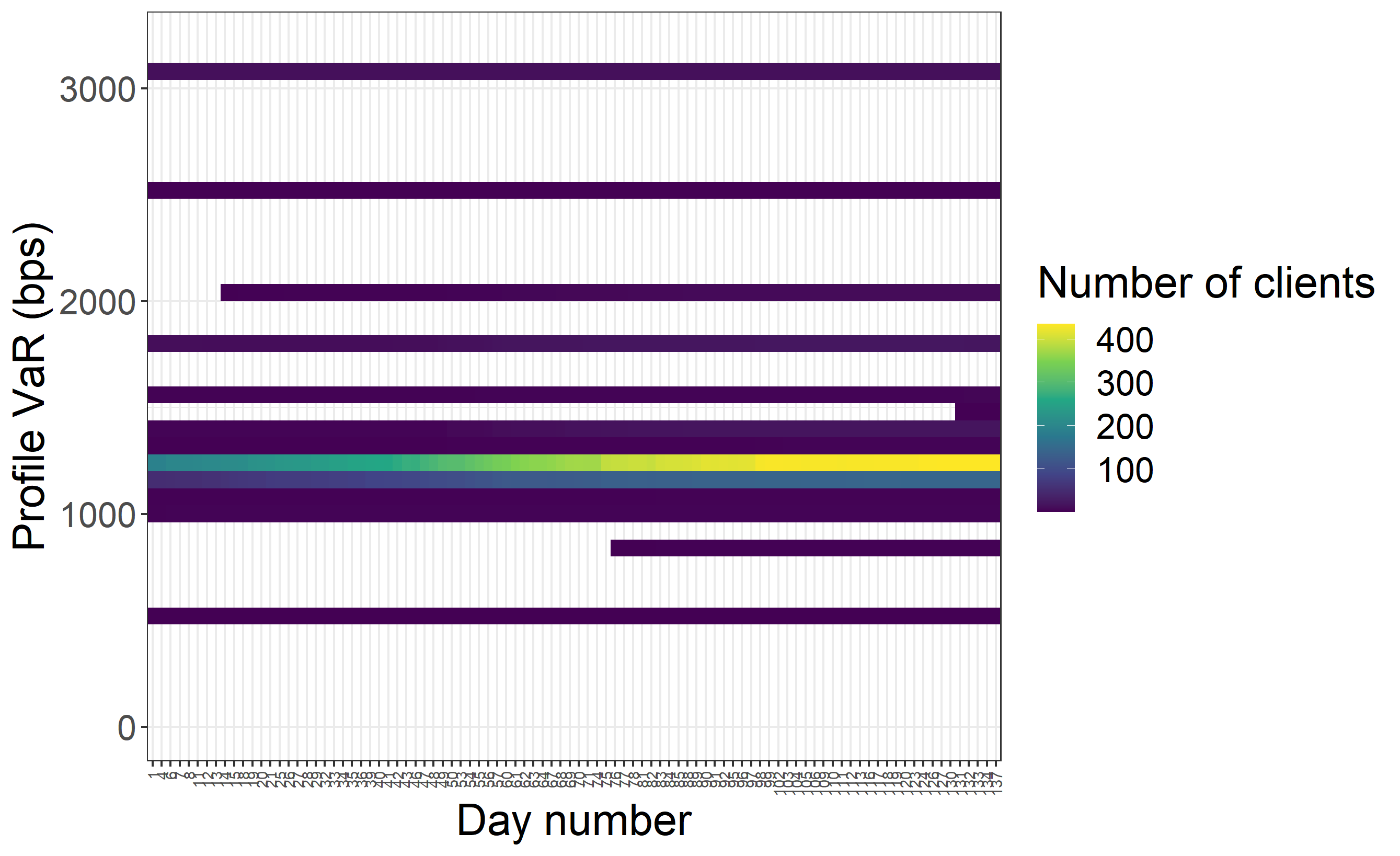}
        \caption{Profile VaR} \label{fig:advisorPrescribedVaRHeatMap}
    \end{subfigure}
    \hfill
    \begin{subfigure}[t]{0.49\textwidth}
        \centering
        \includegraphics[width=\linewidth]{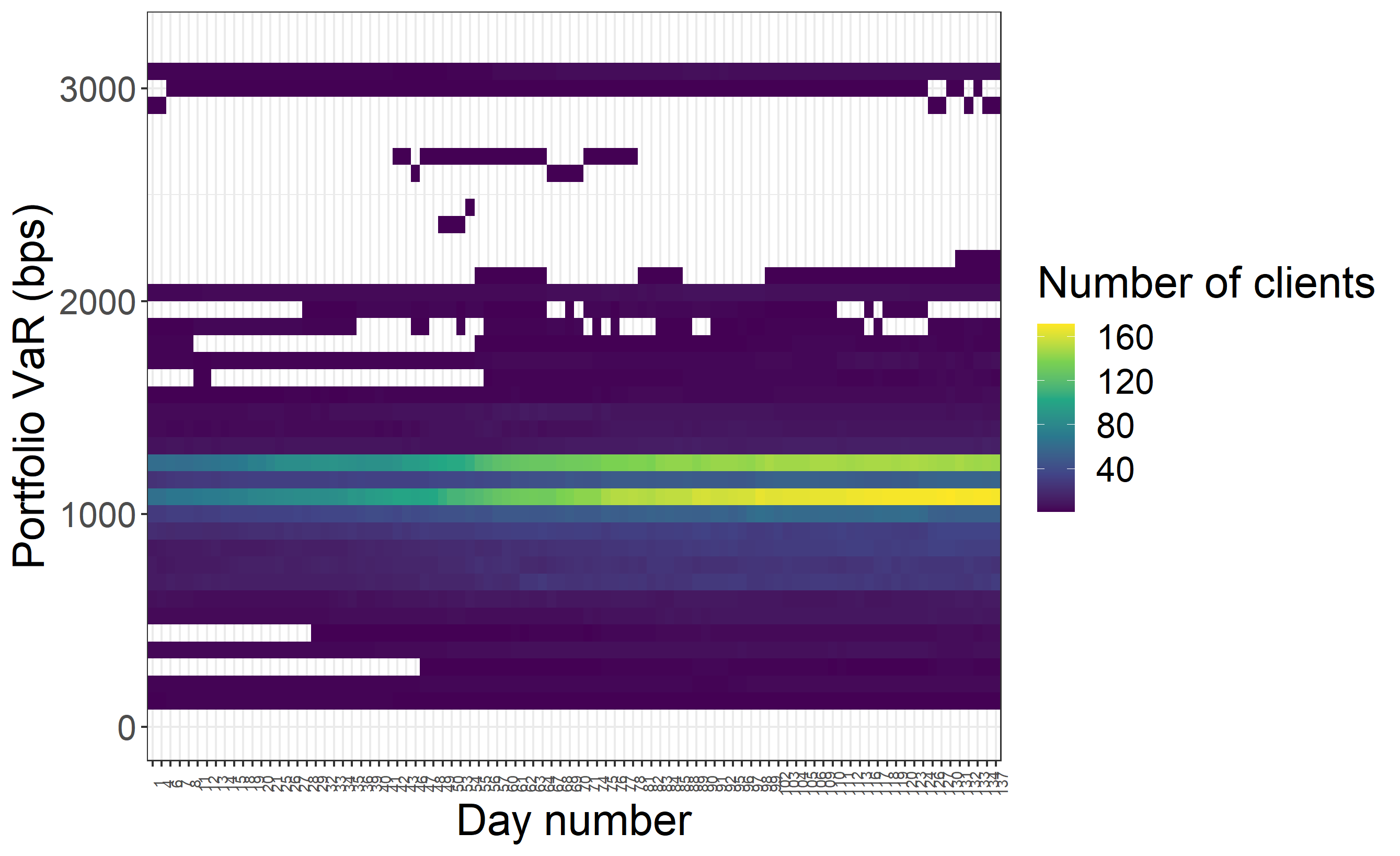} 
        \caption{Portfolio VaR} \label{fig:advisorActualVaRHeatMap}
    \end{subfigure}
    \vspace{1cm}
    \begin{subfigure}[t]{0.45\textwidth}
        \centering
        \includegraphics[width=\linewidth]{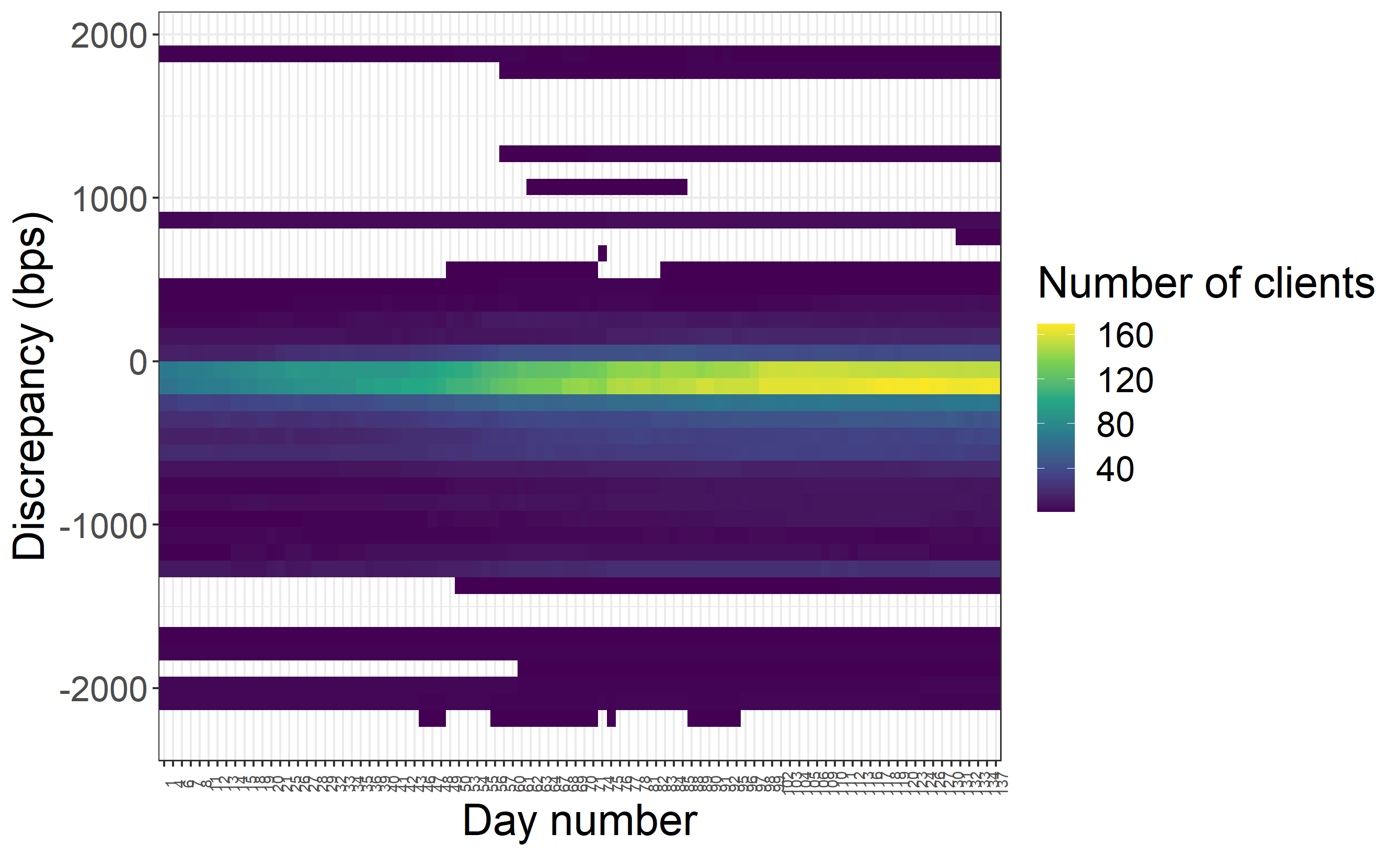} 
        \caption{Discrepancy in VaR} \label{fig:advisorDiscrepancyOverTime2}
    \end{subfigure}
     \caption{Heatmaps showing the distributions of profile VaR (upper left panel), portfolio VaR(upper right panel), and the discrepancy between them (lower panel) over time for a single advisor.}
     \label{fig:singleAdvisorHeatmaps}
\end{figure}

From the individual client view perspective, our methodology allows advisors to gain insight into their clients' risk positions and easily compare to their preferences.  In addition, advisors can gain insights on all clients across their firms by looking at the average metrics across clients and the variation of each metric. Figure \ref{fig:singleAdvisor} shows the advisor with the fifth-most number of clients (670) with 1131 accounts over the time period of the dataset. Figure \ref{fig:advisorPrescribedVaR} shows the mean and median of all the advisor's client's profile VaR over time, where the majority of clients are prescribed less than 1250 bps at risk on any given day. Figure \ref{fig:advisorActualVaR} shows that the majority of clients have less than 1150 bps at risk in the portfolio VaR, and Figure \ref{fig:advisorDiscrepancyOverTime} shows that most clients are under-risked with a discrepancy less than 0. 
A closer look into VaR distributions for clients is shown using heatmap plots in Figure \ref{fig:singleAdvisorHeatmaps}. The distribution of profile VaR is shown in Figure \ref{fig:advisorPrescribedVaRHeatMap} which shows that over time, most clients are being put into a 100\% medium account shown by the bright yellow strip around $1,110$ bps. The distribution of the portfolio VaR in Figure \ref{fig:advisorActualVaRHeatMap} shows that the advisor appears to place each client into two swim lanes--one group is at essentially 100\% medium risk, and the other group is slightly below it. This is reflected in the discrepancy in Figure \ref{fig:advisorDiscrepancyOverTime2} where there is a spread of clients below zero discrepancy. 

\begin{figure}
    \centering
    \includegraphics[width=14cm]{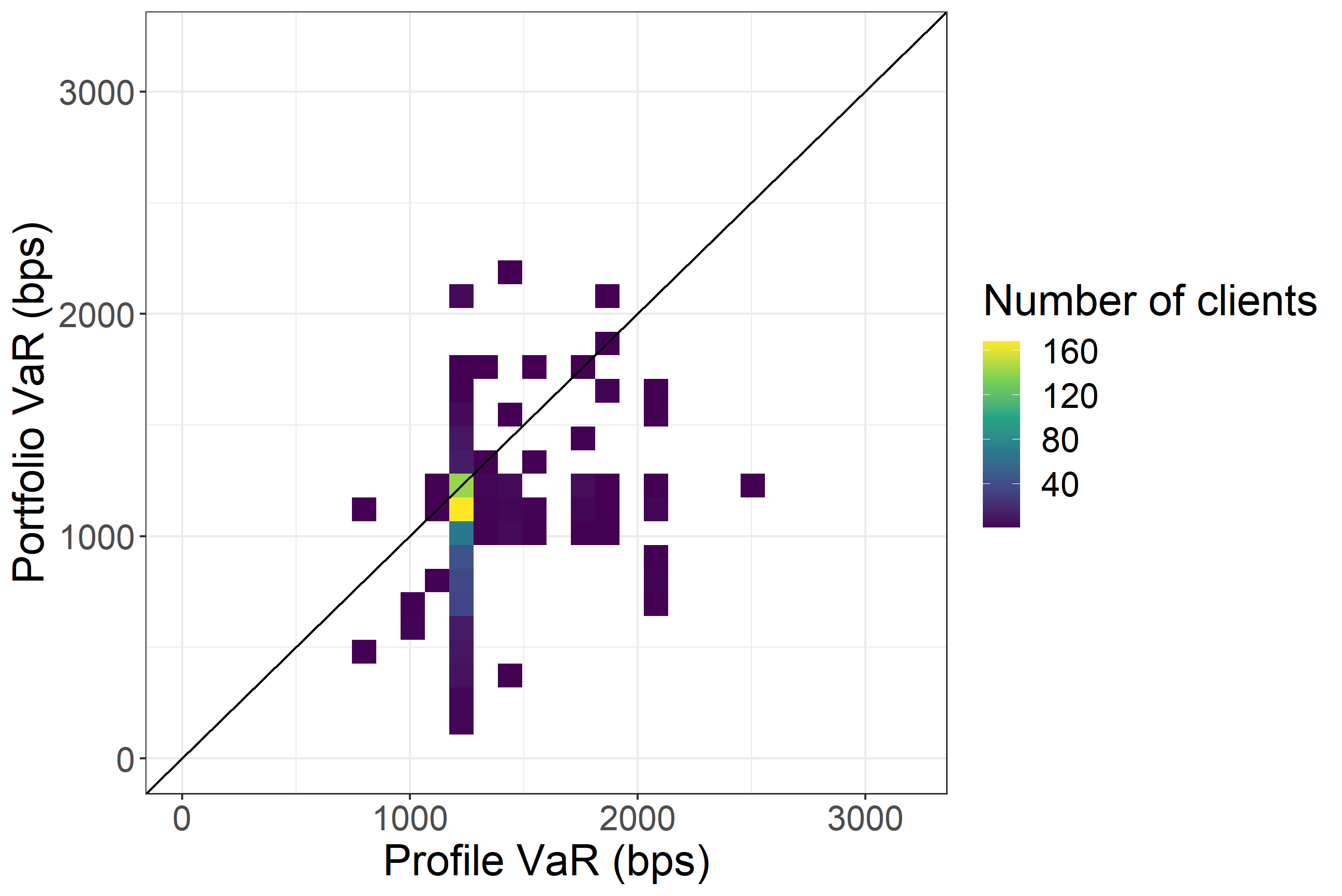}
    \caption{Heatmap of the two-dimensional distribution of portfolio VaR versus profile VaR on August 12th 2019. The diagonal black line represents the points at which a client has equal profile and portfolio VaR.}
    \label{fig:portfolioVsProfile}
\end{figure}

\begin{figure}
    \centering
    \begin{subfigure}[t]{0.49\textwidth}
        \centering
        \includegraphics[width=\linewidth]{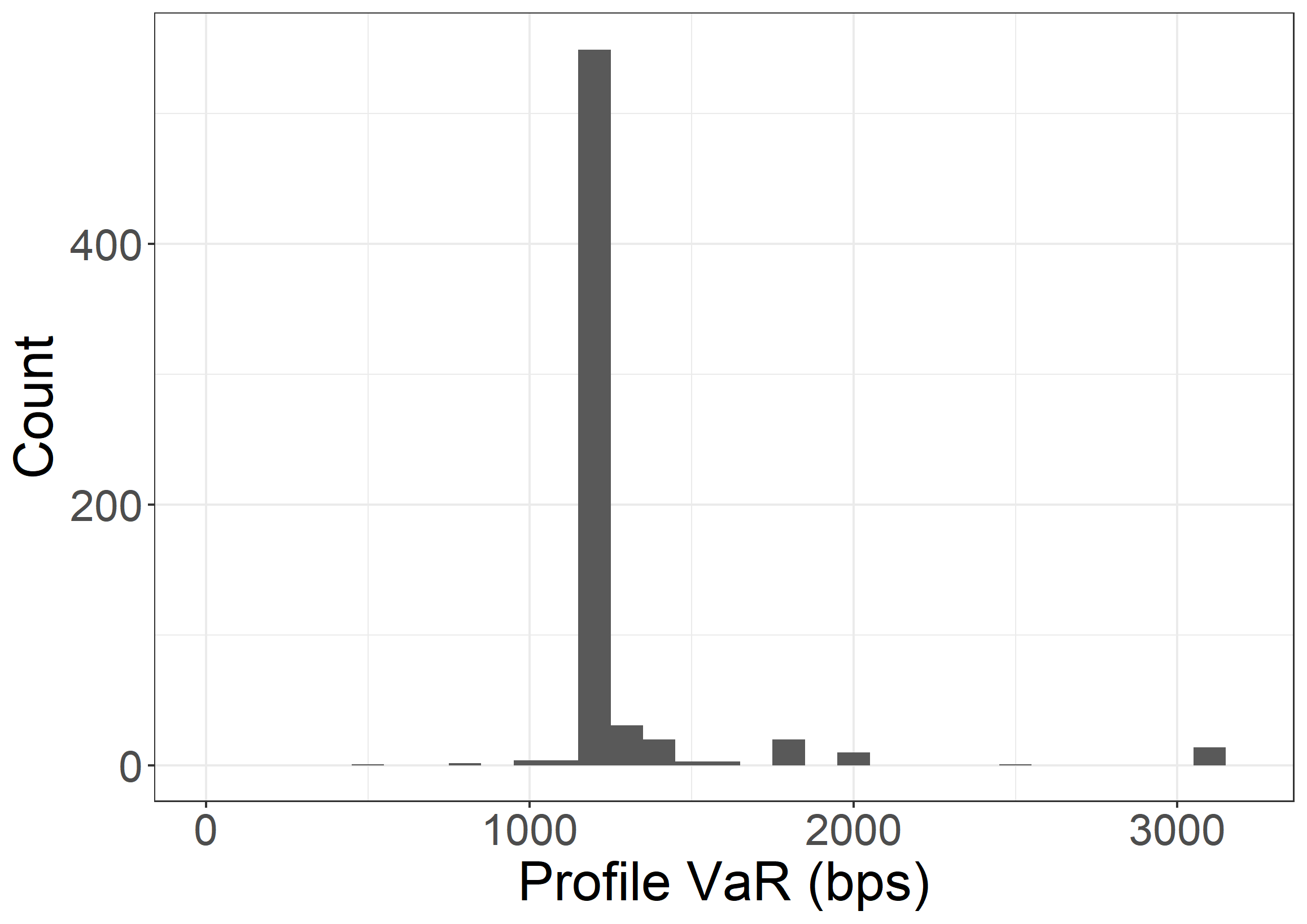}
        \caption{Profile VaR} \label{fig:advisorPrescribedVaRDistro}
    \end{subfigure}
    \hfill
    \begin{subfigure}[t]{0.49\textwidth}
        \centering
        \includegraphics[width=\linewidth]{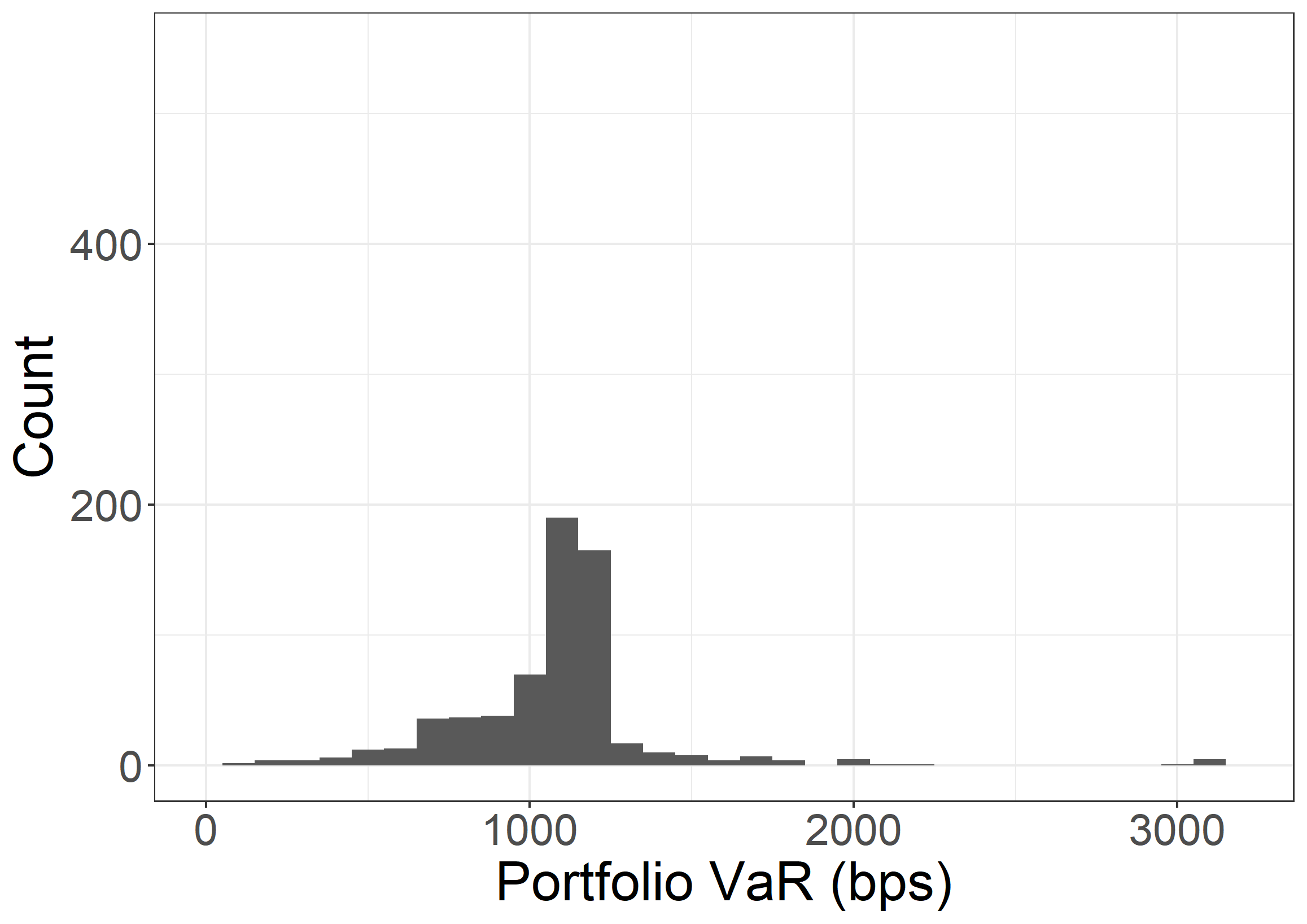} 
        \caption{Portfolio VaR} \label{fig:advisorActualVaRDistro}
    \end{subfigure}
    \vspace{1cm}
    \begin{subfigure}[t]{0.45\textwidth}
        \centering
        \includegraphics[width=\linewidth]{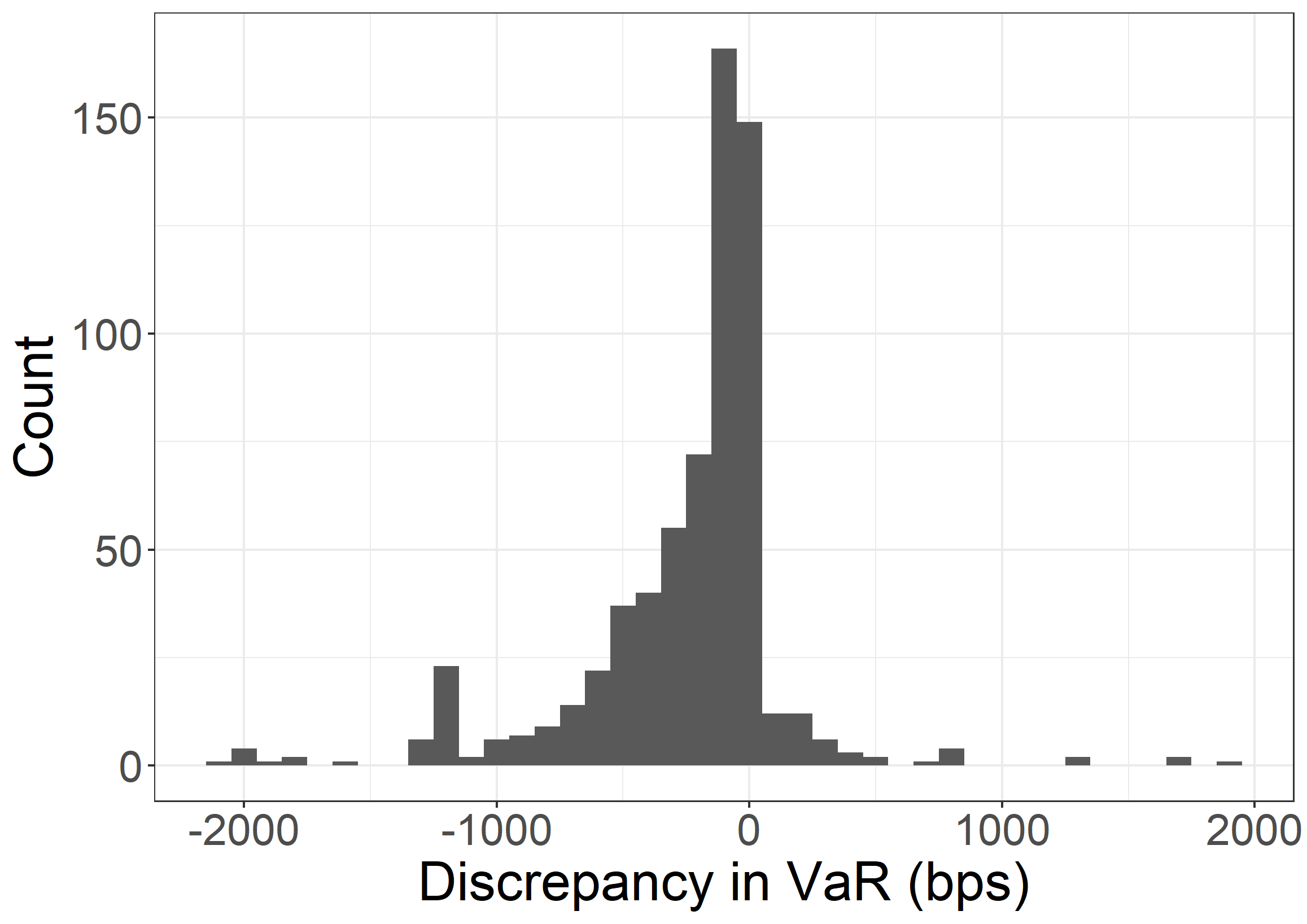} 
        \caption{Discrepancy} \label{fig:advisorDiscrepancyDistro}
    \end{subfigure}
     \caption{The single day distributions of profile VaR (upper left panel), portfolio VaR (upper right panel), and the discrepancy between them (lower panel) for a single advisor's clientele on August 12th 2019.}
     \label{fig:singleDayDistributions}
\end{figure}

Figure \ref{fig:portfolioVsProfile} shows the two-dimensional distribution of portfolio and profile VaR, where a large proportion of clients are close to the VaR equivalency line. However, there exist a significant number of clients who have the same profile risk (100\% as mentioned before), which have a portfolio VaR at (approximately 120 clients) and lower (approximately 180 clients) than the profile VaR.
Figure \ref{fig:singleDayDistributions} shows the distribution of VaR on August 12th 2019, where we can specifically see the clear increase in the standard deviation from the profile VaR in Figure \ref{fig:advisorPrescribedVaRDistro} to the portfolio VaR in Figure \ref{fig:advisorActualVaRDistro}. The bias to being under-risked for the advisor's clientele is shown by the vast majority of the distribution of discrepancies below zero in Figure \ref{fig:advisorDiscrepancyDistro}.
\subsection{Dealership level}
\begin{figure}
    \centering
    \begin{subfigure}[t]{0.45\textwidth}
        \centering
        \includegraphics[width=\linewidth]{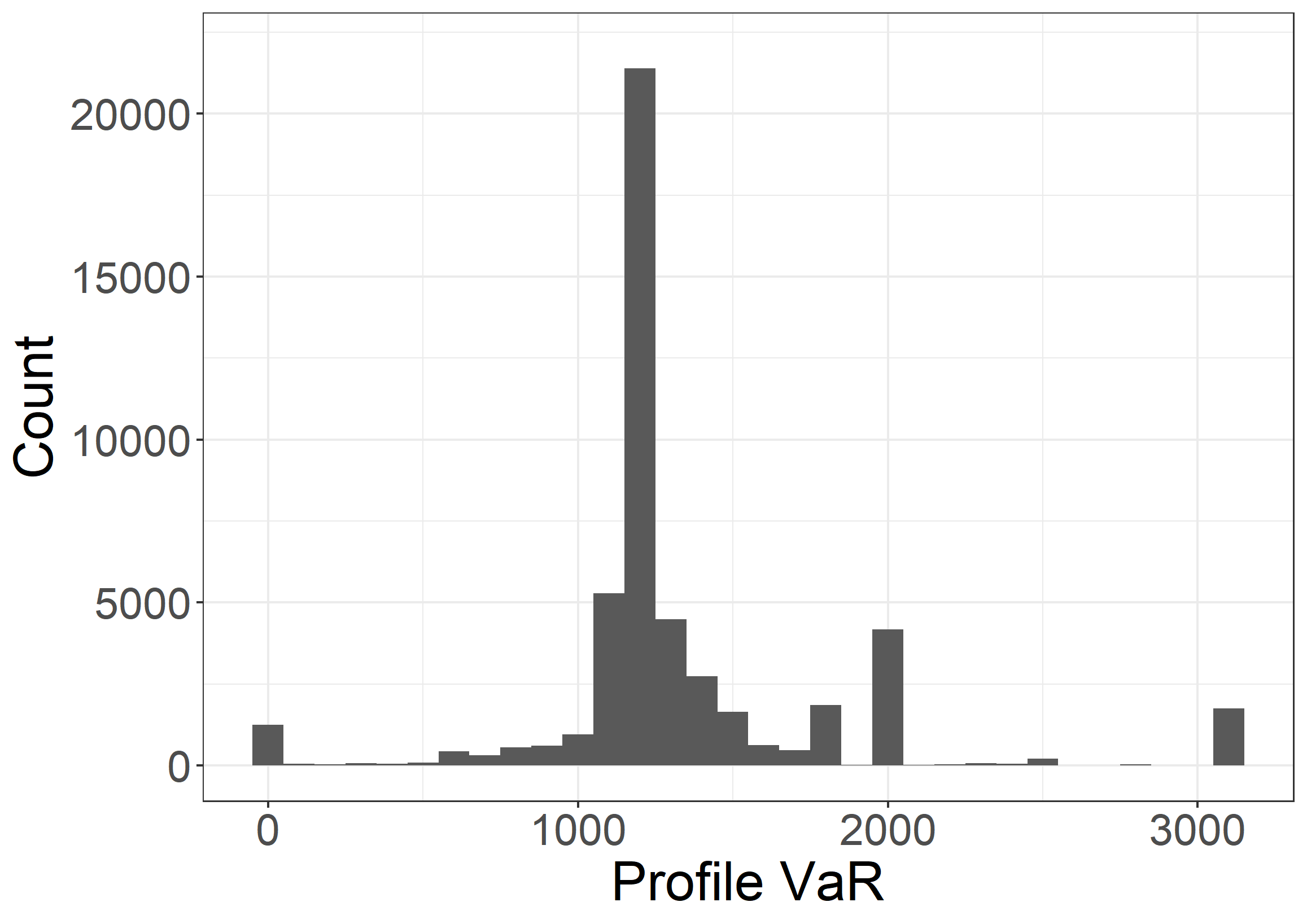}
        \caption{Profile VaR} \label{fig:prescribedVaRcross}
    \end{subfigure}
    \hfill
    \begin{subfigure}[t]{0.45\textwidth}
        \centering
        \includegraphics[width=\linewidth]{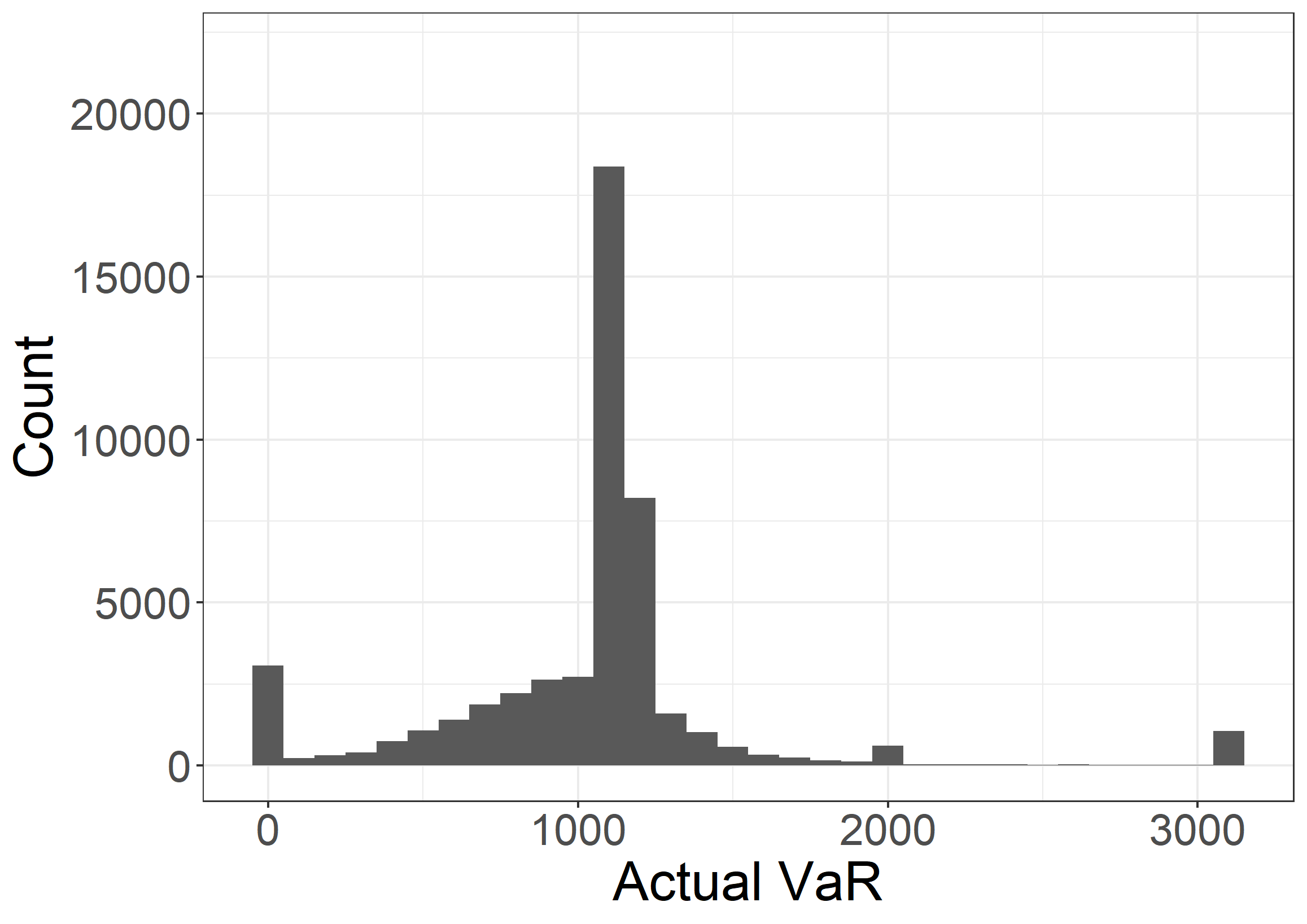} 
        \caption{Portfolio VaR} \label{fig:actualVaRcross}
    \end{subfigure}
    \vspace{1cm}
    \begin{subfigure}[t]{0.45\textwidth}
        \centering
        \includegraphics[width=\linewidth]{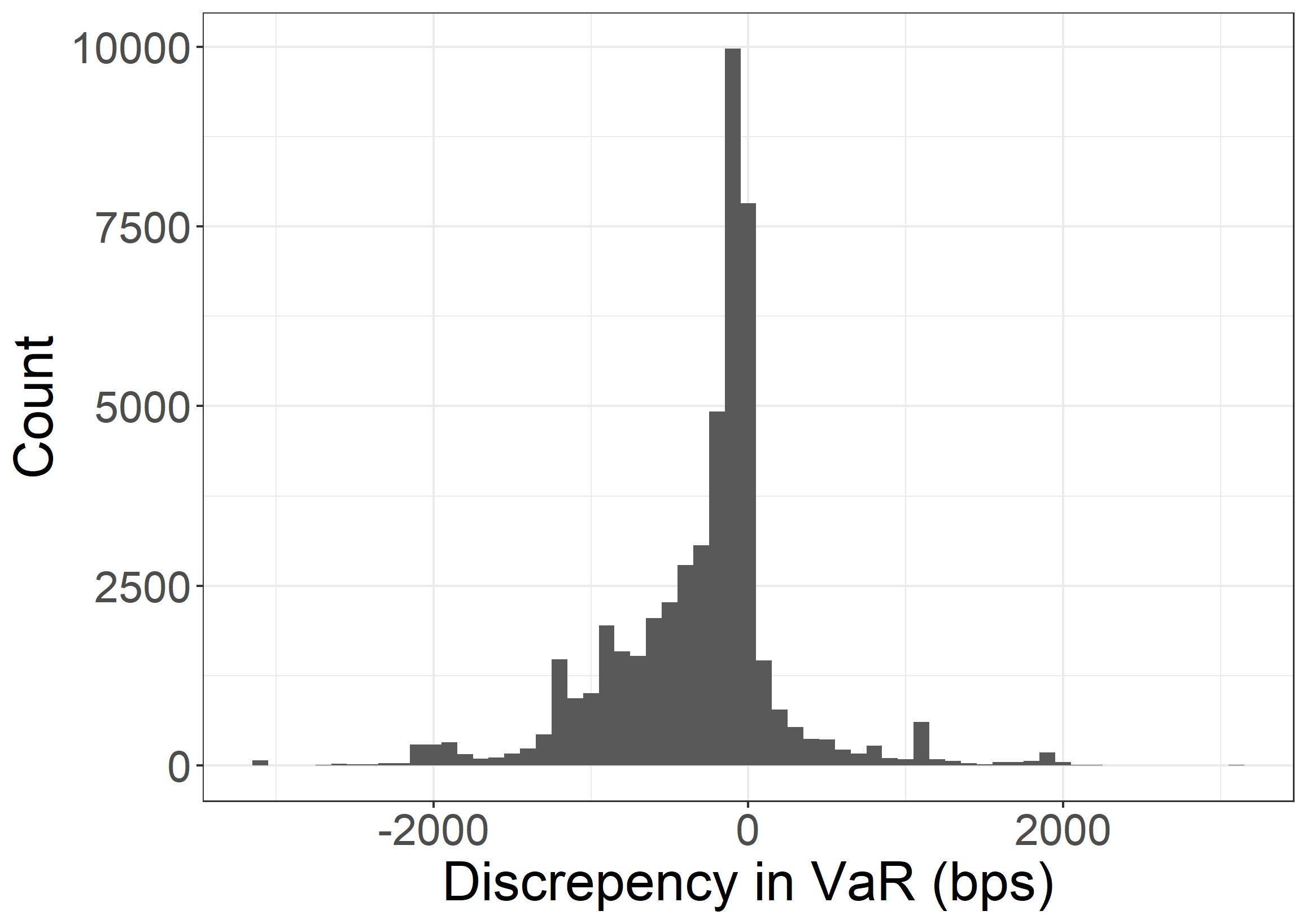} 
        \caption{Discrepancy} \label{fig:discrepancyVaRcross}
    \end{subfigure}
     \caption{The distribution of the profile VaR (upper left panel), portfolio VaR (upper right panel), and the discrepancy between them (lower panel) for all clients on August 12th 2019.}
     \label{fig:crossSection}
\end{figure}

\begin{figure}
    \centering
    \begin{subfigure}[t]{0.45\textwidth}
        \centering
        \includegraphics[width=\linewidth, keepaspectratio]{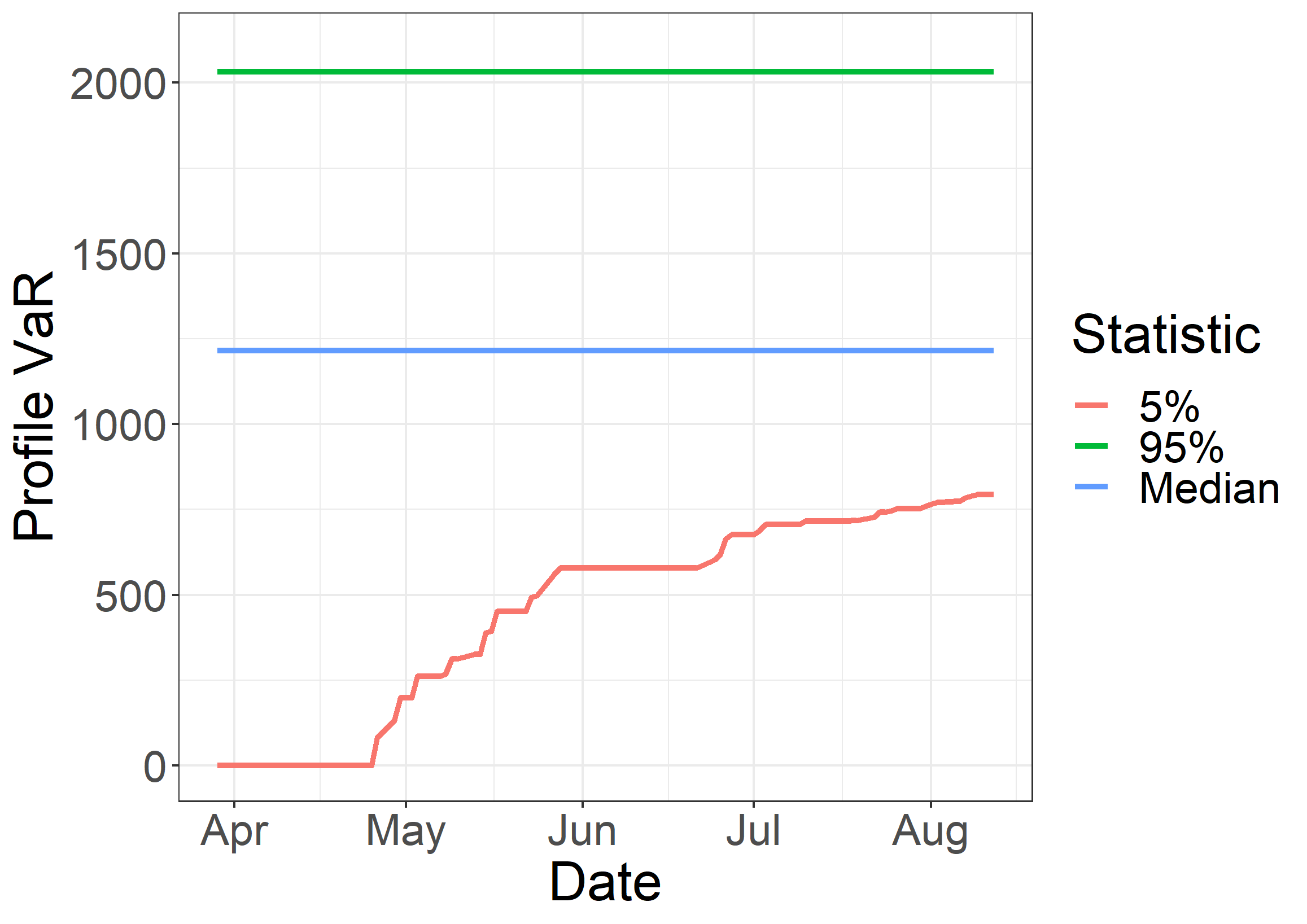}
        \caption{Profile VAR} \label{fig:prescribedVaR}
    \end{subfigure}
    \hfill
    \begin{subfigure}[t]{0.45\textwidth}
        \centering
        \includegraphics[width=\linewidth]{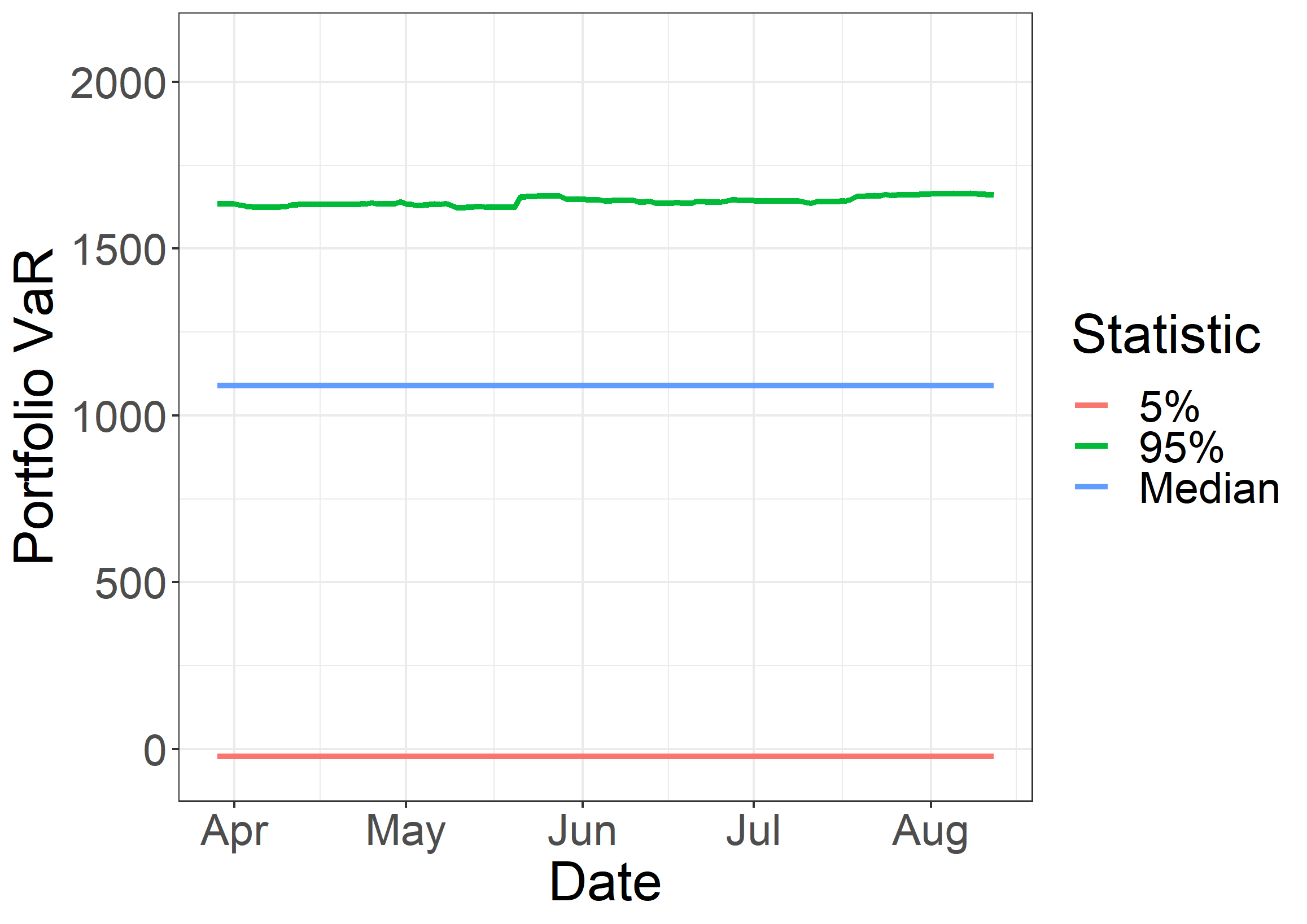} 
        \caption{Portfolio VaR} \label{fig:actualVaR}
    \end{subfigure}
    \vspace{1cm}
    \begin{subfigure}[t]{0.45\textwidth}
        \centering
        \includegraphics[width=\linewidth]{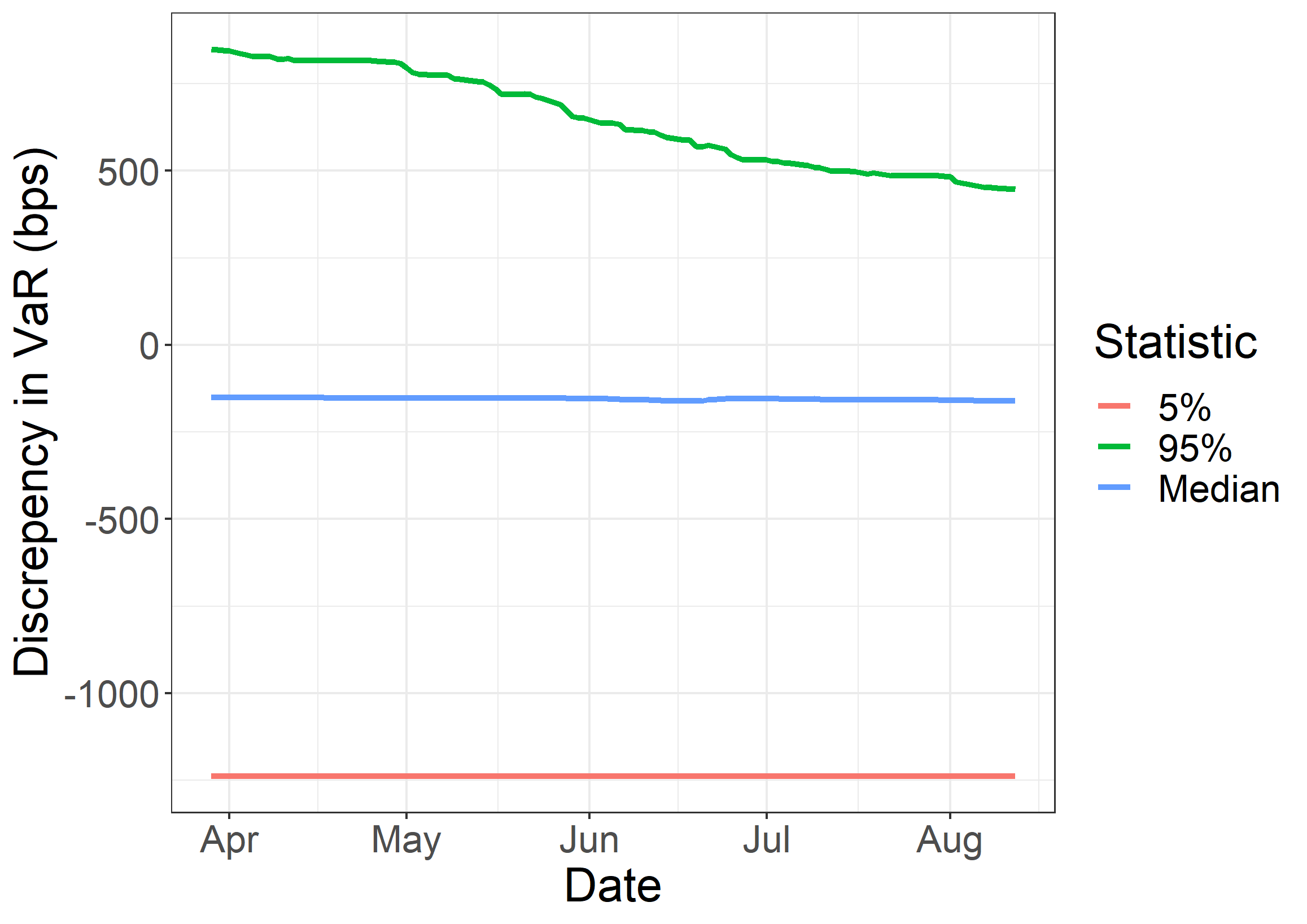} 
        \caption{Discrepancy} \label{fig:discrepancyOverTime}
    \end{subfigure}
     \caption{The 5\%, 50\%, and 95\% quantiles of the profile VaR (upper left panel), portfolio VaR (upper right panel), and the discrepancy between them (lower panel) over time.}
     \label{fig:overTime}
\end{figure}

The dealership cross-section of all client account VaRs and discrepancies on August 12th 2019 are shown in Figure \ref{fig:crossSection}. Across all accounts, we can see the dominant profile VaR in Figure \ref{fig:prescribedVaRcross} is a 100\% medium profile (approximately 1200 bps), with lesser spikes at 100\% low (approximately 0 bps), 100\% medium-high (approximately 2000 bps), 50\% medium and 50\% high (approximately 1750 bps) and 100\% high (approximately 3200 bps). Figure \ref{fig:actualVaRcross} shows that clients are typically set up around 1000 bps, but spikes exist again at 100\% medium, low, med-high, and high. Figure \ref{fig:discrepancyVaRcross} shows that the majority of clients are under-risked, where 86.7\% have a discrepancy at or below zero. 
The VaR cross-section of a single day is indicative of a global pattern over the time period of the dataset shown in Figure \ref{fig:overTime}. Figure \ref{fig:prescribedVaR} shows that the median profile VaR is consistently at 1216 bps or, on the worst day out of one hundred, the minimum loss that is prescribed is 12.16\% of the total portfolio market value. We found that the portfolio risk median in Figure \ref{fig:actualVaR} is consistently smaller at 10.90\% than the profile risk median. When the VaR are viewed together in Figure \ref{fig:discrepancyOverTime}, we see that profile VaR forms an upper boundary that the portfolio VaR stays well below. The gap is considerable at 150.2 to 160.2 bps. In our dataset, the median portfolio market value was \$$113,147$, and therefore 10 bps represents \$$113.15$ of potential capital impairment.    

\begin{figure}
    \centering
    \begin{subfigure}[t]{0.45\textwidth}
        \centering
        \includegraphics[width=\linewidth, keepaspectratio]{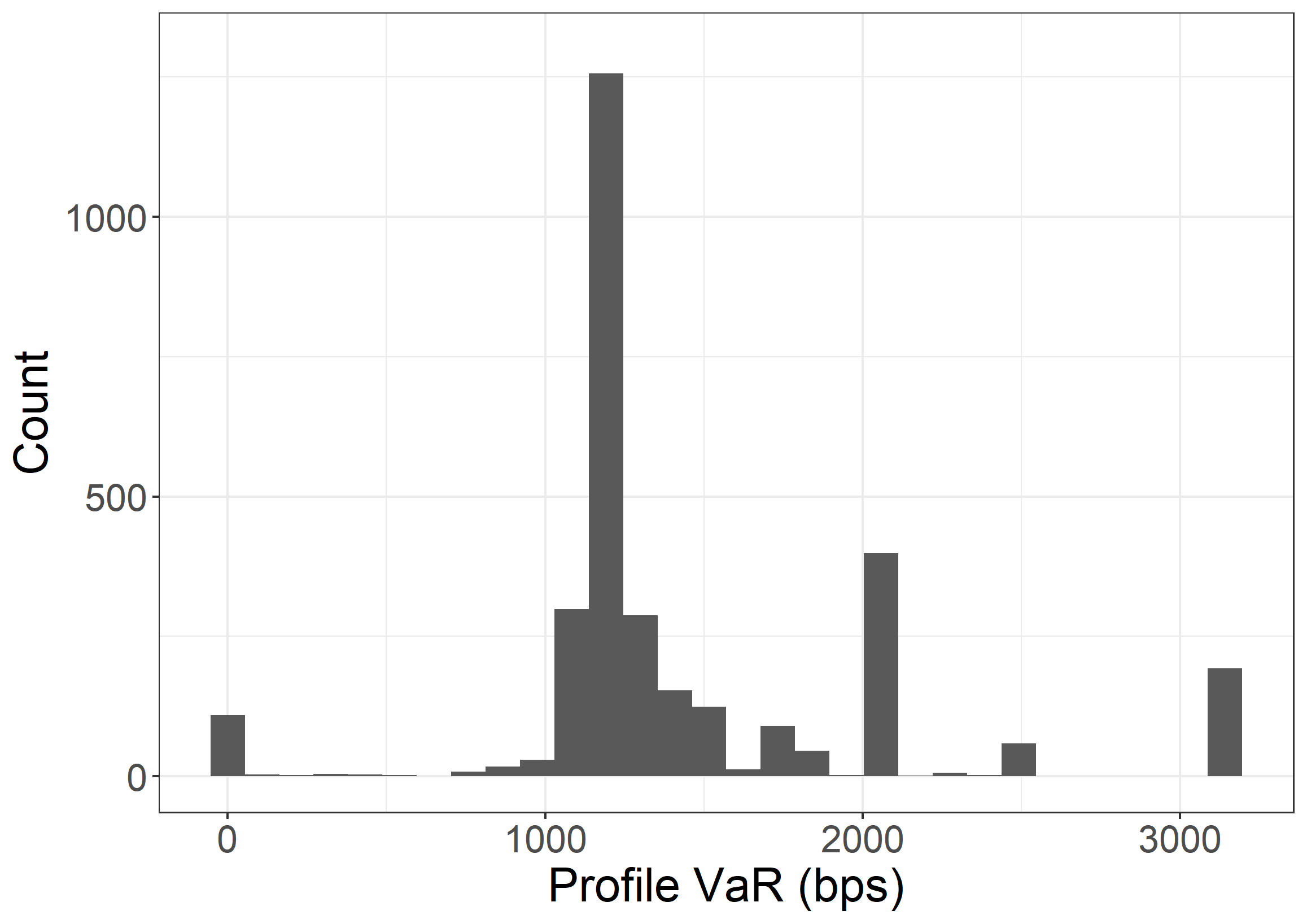}
        \caption{Profile VAR} \label{fig:influxProfile}
    \end{subfigure}
    \hfill
    \begin{subfigure}[t]{0.45\textwidth}
        \centering
        \includegraphics[width=\linewidth]{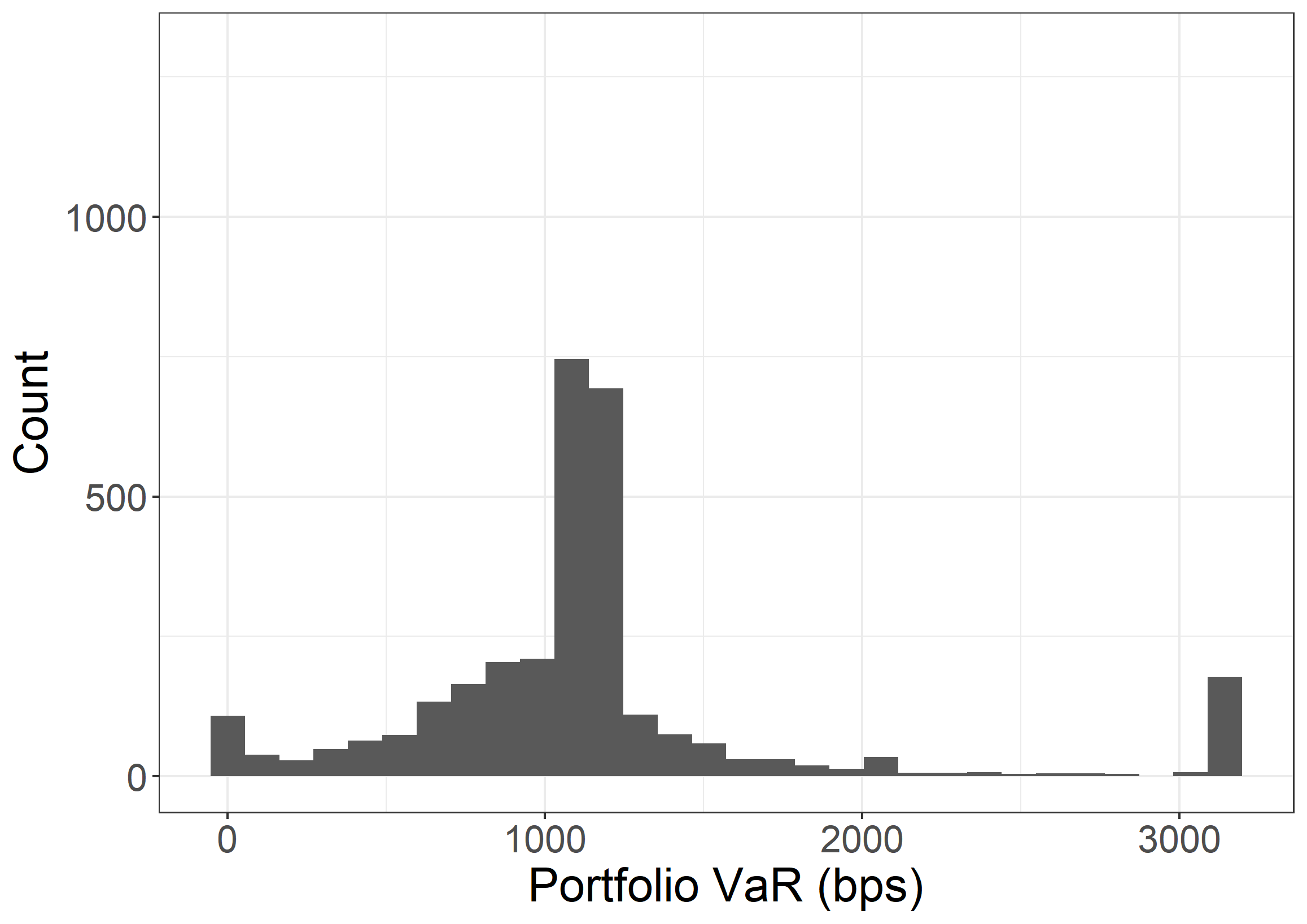} 
        \caption{Portfolio VaR} \label{fig:influxPortfolio}
    \end{subfigure}
    \vspace{1cm}
    \begin{subfigure}[t]{0.45\textwidth}
        \centering
        \includegraphics[width=\linewidth]{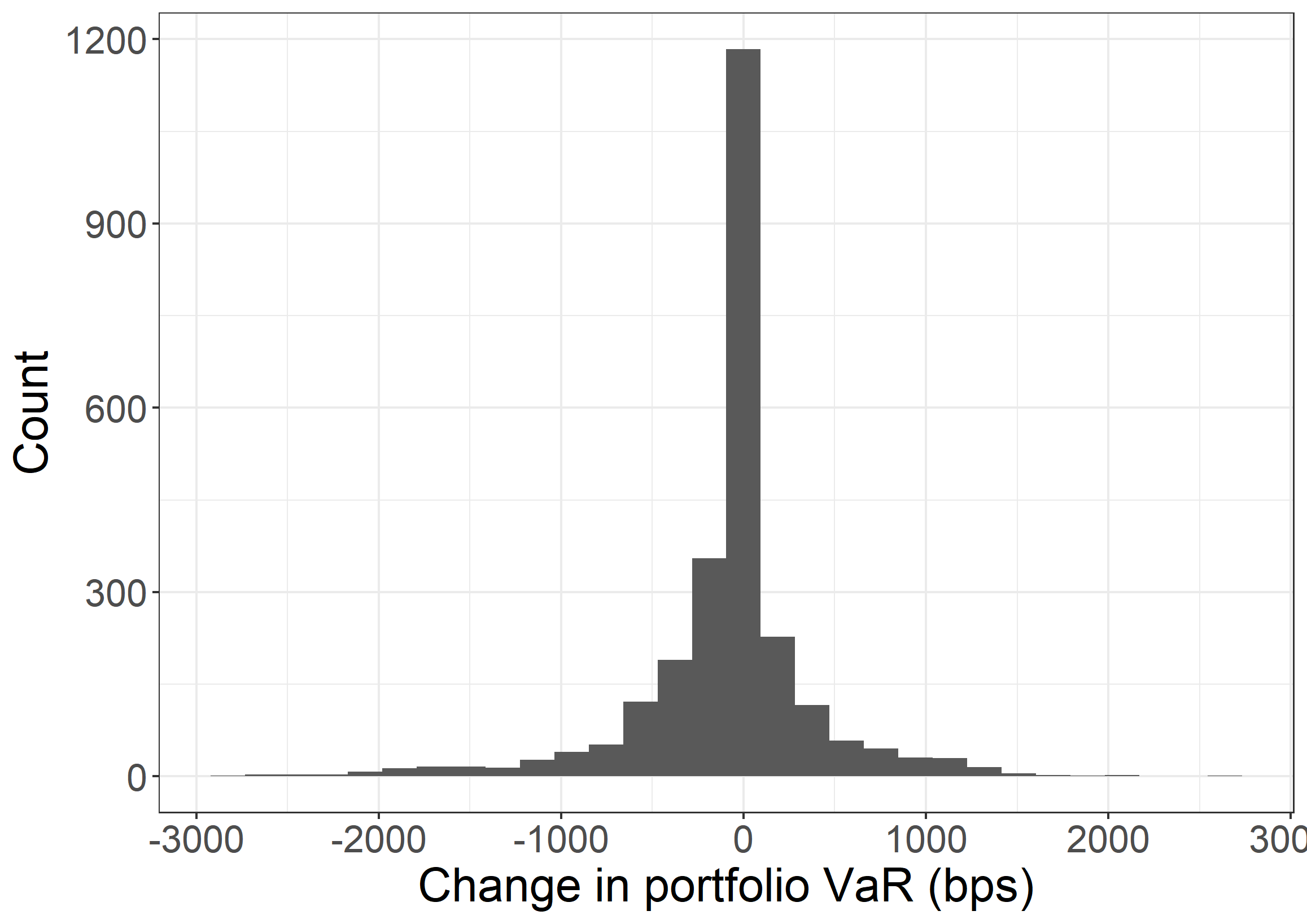} 
        \caption{Change caused by significant investment} \label{fig:influxChangeInPortfolio}
    \end{subfigure}
     \caption{The distribution of the profile VaR (upper left panel) and portfolio VaR (upper right panel) on days when clients added an investment of at least 50\% of their portfolio market value (3109 occurrences). The lower panel shows the effect that the large investment had on the portfolio VaR.}
     \label{fig:cashInflux}
\end{figure}

We further investigated VaR for each of the variables described in Table \ref{tbl:clientDetails} using the box plots found in Appendix C. We found significant evidence of advisors tailoring of portfolios to the specific needs or attributes of individual clients. We found that on average:
\begin{itemize}
\itemsep0em 
    \item As income increases, profile and portfolio VaR increase. 
    \item As age increases, profile and portfolio VaR decreases.
    \item Margin accounts have the highest profile and portfolio VaR, while RIF had the lowest.
    \item As investment knowledge increases, profile and portfolio VaR decreases.
    \item People from British Columbia tend to have the lowest profile and portfolio VaR.
    \item Across all variables except account type, discrepancies had similar distributions.
    \item Retired individuals had lower profile and portfolio VaR.
    \item Men had slightly higher profile and portfolio VaR than women.
    \item Single people had the highest profile VaR, and divorced had the lowest.
    \item 40 to 50 year-old's tended to have higher income, but lower portfolio market value.
    \item Investment knowledge is relatively similar across age. 
\end{itemize} 
Figures \ref{fig:influxProfile} and \ref{fig:influxPortfolio} show the distributions of profile and portfolio VaR after a large (50\%) injection of funds. Figure \ref{fig:influxChangeInPortfolio} shows the effect on the portfolio VaR, where most additions are close to zero and fit into the portfolio risk selection of the client. 

\subsection{Cluster value-at-risk}
For illustration purposes, we might consider the profile risk as the bumper rails in a bowling alley. Once the bumper rails have been engaged, participants are protected from throwing a gutter ball and scoring zero. However, obtaining a good score, or throwing consistent strikes, still requires skill, and perhaps, a little luck. The question then becomes, in setting the risk ``guard rails", do advisors and clients customize the two risks--prescribed and portfolio--to recognize unique trading behaviour preferences. We found that advisors and their clients actively manage differing clients' needs because the VaR differs for each cluster, as does the VaR discrepancy (profile VaR minus portfolio VaR). 

\begin{figure}
    \centering
    \begin{subfigure}[t]{0.45\textwidth}
        \centering
        \includegraphics[width=\linewidth]{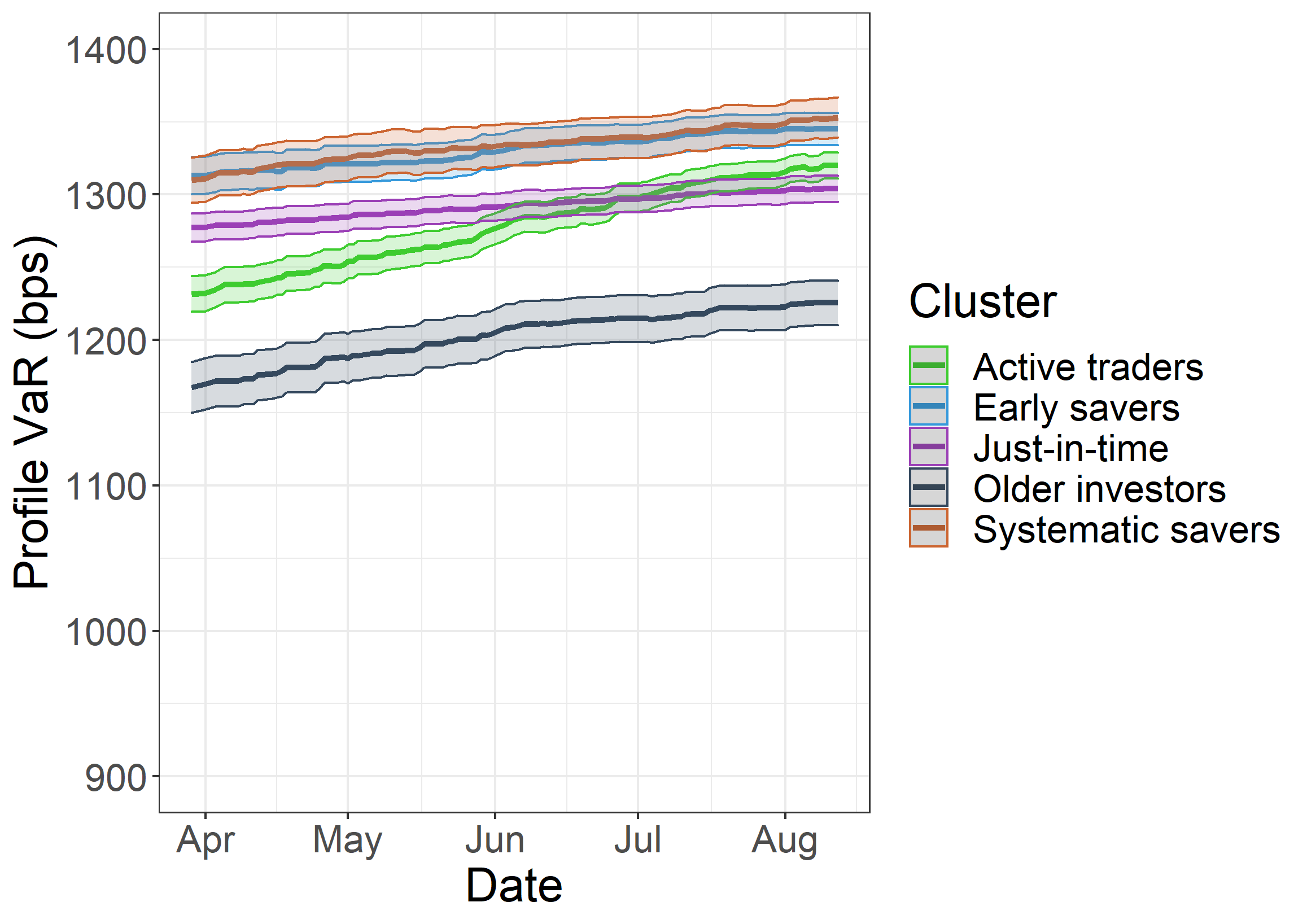}
        \caption{Profile VaR} \label{fig:clusterprescribedVaR}
    \end{subfigure}
    \hfill
    \begin{subfigure}[t]{0.45\textwidth}
        \centering
        \includegraphics[width=\linewidth]{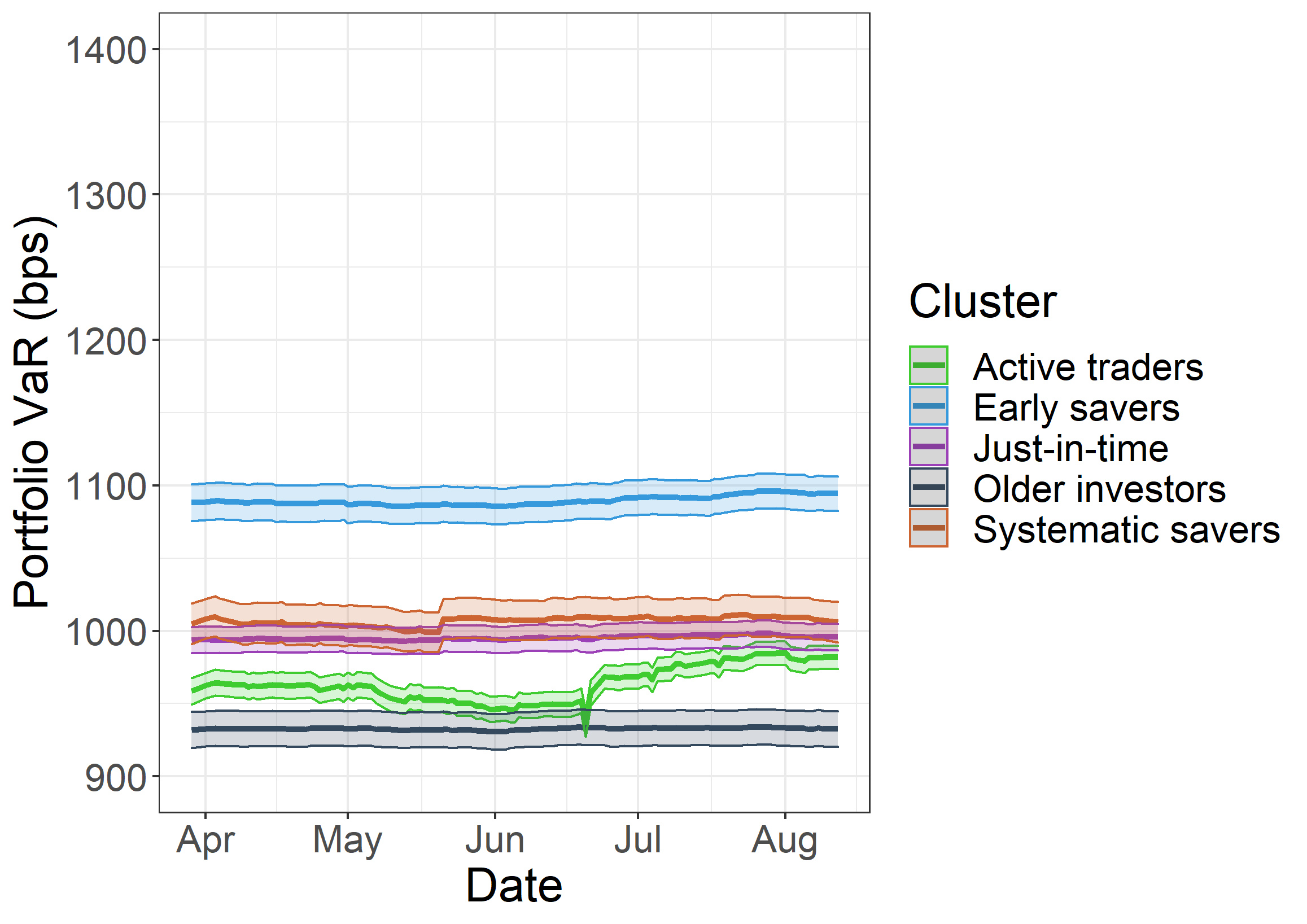} 
        \caption{Portfolio VaR} \label{fig:clusteractualVaR}
    \end{subfigure}
    \vspace{1cm}
    \begin{subfigure}[t]{0.45\textwidth}
        \centering
        \includegraphics[width=\linewidth]{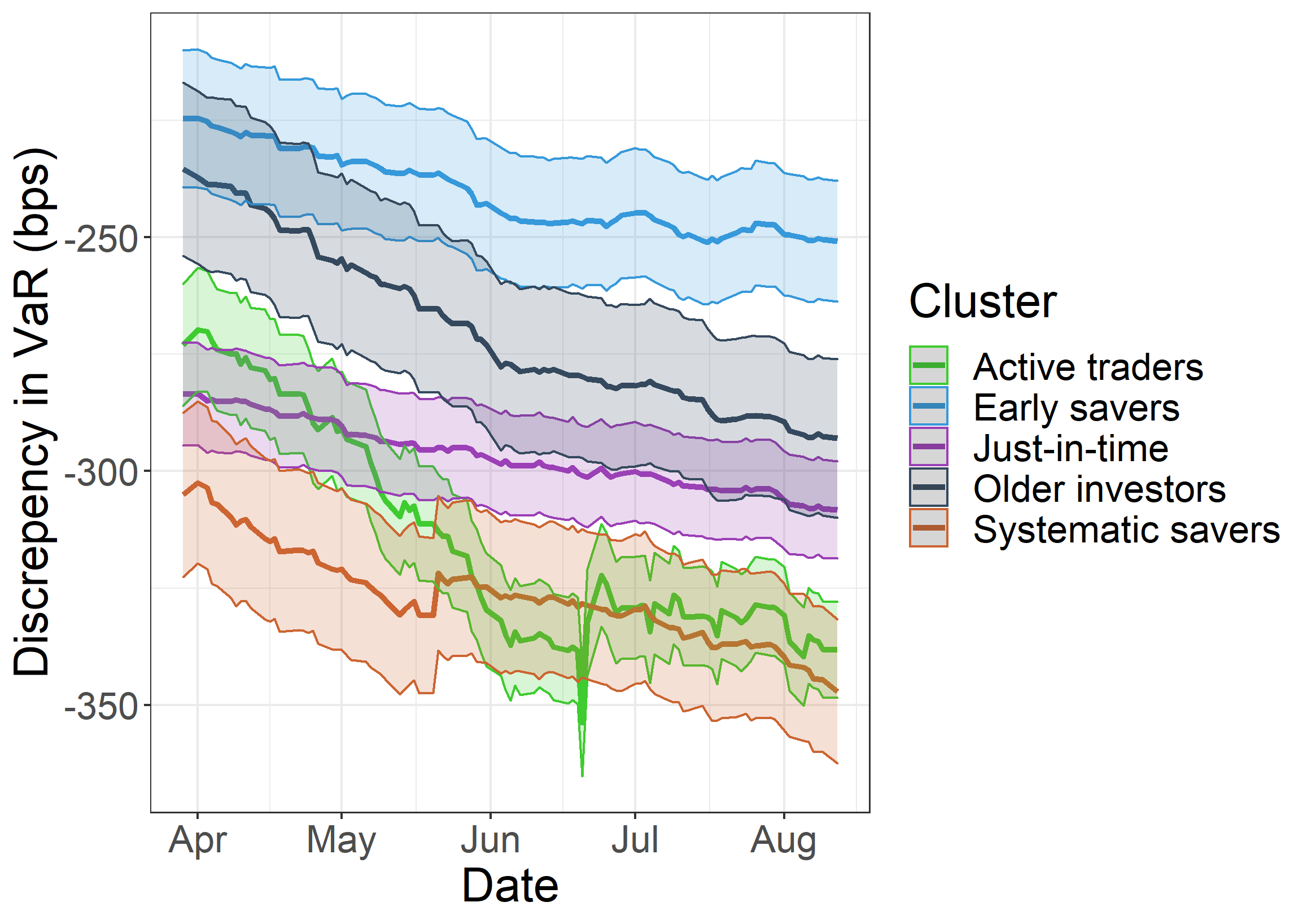} 
        \caption{Discrepancy in VaR} \label{fig:clusterdiscrepancyOverTime}
    \end{subfigure}
     \caption{Mean of the daily VaR for clusters with 95\% bootstrapped confidence intervals ($B=9999$ re-samples). The distribution of the profile VaR is show in the upper left panel and portfolio VaR in the upper right panel, and the discrepancy of the VaRs are shown in the lower panel.}
     \label{fig:clusteroverTime}
\end{figure}

Cluster personas were determined in previous work, and the results of those clusters are discussed in Section \ref{sec:clusters}. In Figure \ref{fig:clusterprescribedVaR}, we can see clear delineations between the profile VaR for each cluster and that the profile VaR generally behaves as we would expect--the highest profile VaR is with the Early Savers (longer time horizons) and Systematic Savers (dollar-cost averaging) while the lowest profile VaR is with the Older Investors. The most significant increase in profile VaR is with the Active Traders--presumably consistent with their preferred behaviour. 

Figure \ref{fig:clusteractualVaR} shows that customized portfolio construction carries through to the actual portfolio VaR--but within the limits set by the guard rails (the profile risk). It should also be noted that portfolio VaR remains relatively flat, indicating active management of the portfolios against \textbf{background beta}. Figure \ref{fig:clusterdiscrepancyOverTime} shows the gap between profile and portfolio risk. We observed that the discrepancy is consistently negative (portfolio risk is safer than profile risk) and small, but not unique to each cluster. Therefore, clusters are managed within their guardrails, and portfolios are tailored to each cluster's unique, presumably preferred, trading behaviours. 

\begin{figure}
    \centering
    \begin{subfigure}[t]{0.45\textwidth}
        \centering
        \includegraphics[width=\linewidth]{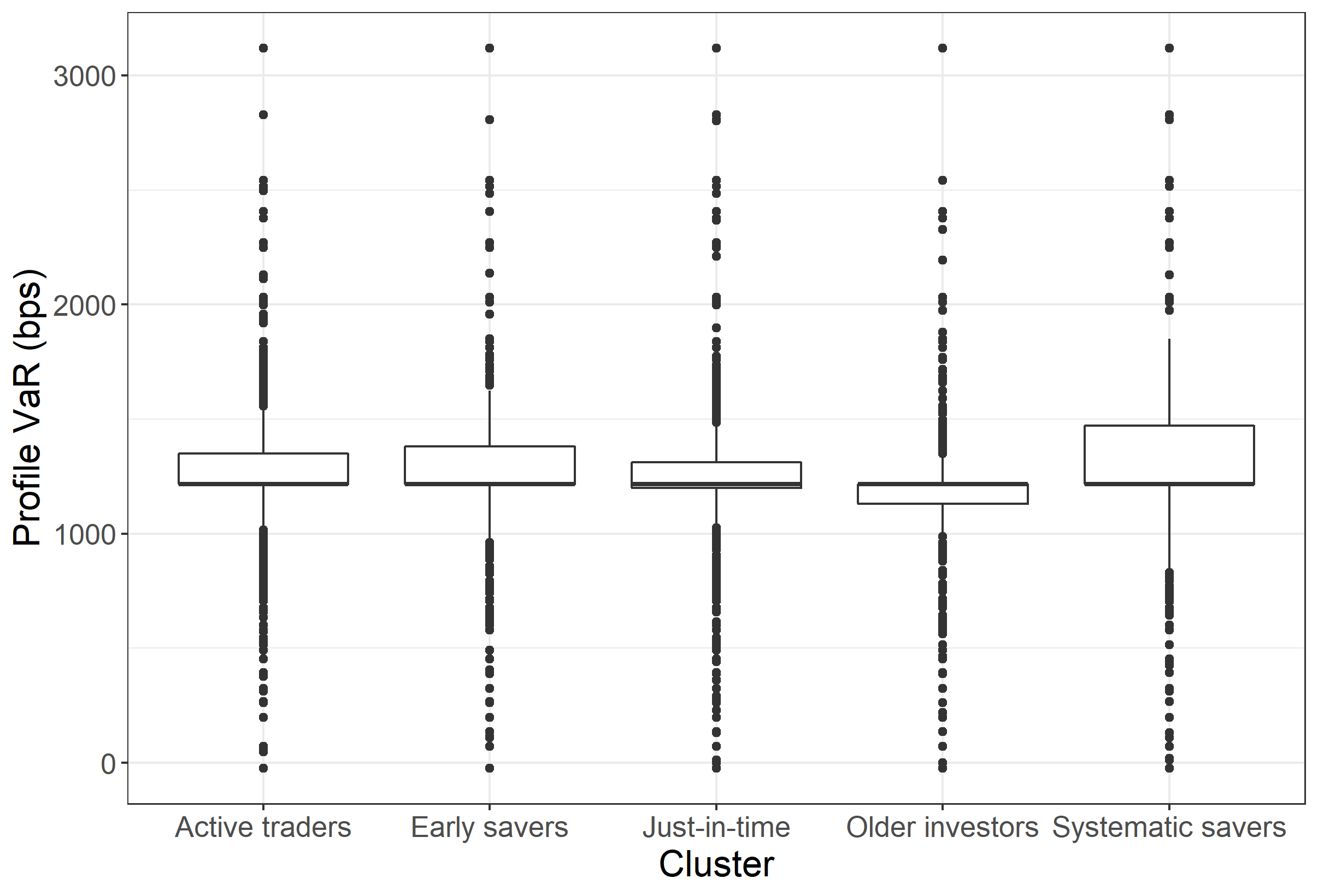}
        \caption{Profile VaR} \label{fig:advisorPrescribedVaRClusterBoxplot}
    \end{subfigure}
    \hfill
    \begin{subfigure}[t]{0.45\textwidth}
        \centering
        \includegraphics[width=\linewidth]{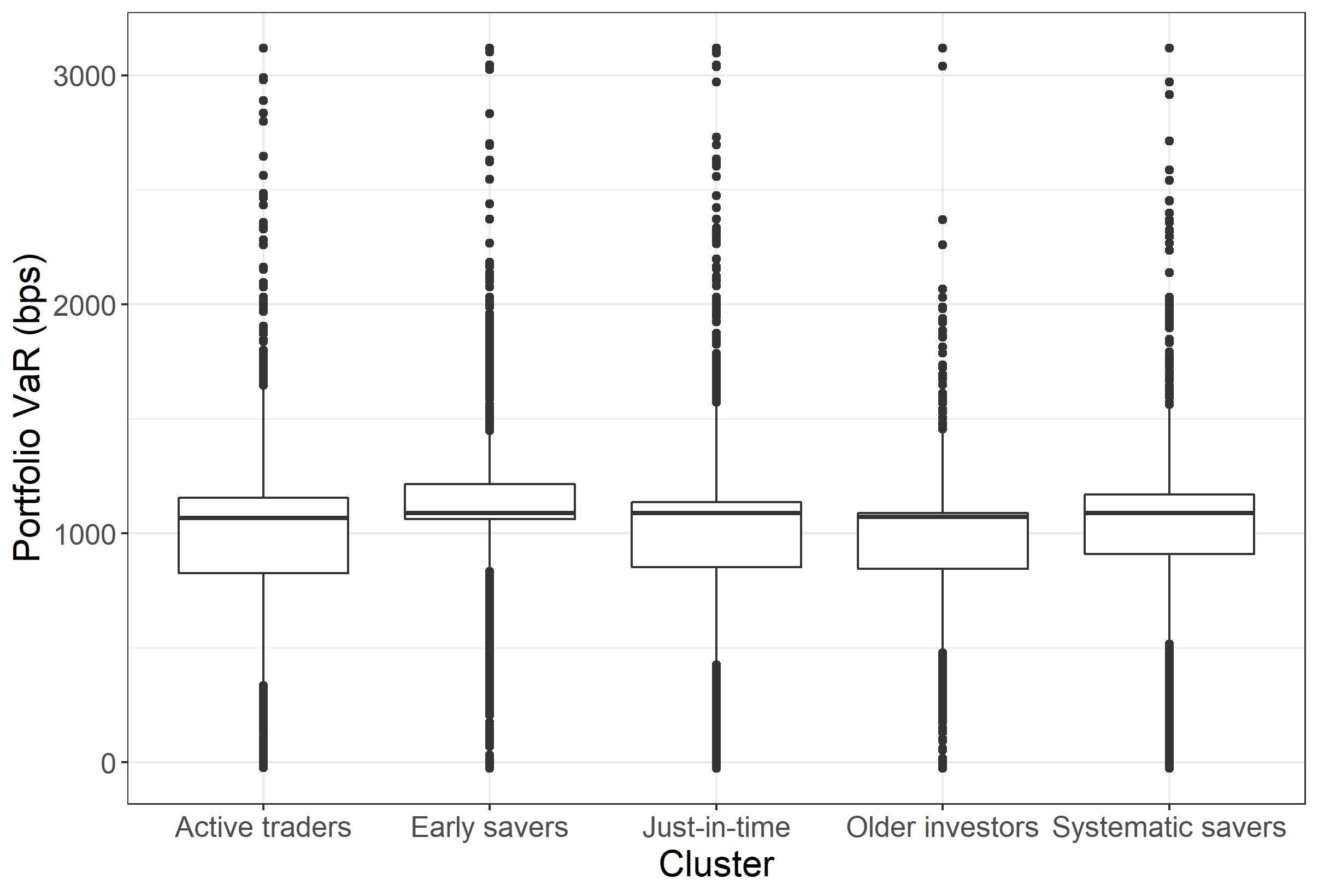} 
        \caption{Portfolio VaR} \label{fig:advisorActualVaRClusterBoxplot}
    \end{subfigure}
    \vspace{1cm}
    \begin{subfigure}[t]{0.45\textwidth}
        \centering
        \includegraphics[width=\linewidth]{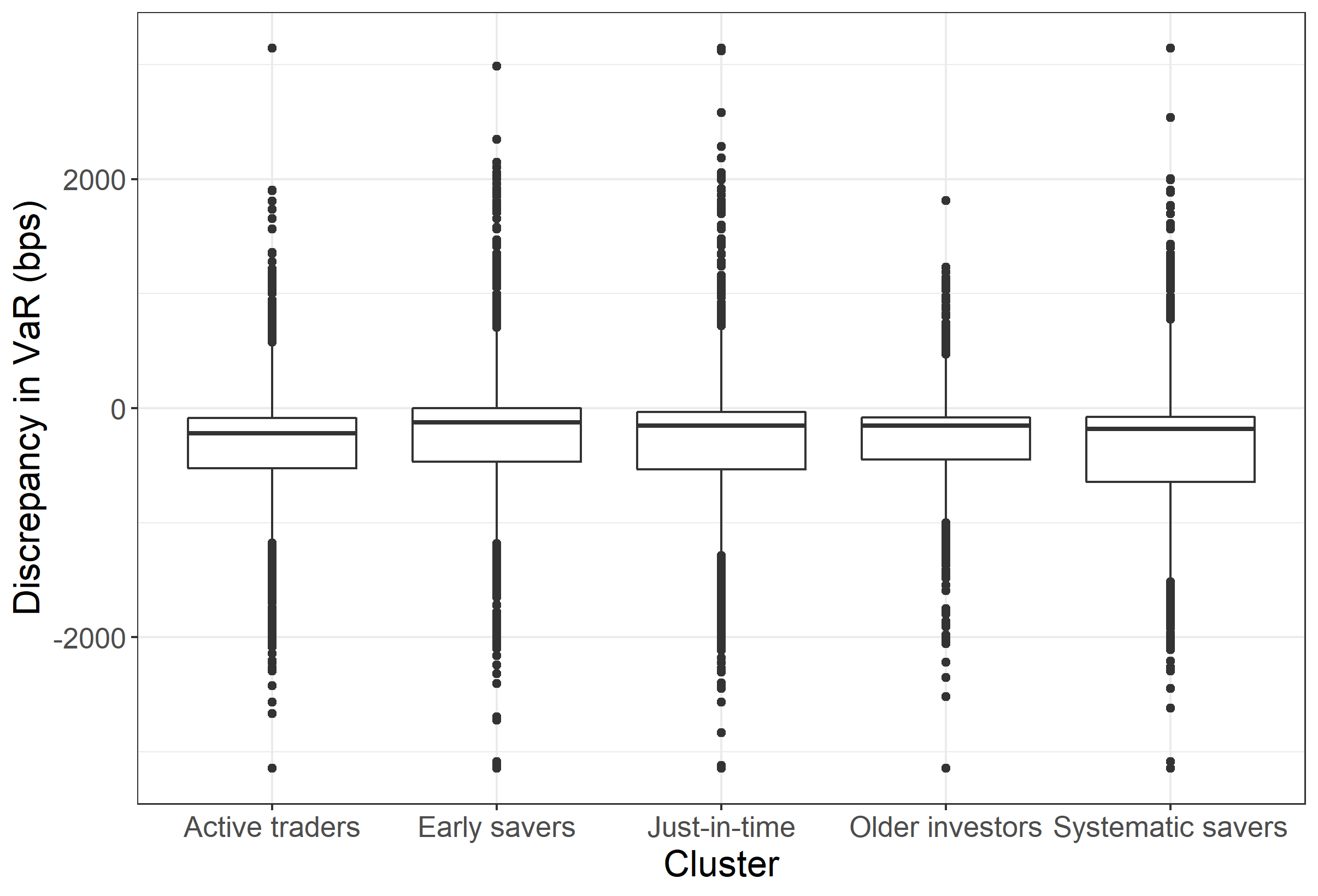} 
        \caption{Discrepancy in VaR} \label{fig:advisorDiscrepancyClusterBoxplot}
    \end{subfigure}
     \caption{Profile VaR (top left panel), portfolio VaR (top right panel) and the discrepancy in VaR (lower panel) on August 12th 2019 by cluster membership.}
     \label{fig:boxplotsCluster}
\end{figure}

Figure \ref{fig:boxplotsCluster} demonstrates each cluster's account VaRs on the last day of the dataset. Figure \ref{fig:advisorPrescribedVaRClusterBoxplot} shows that the majority of the data has a profile VaR between 1600 and 850 bps, where systematic savers tend to have the highest profile risk, and older investors have the lowest profile risk. Figure \ref{fig:advisorActualVaRClusterBoxplot} shows that early savers have a higher overall portfolio risk while older investors again have the lowest overall portfolio risk. Figure \ref{fig:advisorDiscrepancyClusterBoxplot} shows that most investors, regardless of cluster, are under-risked.

\subsection{Discretionary advising}
As further evidence of client customization, we examined the same value-at-risk metrics against advisor licensing regimes. Within our dataset, there were two significant advisor licensing regimes--discretionary and non-discretionary. In Canada, under the IIROC regime, some investment representatives can provide discretionary portfolio management services\footnote{Investment Industry Regulatory Authority of Canada, Rule 2900, Proficiency and Education \url{https://www.iiroc.ca/Rulebook/MemberRules/Rule02900_en.pdf}} and can make wholesale portfolio decisions on behalf of their clients, without the need for explicit client permission before placing a trade. In our dataset, 8.7\% of accounts listed investment representatives as licensed to be discretionary portfolio managers. Non-discretionary advisors, on the other hand, must seek a client’s permission before every trade. Given these structural differences, our operating hypothesis was that discretionary managers would have the capacity to accommodate client behaviour more than non-discretionary advisors with restricted licenses.

\begin{figure}
    \centering
    \begin{subfigure}[t]{0.45\textwidth}
        \centering
        \includegraphics[width=\linewidth, keepaspectratio]{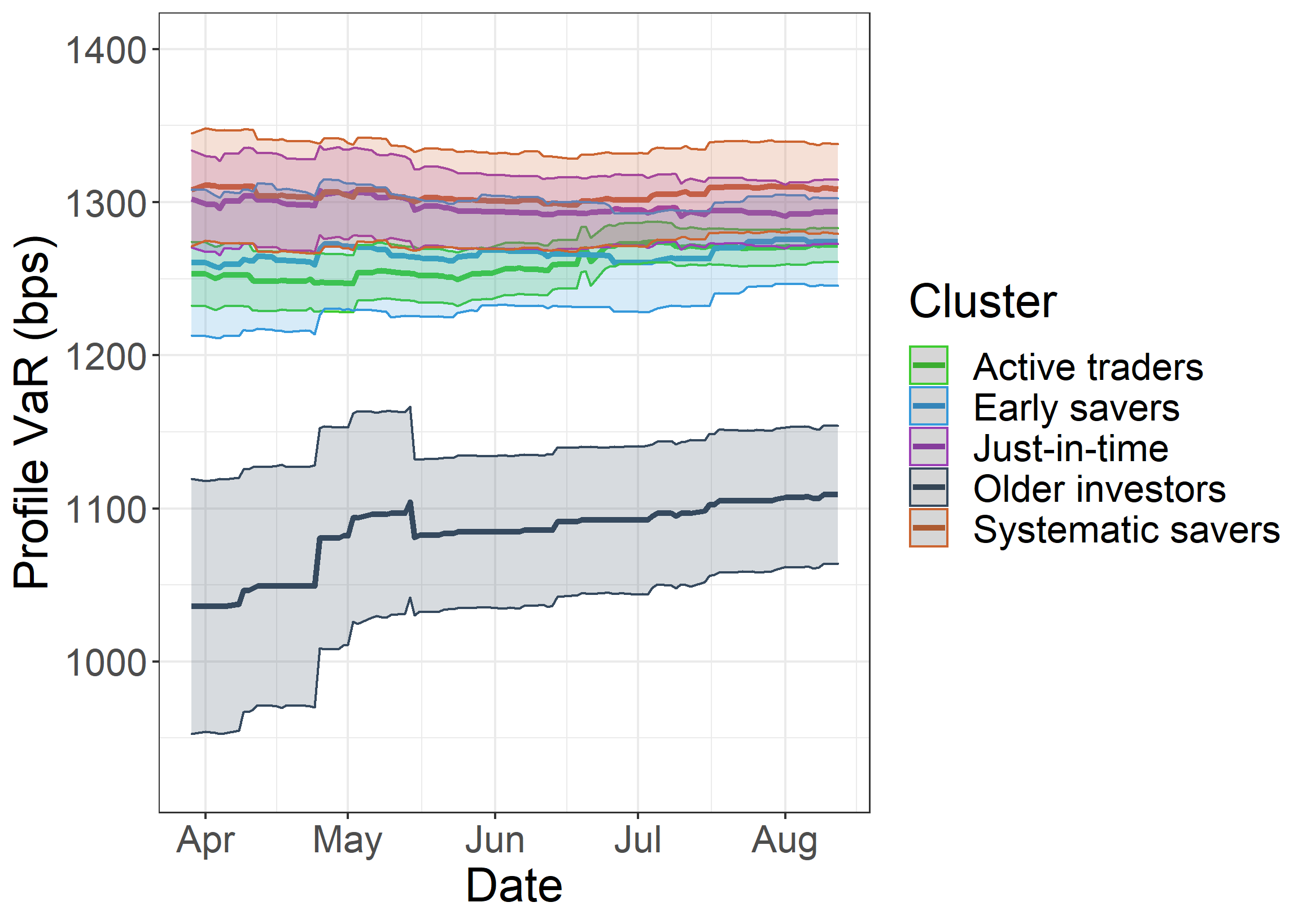}
        \caption{Discretionary profile VaR} \label{fig:discretionaryProfileVaR}
    \end{subfigure}
    \hfill
    \begin{subfigure}[t]{0.45\textwidth}
        \centering
        \includegraphics[width=\linewidth, keepaspectratio]{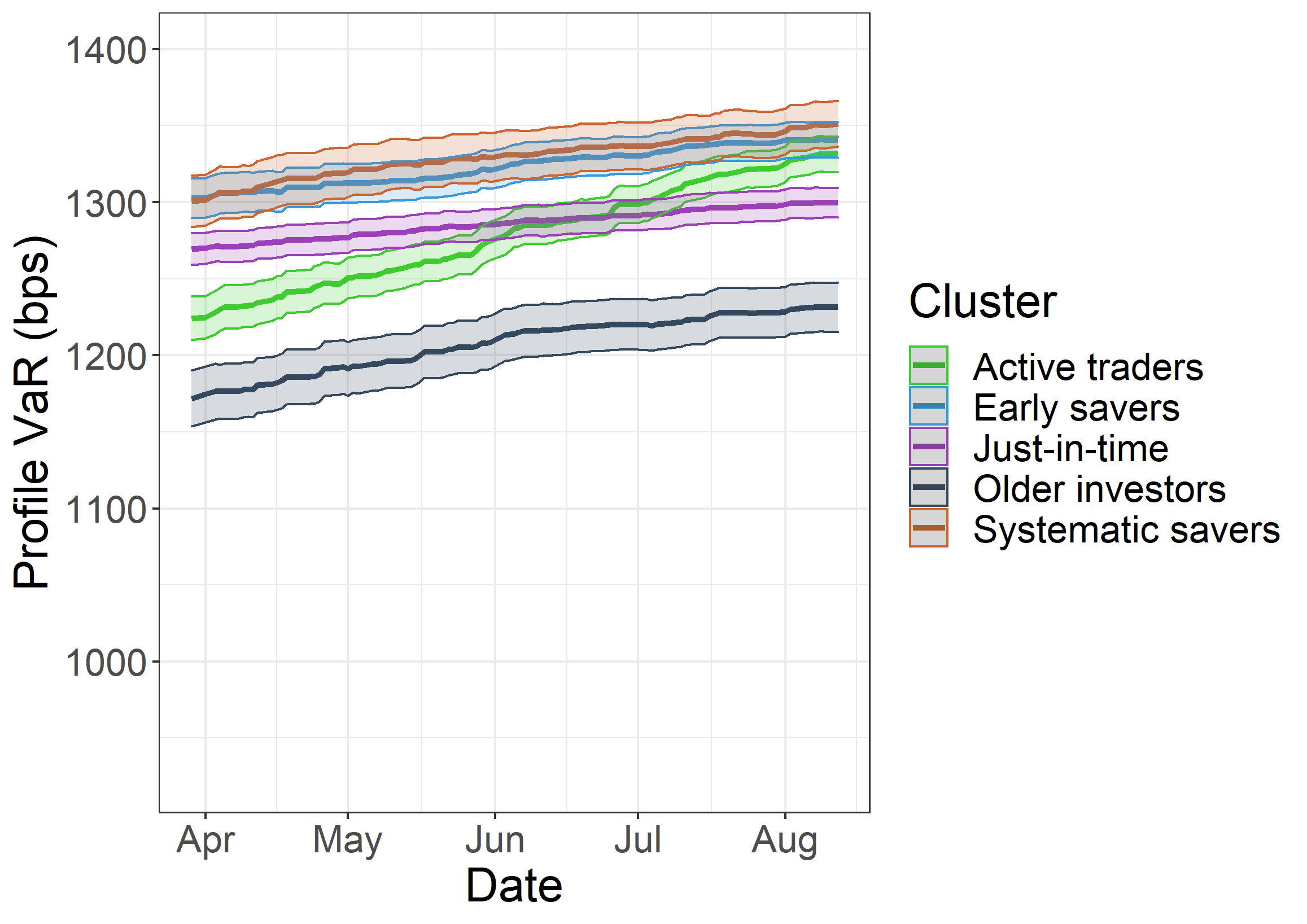}
        \caption{Non-discretionary profile VaR} \label{fig:nonDiscretionaryProfileVaR}
    \end{subfigure}
    \vspace{1cm}
    \begin{subfigure}[t]{0.45\textwidth}
        \centering
        \includegraphics[width=\linewidth, keepaspectratio]{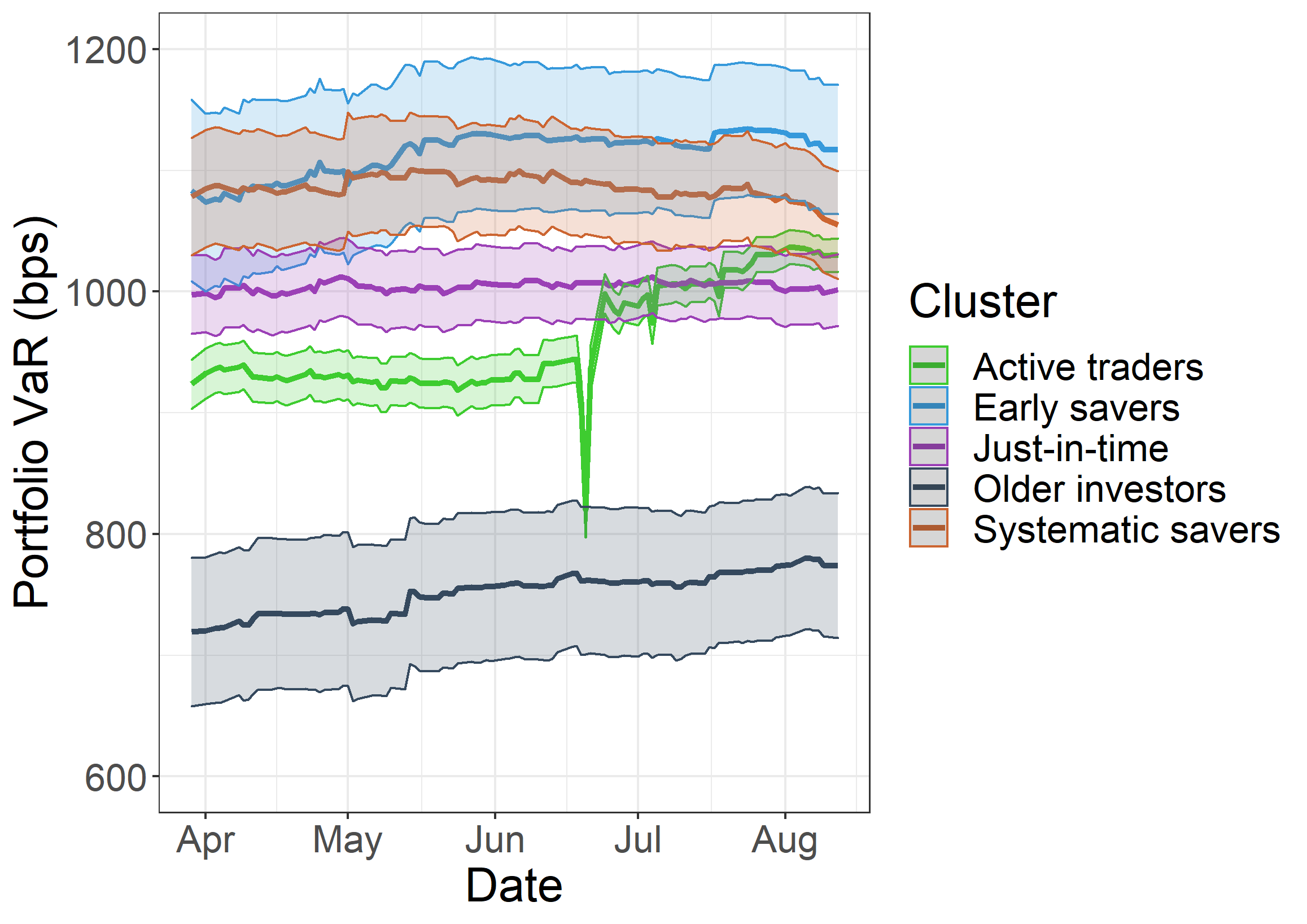}
        \caption{Discretionary portfolio VaR} \label{fig:discretionaryPortVaR}
    \end{subfigure}
    \hfill
    \begin{subfigure}[t]{0.45\textwidth}
        \centering
        \includegraphics[width=\linewidth, keepaspectratio]{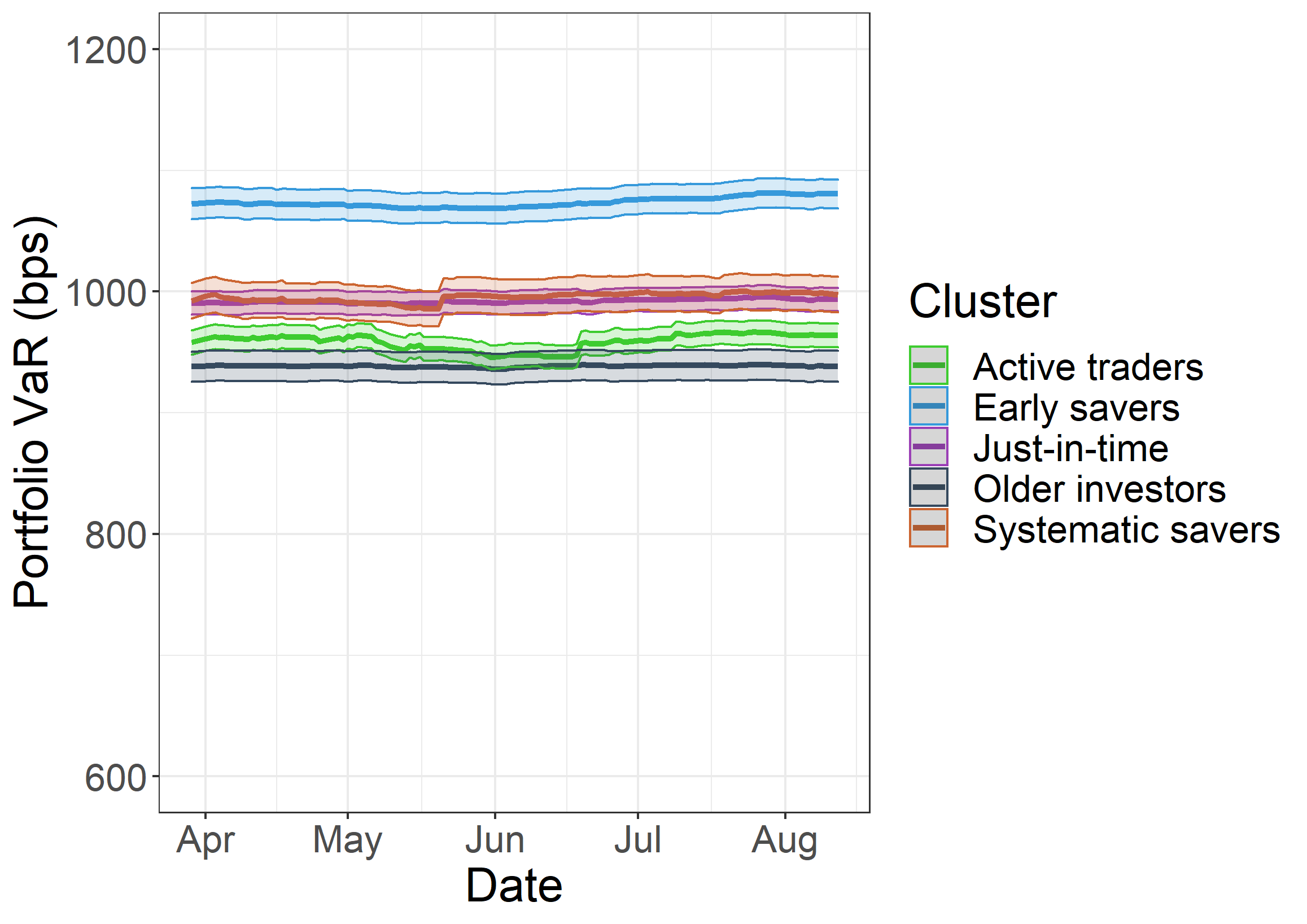}
        \caption{Non-discretionary portfolio VaR} \label{fig:nonDiscretionaryPortVaR}
    \end{subfigure}
    \vspace{1cm}
    \begin{subfigure}[t]{0.45\textwidth}
        \centering
        \includegraphics[width=\linewidth, keepaspectratio]{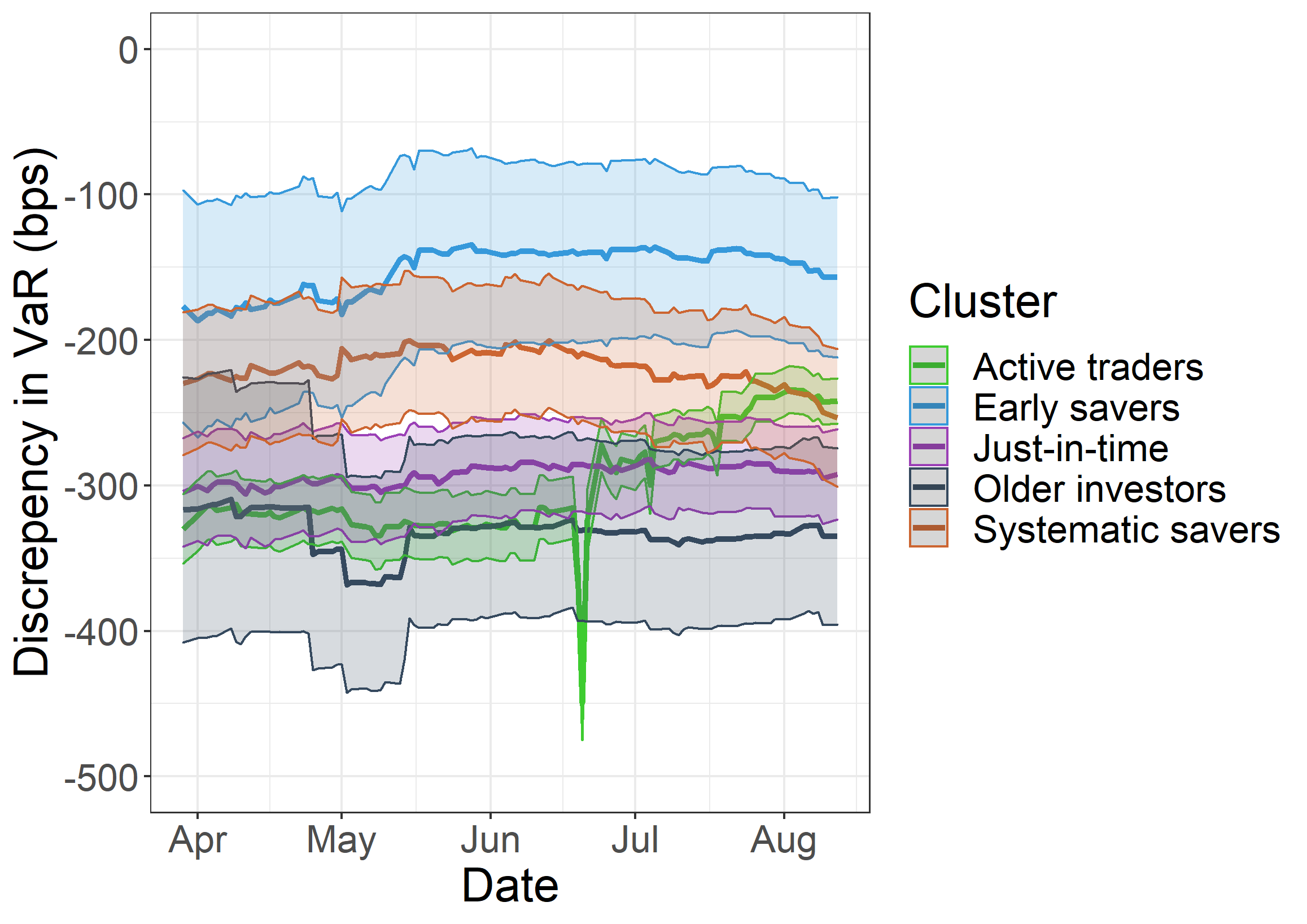}
        \caption{Discretionary discrepancy} \label{fig:discretionaryVaRdisc}
    \end{subfigure}
    \hfill
    \begin{subfigure}[t]{0.45\textwidth}
        \centering
        \includegraphics[width=\linewidth]{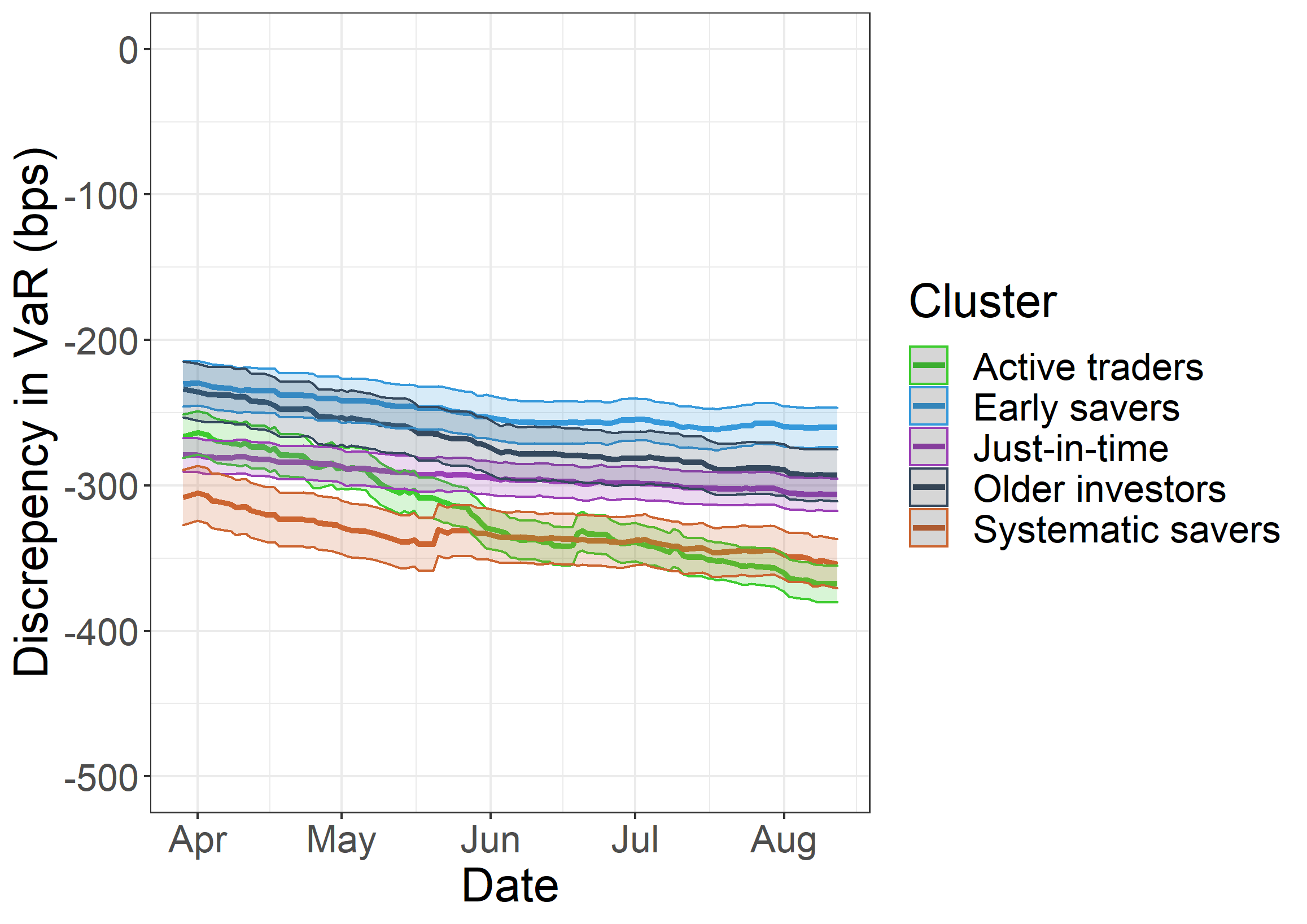} 
        \caption{Non-discretionary discrepancy} \label{fig:nonDiscretionaryVaRdisc}
    \end{subfigure}
 \caption{The profile VaR, portfolio VaR, and discrepancy between VaRs for accounts with advisors that have a discretionary licence (top-to-bottom of left panels) and that have a non-discretionary licence (top-to-bottom of right panels).}
     \label{fig:discretionaryType}
\end{figure}

We found that both licensing regimes followed the same patterns noted above regarding the behaviour of the profile VaR versus the portfolio VaR but that the discretionary advisors (Figures \ref{fig:discretionaryProfileVaR}, \ref{fig:discretionaryPortVaR}, \ref{fig:discretionaryVaRdisc}) appear to be more effective at maintaining consistent profile VaR and discrepancy gap over time (except the active traders whom we would expect to look to take advantage of market outlooks). Non-discretionary advisors, on the other hand, appeared to allow client profile VaR to systematically drift upwards with the markets, but portfolios are relatively consistent which results in a growing discrepancy (Figures \ref{fig:nonDiscretionaryProfileVaR}, \ref{fig:nonDiscretionaryPortVaR}, \ref{fig:nonDiscretionaryVaRdisc}).

%Do our findings suggest positive or negative consequences for investors? Given the limitations of our dataset, we would have to conclude that our observations are inconclusive. The definitive evidence of whether advisors respond to unique client characteristics is captured in the minutes of an advisor’s direct interaction with a client, to which we are not privy.

It appears that advisors and their clients are systemically safe and conservative, which can be in the best interest of many, but not all, clients. For clients seeking to preserve capital, it is good news--no gutter balls. However, for clients seeking to maximize growth, it may be inconsistent with their objectives. An example might be older clients who can preserve their capital but cannot achieve investment incomes that allow them to maintain their lifestyles. Maintaining investment income has become a priority in an era of low interest rates and growing longevity risk. 

\begin{figure}
    \centering
    \begin{subfigure}[t]{0.45\textwidth}
        \centering
        \includegraphics[width=\linewidth, keepaspectratio]{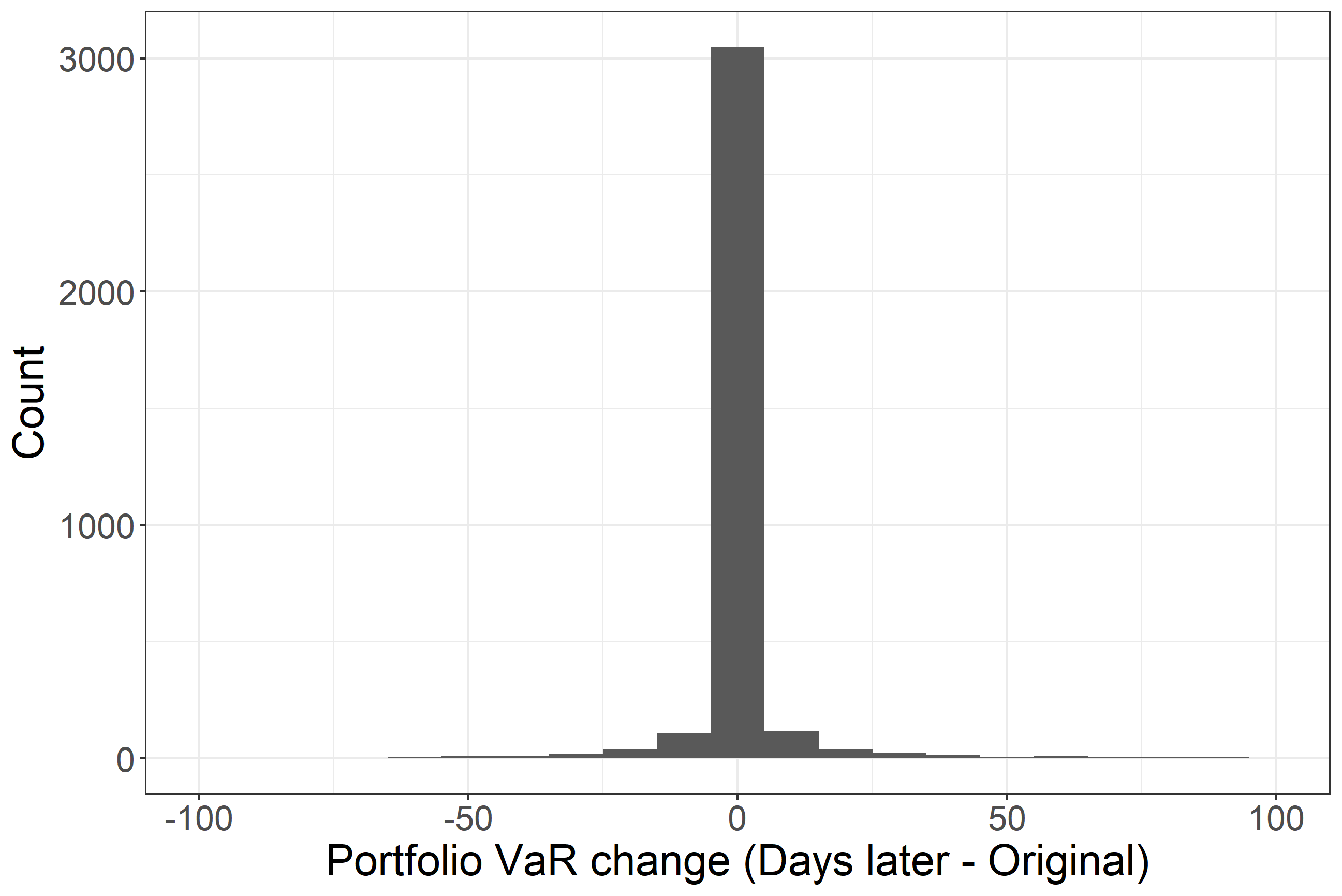}
        \caption{Incidence of portfolio changes after a KYC change} 
        \label{fig:portVarChangeAfterKYC}
    \end{subfigure}
    \hfill
    \begin{subfigure}[t]{0.45\textwidth}
        \centering
        \includegraphics[width=\linewidth]{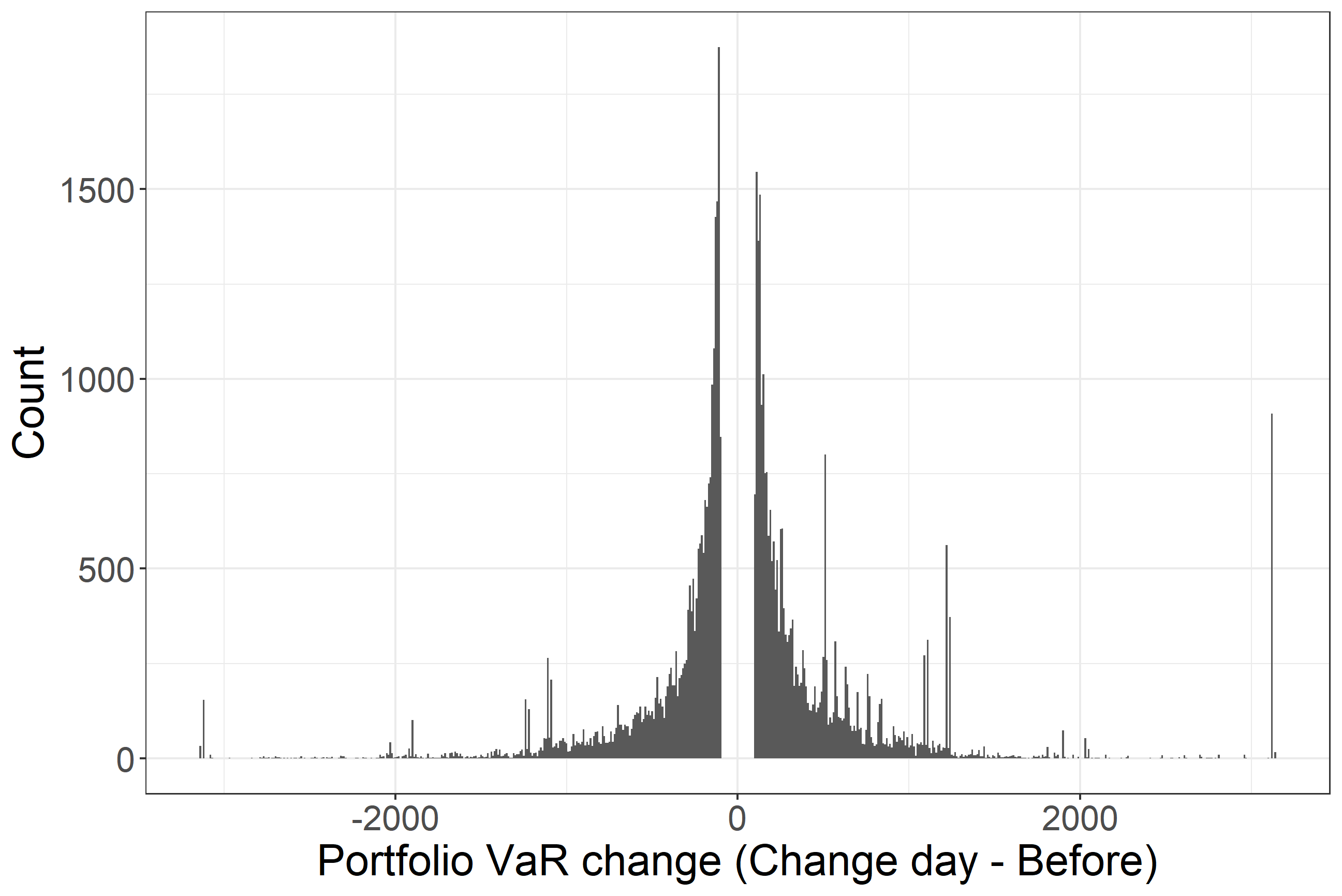} 
        \caption{Incidence of portfolio changes leading up to a significant KYC change} \label{fig:bigPortVarChange}
    \end{subfigure}
 \caption{The left panel is the change in portfolio VaR preceded by a change in profile VaR two weeks before. The right panel are changes to the portfolio VaR over a two-week period leading up to a change in the profile VaR.}
     \label{fig:portprochange}
\end{figure}

We also noted that there was little evidence that advisors and clients actively manage the profile risk once it is set--presumably at the time the account was opened. We noted that it is relatively rare for advisors to change the portfolio risk after a change in the KYC (Figure \ref{fig:portVarChangeAfterKYC}) and vice versa, to change the KYC after a significant change in portfolio risk (Figure \ref{fig:bigPortVarChange}). It appears the gap between profile and portfolio VaR is large enough that subsequent changes provide little incentive to nudge the two portfolios. In other words, as long as the gutter is protected ("we are fine") but this observation would appear to question the inherent utility of the profile risk. If it is not used to inform or respond to changes in actual portfolio risk, why bother maintaining it?

\section{Summary of results and benefits to financial agents} \label{sec:discussion}

%Paragraph on summary of paper
This paper found that advisors at the dealership generally similarly under-risk their clients' portfolios relative to their stated risk preferences across the board, regardless of the profile risk. Advisors tend to put clients into medium or higher profile risks, but consistently medium or lower portfolios. We found that the clustering methodology from our previous paper \citep{thompson20} showed that regardless of the cluster, profile VaRs increased over time, portfolio VaRs remained consistent except for the active traders, and discrepancies were similarly under-risked and decreasing over time. We found that VaR aligned well with our cluster personas--older investors typically had the lowest portfolio VaR while early savers had the highest portfolio risk, and active traders changed their portfolio risk often. Discretionary advisors set up their clients in profile and portfolio VaR consistently over time, while non-discretionary profile risk increased with time. Discretionary advisors tended to take lower portfolio risks with their older investors, but higher risks with all other clusters when compared to non-discretionary advisors. 

%%%Adams
To compare to similar research, \cite{corter06} took a weighted average of the allocation to each bucket, with higher weights applied to riskier buckets. The weights were determined via consultation with a panel of experts.  Though well-founded and entirely reasonable, this approach is designed to determine if risk questionnaires correlate with hypothetical asset risk allocations, and is not directly customizable for application to our data.  An alternative option that may meet these goals (design and customization) would be to consider portfolio volatility.  Drawbacks of using portfolio volatility include the fact that its units are not overly meaningful to clients (what exactly does a 40\% volatility mean, compared to a 20\% volatility) and that it does not reflect the ever-present tension between risk and return.  

We assessed the risk of a particular allocation using the well-known VaR metric to address those issues.  Any VaR calculation requires assumptions about the statistical properties of the underlying asset returns standard deviations and correlations. In the present context, a VaR calculation requires assumptions about how investments are made within each risk bucket.  We assume that returns are multivariate normally distributed, an assumption that is very quickly relaxed to include thicker tails (such as multivariate $t$-distribution) and asymmetries (such as mixture models).  We also assume that every dollar a client invests in a particular risk bucket is invested in a representative ETF from that class. The representative asset for a given class was selected from iShares ETFs. Given more detailed information on an individual client's holdings, and an adequate amount of historical data on those assets to get estimates of expected returns, standard deviations, and correlations, our approach is even more directly applicable and customizable for advisors.
%%%

We conclude this paper by discussing how we have shown that VaR benefits four financial agents: clients, advisors, dealers, and regulators. Clients can monitor their risk using VaR to understand better how their asset portfolios' changes affect risk at the account level. VaR is a particularly useful communication tool so that clients can understand risk in dollar amounts relative to their total market value to conceptualize how well different asset portfolios match up with stated profile risk preferences.

The advisors in this study appeared to manage risk actively on behalf of their clients. They did so in the context of their clients' explicit or derived risk tolerance or suitability as measured by value at risk. Researchers have theorized, studied and written about the role of financial advisors \citep{foerster17,linnainmaa18}. Much of that work has been grounded in the assumption that advice should generate returns and, in particular, excess returns or alpha. Our work would appear to suggest that advisors play an essential role in terms of the perennial balance between return and its dependant variable risk, and that an advisor's role is, therefore, more complex than the simple pursuit of alpha.

By studying this behaviour empirically and using a relatively straightforward measure, our work can be extended from the chalkboard to the floorboards through client-facing platforms such as robo-advisors, financial planning software or client statements. Moreover, it can be done while respecting existing regulatory frameworks and a client's best-interest standard of care. Using the same measure for both profile risk and revealed risk--particularly the difference between the two--in real-time, we believe VaR could become an important tool in helping clients, advisors, dealers, and regulators monitor a careful balance between all the variables and stakeholders. 

Advisors can use this methodology to understand each client's current risk in terms of a dollar amount or portfolio percentage, and understand how each client's stated risk tolerance aligns with any prefabricated investor portfolios. We believe this will improve communications with the client and help clients understand better how much of their wealth is at risk and what a ``significant" loss looks like for their worst day out of one hundred. Additionally, an advisor can quickly understand how a change in the overall assets across a firm will affect all of their client's risk positions and easily detect any clients that may be put into a risky position that they are not comfortable with stated goals. Advisors can also use VaR to understand all the assets under their entire firm's ownership, similar to how VaR is used for larger financial institutions. 

%Paragraph on how this benefits dealers
Dealerships and regulators will benefit from VaR by having a methodology for either advisors to report on how they manage suitability or real-time monitoring of advisor behaviours. Dealers, and their advisors, can also demonstrate an additional measure of their value for clients. Moreover, regulators can use VaR as a tool to monitor suitability or test for intrinsic stress at the client, dealer and industry level.  

\section*{Acknowledgements}

The authors would like to thank Nathan Phelps (Wilfrid Laurier University), Andrew Sarta (Ivey Business School), Poornima Vinoo (Ivey Business School), Matt Davison (Western University), Lori Weir (Four Eyes Financial), Kendall McMenamon (Four Eyes Financial), Philip Patterson (Four Eyes Financial), Lucas Loughead (Four Eyes Financial), and the many members of our data donor team for their valuable input and insights that improved the content and writing of this document.

% and the many members of our data donor team

\newpage

% \printbibliography
\bibliography{references}

\begin{thebibliography}{29}
\providecommand{\natexlab}[1]{#1}
\providecommand{\url}[1]{\texttt{#1}}
\expandafter\ifx\csname urlstyle\endcsname\relax
  \providecommand{\doi}[1]{doi: #1}\else
  \providecommand{\doi}{doi: \begingroup \urlstyle{rm}\Url}\fi

\bibitem[Aspara(2013)]{aspara13}
Jaakko Aspara.
\newblock The role of product and brand perceptions in stock investing: effects
  on investment considerations, optimism and confidence.
\newblock \emph{Journal of Behavioral Finance}, 14\penalty0 (3):\penalty0
  195--212, 2013.

\bibitem[Corter and Chen(2006)]{corter06}
James~E Corter and Yuh-Jia Chen.
\newblock Do investment risk tolerance attitudes predict portfolio risk?
\newblock \emph{Journal of business and psychology}, 20\penalty0 (3):\penalty0
  369, 2006.

\bibitem[Diacon and Ennew(2001)]{diacon01}
Stephen Diacon and Christine Ennew.
\newblock Consumer perceptions of financial risk.
\newblock \emph{The Geneva Papers on Risk and Insurance. Issues and Practice},
  26\penalty0 (3):\penalty0 389--409, 2001.

\bibitem[Dybvig and Polemarchakis(1981)]{dybvig81}
Philip Dybvig and Heraklis Polemarchakis.
\newblock Recovering cardinal utility.
\newblock \emph{The Review of Economic Studies}, 48\penalty0 (1):\penalty0
  159--166, 1981.

\bibitem[Dybvig and Rogers(1997)]{dybvig97}
Philip~H Dybvig and Leonard~CG Rogers.
\newblock Recovery of preferences from observed wealth in a single realization.
\newblock \emph{The Review of Financial Studies}, 10\penalty0 (1):\penalty0
  151--174, 1997.

\bibitem[Finke and Guillemette(2016)]{finke16}
Michael~S Finke and Michael~A Guillemette.
\newblock Measuring risk tolerance: A review of literature.
\newblock \emph{Journal of Personal Finance}, 15\penalty0 (1):\penalty0 63,
  2016.

\bibitem[Foerster et~al.(2017)Foerster, Linnainmaa, Melzer, and
  Previtero]{foerster17}
Stephen Foerster, Juhani~T. Linnainmaa, Brian~T. Melzer, and Alessandro
  Previtero.
\newblock Retail financial advice: does one size fit all?
\newblock \emph{The Journal of Finance}, 72\penalty0 (4):\penalty0 1441--1482,
  2017.

\bibitem[Ganzach(2000)]{ganzach00}
Yoav Ganzach.
\newblock Judging risk and return of financial assets.
\newblock \emph{Organizational behavior and human decision processes},
  83\penalty0 (2):\penalty0 353--370, 2000.

\bibitem[Grable and Lytton(2001)]{grable01}
John~E Grable and Ruth~H Lytton.
\newblock Assessing the concurrent validity of the scf risk tolerance question.
\newblock \emph{Journal of Financial Counseling and Planning}, 12\penalty0
  (2):\penalty0 43, 2001.

\bibitem[Guillemette et~al.(2012)Guillemette, Finke, and
  Gilliam]{guillemette12}
Michael Guillemette, Michael~S Finke, and John Gilliam.
\newblock Risk tolerance questions to best determine client portfolio
  allocation preferences.
\newblock \emph{Journal of Financial Planning}, 25\penalty0 (5):\penalty0
  36--44, 2012.

\bibitem[Guillemette et~al.(2015)Guillemette, Browning, and
  Payne]{guillemette15}
Michael Guillemette, Chris Browning, and Patrick Payne.
\newblock Don’t like the picture? change the frame: the impact of cognitive
  ability and framing on risky choice.
\newblock \emph{Applied Economics Letters}, 22\penalty0 (18):\penalty0
  1515--1518, 2015.

\bibitem[Jorion(2007)]{jorion07}
Philippe Jorion.
\newblock \emph{Value at risk: the new benchmark for managing financial risk}.
\newblock The McGraw-Hill Companies, Inc., 2007.

\bibitem[Kahneman and Tversky(2013)]{kahneman13}
Daniel Kahneman and Amos Tversky.
\newblock Prospect theory: An analysis of decision under risk.
\newblock In \emph{Handbook of the fundamentals of financial decision making:
  Part I}, pages 99--127. World Scientific, 2013.

\bibitem[Kuester et~al.(2006)Kuester, Mittnik, and Paolella]{kuester06}
Keith Kuester, Stefan Mittnik, and Marc~S Paolella.
\newblock Value-at-risk prediction: A comparison of alternative strategies.
\newblock \emph{Journal of Financial Econometrics}, 4\penalty0 (1):\penalty0
  53--89, 2006.

\bibitem[Kullback and Leibler(1951)]{kullback51}
Solomon Kullback and Richard~A Leibler.
\newblock On information and sufficiency.
\newblock \emph{The annals of mathematical statistics}, 22\penalty0
  (1):\penalty0 79--86, 1951.

\bibitem[Likert(1932)]{likert32}
Rensis Likert.
\newblock A technique for the measurement of attitudes.
\newblock \emph{Archives of psychology}, 1932.

\bibitem[Linciano and Soccorso(2012)]{linciano12}
Nadia Linciano and Paola Soccorso.
\newblock Assessing investors' risk tolerance through a questionnaire.
\newblock \penalty0 (4), 2012.
\newblock URL \url{https://ssrn.com/abstract=2207958}.

\bibitem[Linnainmaa et~al.(2018)Linnainmaa, Melzer, and
  Previtero]{linnainmaa18}
Juhani~T. Linnainmaa, Brian~T. Melzer, and Alessandro Previtero.
\newblock The misguided beliefs of financial advisors.
\newblock \emph{Kelley School of Business Research Paper}, \penalty0 (18-9),
  2018.

\bibitem[Loewenstein et~al.(2001)Loewenstein, Weber, Hsee, and
  Welch]{loewenstein01}
George~F Loewenstein, Elke~U Weber, Christopher~K Hsee, and Ned Welch.
\newblock Risk as feelings.
\newblock \emph{Psychological bulletin}, 127\penalty0 (2):\penalty0 267, 2001.

\bibitem[Michael et~al.(2015)Michael, Yao, and James~III]{michael15}
A~Michael, Rui Yao, and Russell~N James~III.
\newblock An analysis of risk assessment questions based on loss-averse
  preferences.
\newblock \emph{Journal of Financial Counseling and Planning Volume},
  26\penalty0 (1):\penalty0 17--29, 2015.

\bibitem[{Microsoft Corporation}(2021)]{msexcel}
{Microsoft Corporation}.
\newblock Microsoft excel, 2021.
\newblock URL \url{https://office.microsoft.com/excel}.

\bibitem[{Ontario Securities Commission}(2009)]{ont09}
{Ontario Securities Commission}.
\newblock {CSA} staff notice: 33-315 - suitability obligation and know your
  product.
\newblock September 2009.

\bibitem[{Ontario Securities Commission}(2014)]{ont14}
{Ontario Securities Commission}.
\newblock {CSA} staff notice 31-336 – guidance for portfolio managers, exempt
  market dealers and other registrants on the know-your-client,
  know-your-product and suitablility obligations.
\newblock Jan 2014.

\bibitem[{Ontario Securities Commission}(2019)]{ont19}
{Ontario Securities Commission}.
\newblock Reforms to enhance the client-registrant relationship (client focused
  reforms). notice of amendments to national instruments 31-103 and companion
  policy {31-103CP}.
\newblock Oct 2019.

\bibitem[Samuelson(1948)]{samuelson48}
Paul~A Samuelson.
\newblock Consumption theory in terms of revealed preference.
\newblock \emph{Economica}, 15\penalty0 (60):\penalty0 243--253, 1948.

\bibitem[Samuelson(1975)]{samuelson75}
Paul~A Samuelson.
\newblock Lifetime portfolio selection by dynamic stochastic programming.
\newblock \emph{Stochastic Optimization Models in Finance}, pages 517--524,
  1975.

\bibitem[Sharma and Schoengold(2016)]{sharma16}
Sankalp Sharma and Karina Schoengold.
\newblock Comparison of stated and revealed risk preferences using
  safety-first.
\newblock In \emph{{AgEcon} 2016 Annual Meeting}. Research in Agricultural
  Economics, 2016.

\bibitem[Thompson et~al.(2021)Thompson, Feng, Reesor, and Grace]{thompson20}
John R.~J. Thompson, Longlong Feng, R.~Mark Reesor, and Chuck Grace.
\newblock Know your clients’ behaviours: A cluster analysis of financial
  transactions.
\newblock \emph{Journal of Risk and Financial Management}, 14\penalty0 (2),
  2021.
\newblock ISSN 1911-8074.
\newblock \doi{10.3390/jrfm14020050}.
\newblock URL \url{https://www.mdpi.com/1911-8074/14/2/50}.

\bibitem[Wahl and Kirchler(2020)]{wahl20risk}
Ingrid Wahl and Erich Kirchler.
\newblock Risk screening on the financial market (risc-fm): A tool to assess
  investors’ financial risk tolerance.
\newblock \emph{Cogent Psychology}, 7\penalty0 (1):\penalty0 1714108, 2020.

\end{thebibliography}

\newpage

\section*{Appendix A -- Generalized difference between profile and portfolio risk}
Before comparing between profiles and portfolios, we first consider a natural evaluation of an individual risk allocation--the eye-test. Table \ref{tbl:example1} shows realistic allocations found in our dataset. Suppose a financial advisor with a client has prescribed the profile risk allocation in the first row of Table \ref{tbl:example1}, and the client's actual portfolio risk selection in the second row. The eye-test reveals they are clearly different, where the client is under-risked. A second portfolio selection is shown in the third row, which the eye-test shows that the portfolio is very under-risked and more under-risked than above portfolio is over-risked.  The eye-test is a natural method of comparing categories, but an advisor cannot conduct daily eye-tests on, say, over 500 accounts. 

\begin{table}[hbtp!]
\centering
\caption{An example of profile and portfolio risks before and after a re-balancing of assets. The discrepancy is calculated using Equation (\ref{eqn:discrepancy}) with $P_{i,i}=i$ and $P_{i,j}=0,$ $i\neq j$.} 
\label{tbl:example1}
\begin{tabular}{c|ccccc|c}
Risk & Low & Low-Medium & Medium & Medium-High & High & Discrepancy \\ \hline
Profile & 0 & 0 & 0 & 80 & 20 & \\
Portfolio (Before) & 0 & 16 & 84 & 0 & 0 & 49280 \\
Portfolio (After) & 0 & 94 & 6 & 0 & 0 & 45380 \\ 
\end{tabular}
\end{table}
Next, we shift to mathematical language so that we can implement financial concepts into our customizable categorical comparisons. Let $k\in\{1,2,3,4,5\}$ be a Likert scale index \citep{likert32} for the profile risk categories and $j\in\{1,2,3,4,5\}$ for portfolio risk. Consider profile and portfolio risk selection by percentage $x=(x_{1},x_{2},x_{3},x_{4},x_{5})$ and $y=(y_{1},y_{2},y_{3},y_{4},y_{5})$, which means $x_k,y_{j}\in [0,1]$ and $\sum_{k=1}^5x_k=1,\sum_{j=1}^5y_j=1$.

Consider a generalized comparison between allocation and selection vectors by using 
\begin{eqnarray}\label{eqn:discrepancy}
d(x,y)=(x-y)^{T}P(x-y).
\end{eqnarray}
where $(x-y)$ is a element-by-element comparison of the percent allocations (same as our first comparison), and $P$ is a $5\times5$ matrix that controls customization. Before we choose our perfect customized model, let's consider some basic rules on how the function $d(x,y)$ should behave. If the profile allocation and portfolio selection are the same ($x=y$), then $d(x,y)=0$. The larger the comparison, the more different the allocation and selection of risk are. 

If we view allocations and selections as points in Euclidean space, it most natural to compute discrepancy via the now Euclidean metric in \ref{eqn:discrepancy}, where $P$ is now a positive definite  matrix. Some of the issues of this approach are as follows:
\begin{itemize}
\item
If $P$ is the identity matrix given as 
\begin{eqnarray*}
  P := \begin{bmatrix} 1 & 0 & 0& 0&0 \\0 & 1 & 0& 0&0 \\0 & 0 & 1& 0&0 \\0 & 0 & 0& 1&0 \\0 & 0 & 0& 0&1 \end{bmatrix}.
\end{eqnarray*}
then $d(x,y)$ is now a sum of the squared deviation between each risk category. The two portfolio selections of 100\% low and 100\% low-medium, are equidistant from the profile allocation in Table \ref{tbl:example1}.  This is clearly unreasonable from a financial perspective, as the difference is much larger than for the 100\% low allocation than the 100\% low-medium allocation.
\item
If $P$ is a diagonal matrix we can alleviate the problem indicated above.  For example, setting $P_{i,i}=i$ gives 
\begin{eqnarray*}
  P := \begin{bmatrix} 1 & 0 & 0& 0&0 \\0 & 2 & 0& 0&0 \\0 & 0 & 3& 0&0 \\0 & 0 & 0& 4&0 \\0 & 0 & 0& 0&5 \end{bmatrix}.
\end{eqnarray*}
where $d(x,y)=\sum_{i=1}^5i(x_i-y_i)^2$ heavily penalizes discrepancies in higher risk categories. This ensures that the distance between 100\% high and 100\% low-medium is twice as large as that between 100\% low and 100\% low-medium. Unfortunately, this approach introduces new complexities that we discovered when applying it to our data.  The client in Table \ref{tbl:example1} made a trade on particular date, that clearly moved them further from their profile allocation.  According to the metric, however, the trade moved them \textit{closer}, which is clearly unreasonable.
\item The previous two matrices have only zeros as off diagonal terms and therefore the discrepancy that uses the those matrices only compare the same categories. Consider the penalization matrix that penalizes off diagonal terms given by
\begin{eqnarray*}
  P := \begin{bmatrix} 1 & 0 & 0& 0&1 \\0 & 1 & 0& 0&0 \\0 & 0 & 1& 0&0 \\0 & 0 & 0& 1&0 \\0 & 0 & 0& 0&1 \end{bmatrix}.
\end{eqnarray*}
which yields $d(x,y)=\sum_{i=1}^5(x_i-y_i)^2+(x_1-y_1)*(x_5-y_5)$ The new term in the sum--a penalization of misallocations in the low category has been added, but only if there is also a misallocation in the high category (and vice versa). As it turns out, off-diagonal terms place a heavier penalty on not just \textit{if} there is a misallocation, but \textit{how} they are misallocated. This comparison will show us that not all misallocations are equal. A penalization matrix that penalizes how far the misallocations are from the stated goals is
\begin{eqnarray*}
  P := \begin{bmatrix} 0 & 1 & 2& 3&4 \\1 & 0 & 1& 2&3 \\2 & 1 & 0& 1&2 \\3 & 2 & 1& 0& 1\\4 & 3 & 2& 1&0 \end{bmatrix}.
\end{eqnarray*}
Notice that the diagonal terms are zero, which works in this configuration since we are considering all misallocations in the off-diagonal terms. We could also relax the assumption that the penalization matrix is symmetric, and allow for higher penalties for over-risked profiles. A penalization matrix where we penalize an over-risked profile more than under-risked could be
\begin{eqnarray*}
  P := \begin{bmatrix} 0 & 1 & 2& 3&4 \\1 & 0 & 1& 2&3 \\1 & 1 & 0& 1&2 \\1 & 1 & 1& 0& 1\\1 & 1 & 1& 1&0 \end{bmatrix}.
\end{eqnarray*}
\end{itemize}

Consider another perspective that views profile and portfolio risk  allocations and selections as discrete probability distributions. It is most natural to compute discrepancy via an information theoretic divergence such as Kullback-Liebler (K-L) divergence \citep{kullback51}. Unfortunately this approach also leads to non-useful results.  For example, consider a profile of 100\% low-medium, and two portfolio selections of 100\% low and 100\% high give the same K-L divergence with the same sign, where we lose magnitude of the difference and whether the portfolio is under- or over-risked. Other divergences were explored, but these suffered from the same problems as the K-L divergences. 

These metrics and divergences are designed to directly compare risk categories to evaluate profile and portfolio risk alignment. The advantage of the proposed metric is that they can be customized to calculate profile and portfolio risk alignment. We introduced equal and unequal successive weightings for each category to inject ordering of the categories, and allow for the natural conceptualization of risk to be included in the calculation. However, there are ever-present problems with each calculation method that ruin the overall interpretations of results across a dealership. In conclusion, we suggest the financial concept of value-at-risk is a much more feasible method to compare profile risk allocations and portfolio risk selections.
\newpage 

\section*{Appendix B - Value-at-Risk methodology}

Consider $x$ be the profile risk allocation. Let $\mu$ denote the mean return vector and $\Sigma$ be the covariance matrix of the representative risk category ETFs. The $\alpha$-level VaR on a KYC risk profile is given by
\begin{equation*}
\mathrm{VaR}_{\alpha}(x) = x^{T}\mu+\sqrt{x^{T}\Sigma x} \cdot z_{\alpha}\;,
\end{equation*}
where $z_{\alpha}$ is an appropriate $\alpha$-quantile from a standard normal distribution.  In this report, we let $\alpha=0.01$. We have chosen five iShares ETFs with $\mu=[0.52,1.97,2.21,2.93,4.23]$, $\sigma = [0.13,5.53,6.48,9.68,15.22]$, and
\begin{eqnarray*}
    \rho = \begin{bmatrix} 1 & -0.22 & -0.16 & -0.23 & 0.07 \\
                           -0.22 & 1 & 0.79 & 0.59 & 0.12\\
                           -0.16 & 0.79 & 1 & 0.78 & 0.31 \\
                           -0.23 & 0.59 & 0.78 & 1 & 0.06\\
                           0.07 & 0.12 & 0.31 & 0.06 & 1 \end{bmatrix}
\end{eqnarray*}
which yield
\begin{eqnarray*}
    \Sigma = \sigma^T\rho\sigma=  \begin{bmatrix} 0.016900 & -0.158158& -0.134784 &-0.289432 &  0.138502\\
-0.158158 & 30.580900& 28.309176& 31.582936  &10.099992\\
 -0.134784 &28.309176 &41.990400& 48.926592&  30.573936\\
 -0.289432& 31.582936& 48.926592& 93.702400&   8.839776\\
  0.138502 &10.099992& 30.573936&  8.839776& 231.648400 \end{bmatrix}
\end{eqnarray*}

\section*{Appendix C - Distribution of VaRs across KYC information}

\begin{figure}
    \centering
    \begin{subfigure}[t]{0.45\textwidth}
        \centering
        \includegraphics[width=\linewidth]{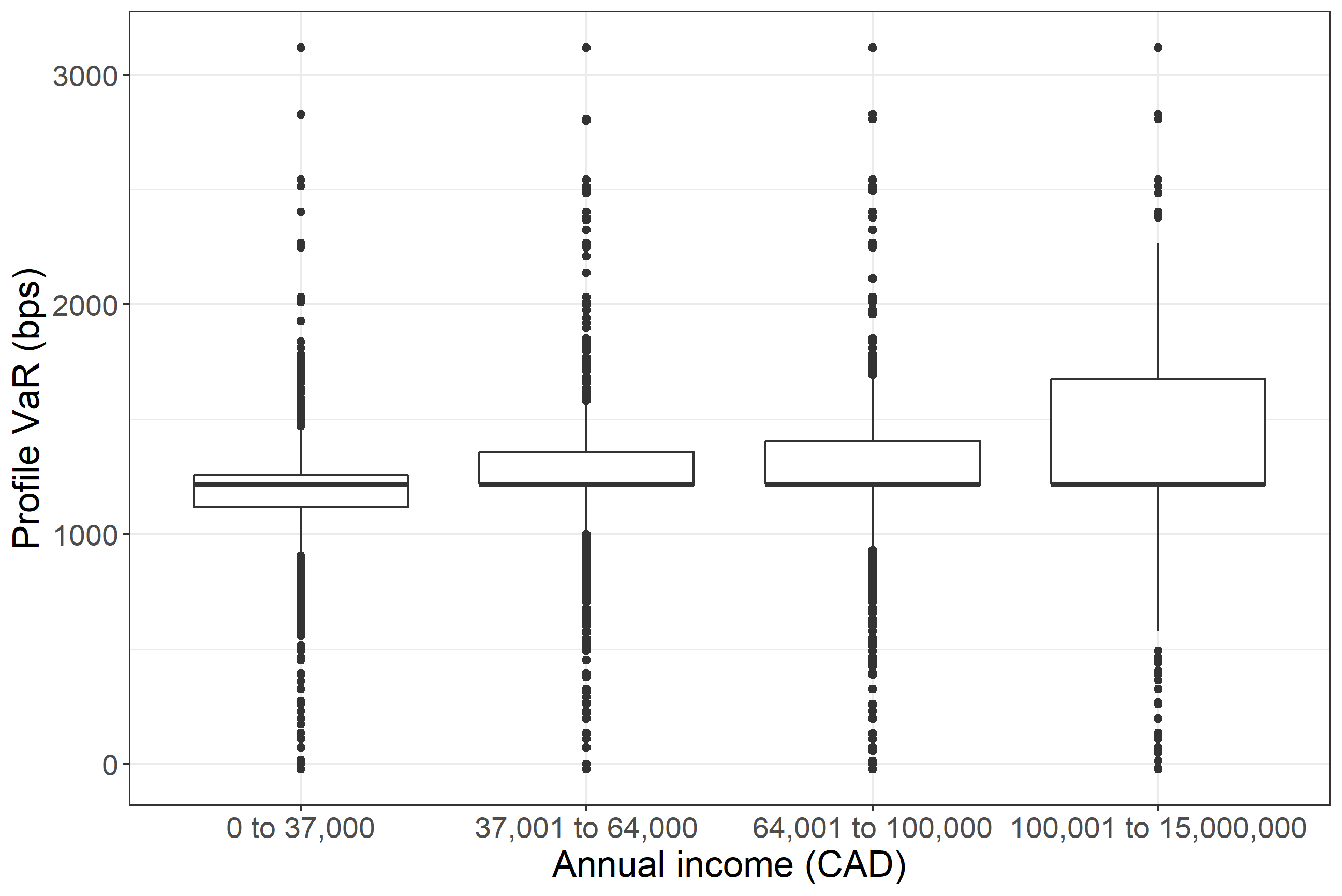}
        \caption{Profile VaR} \label{fig:advisorPrescribedVaRIncomeBoxplot}
    \end{subfigure}
    \hfill
    \begin{subfigure}[t]{0.45\textwidth}
        \centering
        \includegraphics[width=\linewidth]{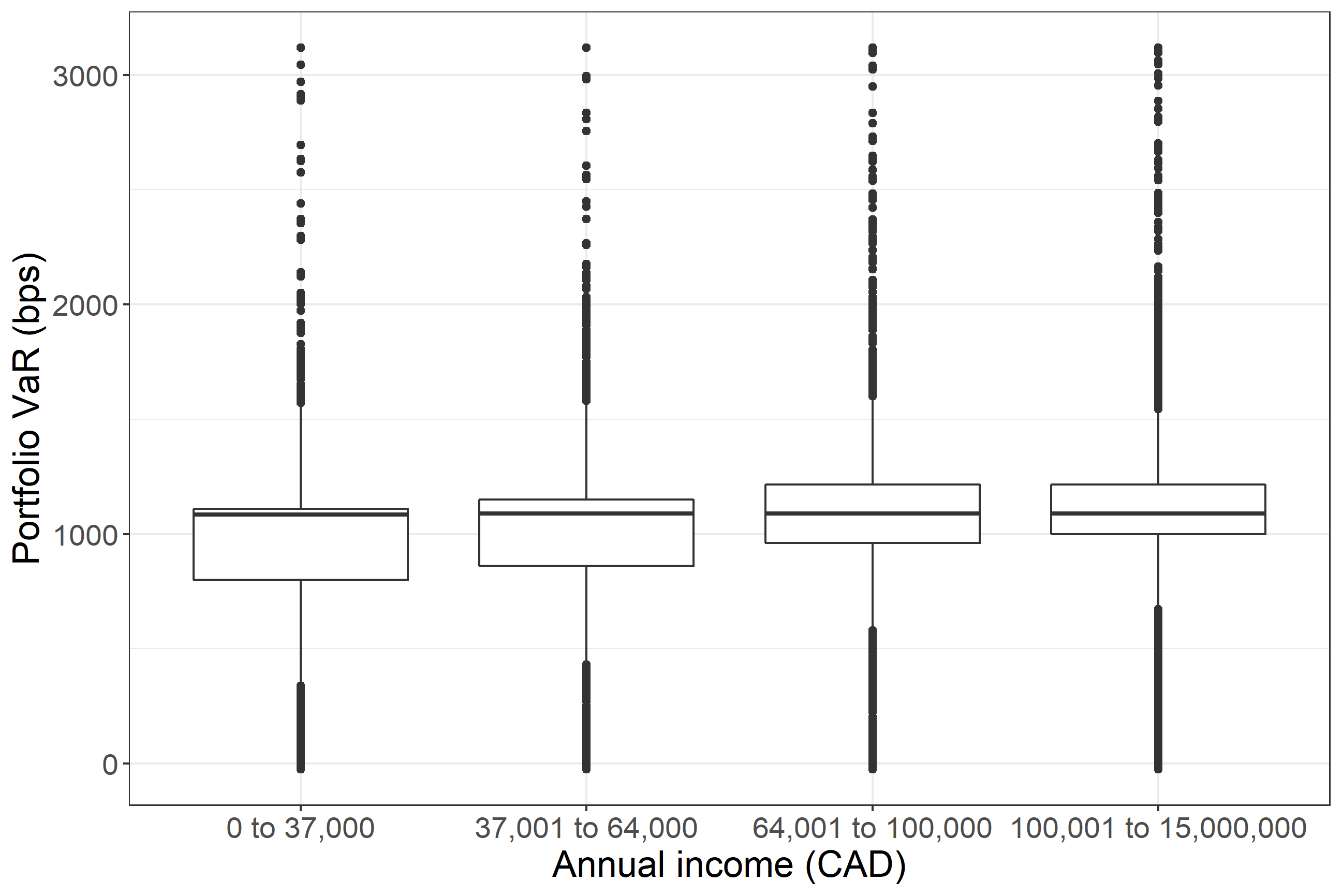} 
        \caption{Portfolio VaR} \label{fig:advisorActualVaRIncomeBoxplot}
    \end{subfigure}
    \vspace{1cm}
    \begin{subfigure}[t]{0.45\textwidth}
        \centering
        \includegraphics[width=\linewidth]{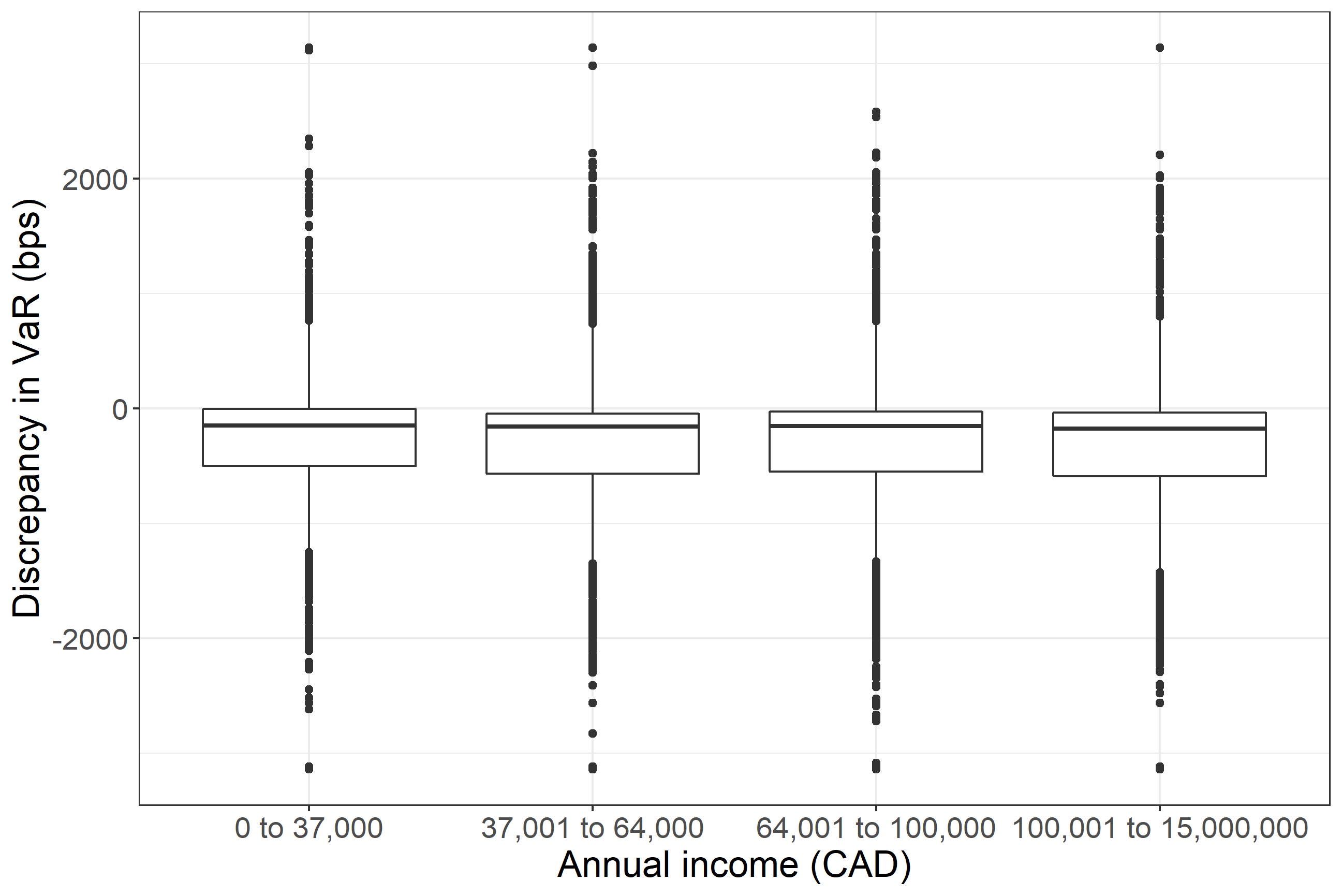} 
        \caption{Discrepancy in VaR} \label{fig:advisorDiscrepancyIncomeBoxplot}
    \end{subfigure}
     \caption{Boxplots of profile VaR (top left panel), portfolio VaR (top right panel) and the discrepancy in VaR (lower panel) on August 12th 2019 by annual income. Each boxplot represents 25\% of the data, with the boxplots in each panel from left to right represent the first, second, third, and fourth quartiles of annual incomes.}
     \label{fig:boxplotsIncome}
\end{figure}

Next, we investigate a series of boxplots for the variables in Table \ref{tbl:clientDetails}. Figure \ref{fig:boxplotsIncome} shows boxplots of VaR at for each of the income levels quartiles. We can see that as income increases, both profile (Figure \ref{fig:advisorPrescribedVaRIncomeBoxplot}) and portfolio (Figure \ref{fig:advisorActualVaRIncomeBoxplot}) show that as annual income increases, VaR increases. The discrepancy in Figure \ref{fig:advisorDiscrepancyIncomeBoxplot} shows that regardless of annual income, most clients are similarly under-risked.

\begin{figure}
    \centering
    \begin{subfigure}[t]{0.45\textwidth}
        \centering
        \includegraphics[width=\linewidth]{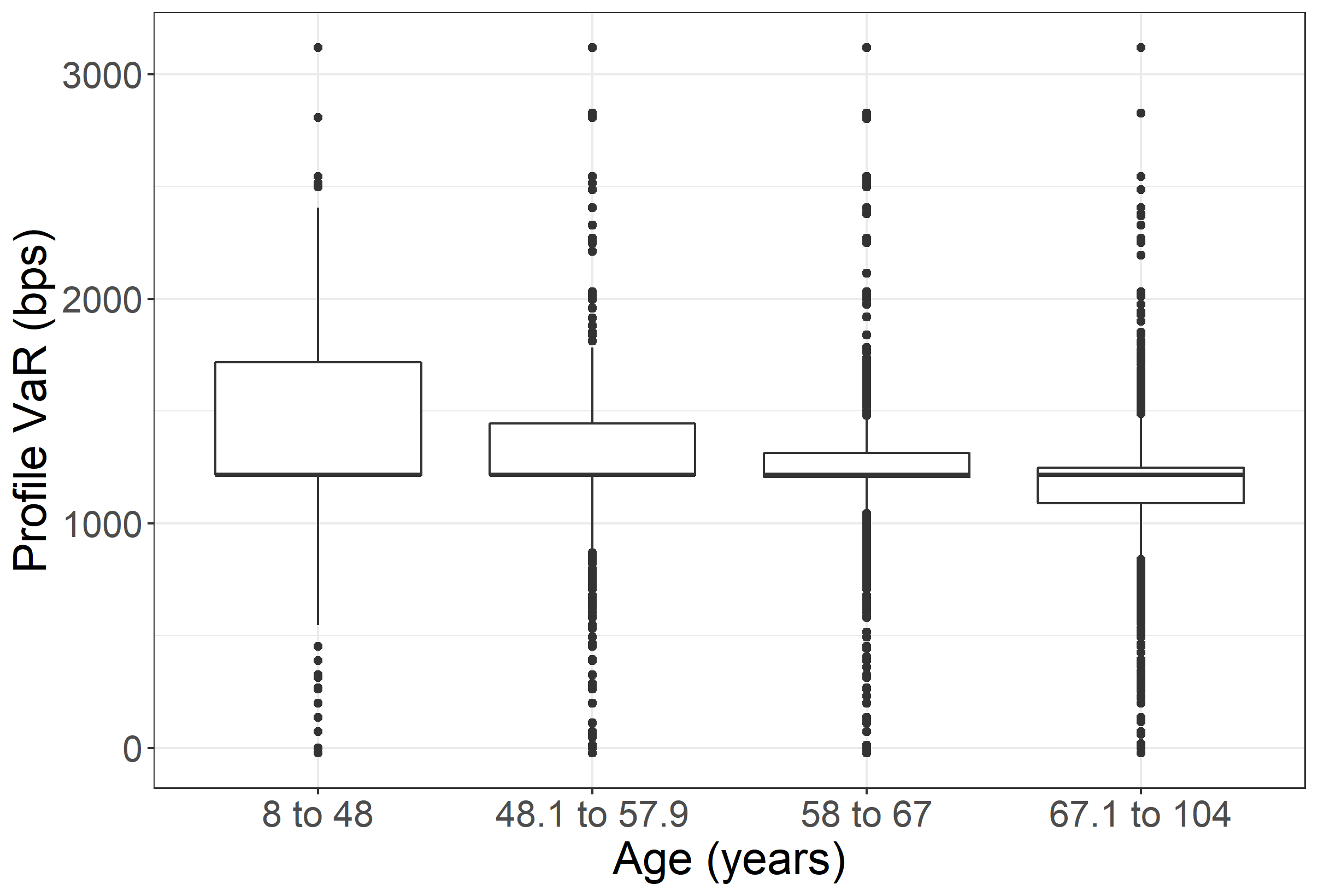}
        \caption{Profile VaR} \label{fig:advisorPrescribedVaRAge}
    \end{subfigure}
    \hfill
    \begin{subfigure}[t]{0.45\textwidth}
        \centering
        \includegraphics[width=\linewidth]{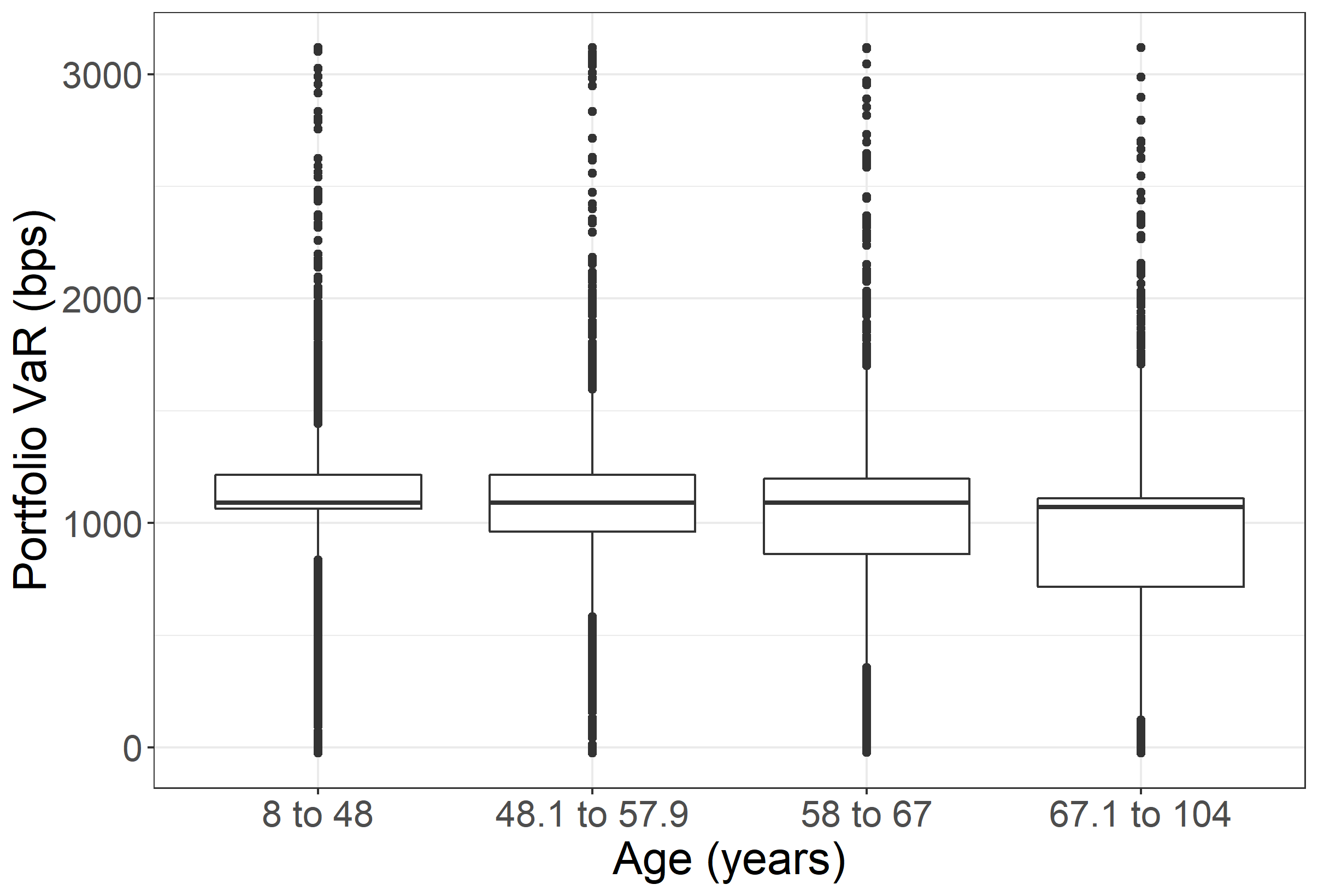} 
        \caption{Portfolio VaR} \label{fig:advisorActualVaRAge}
    \end{subfigure}
    \vspace{1cm}
    \begin{subfigure}[t]{0.45\textwidth}
        \centering
        \includegraphics[width=\linewidth]{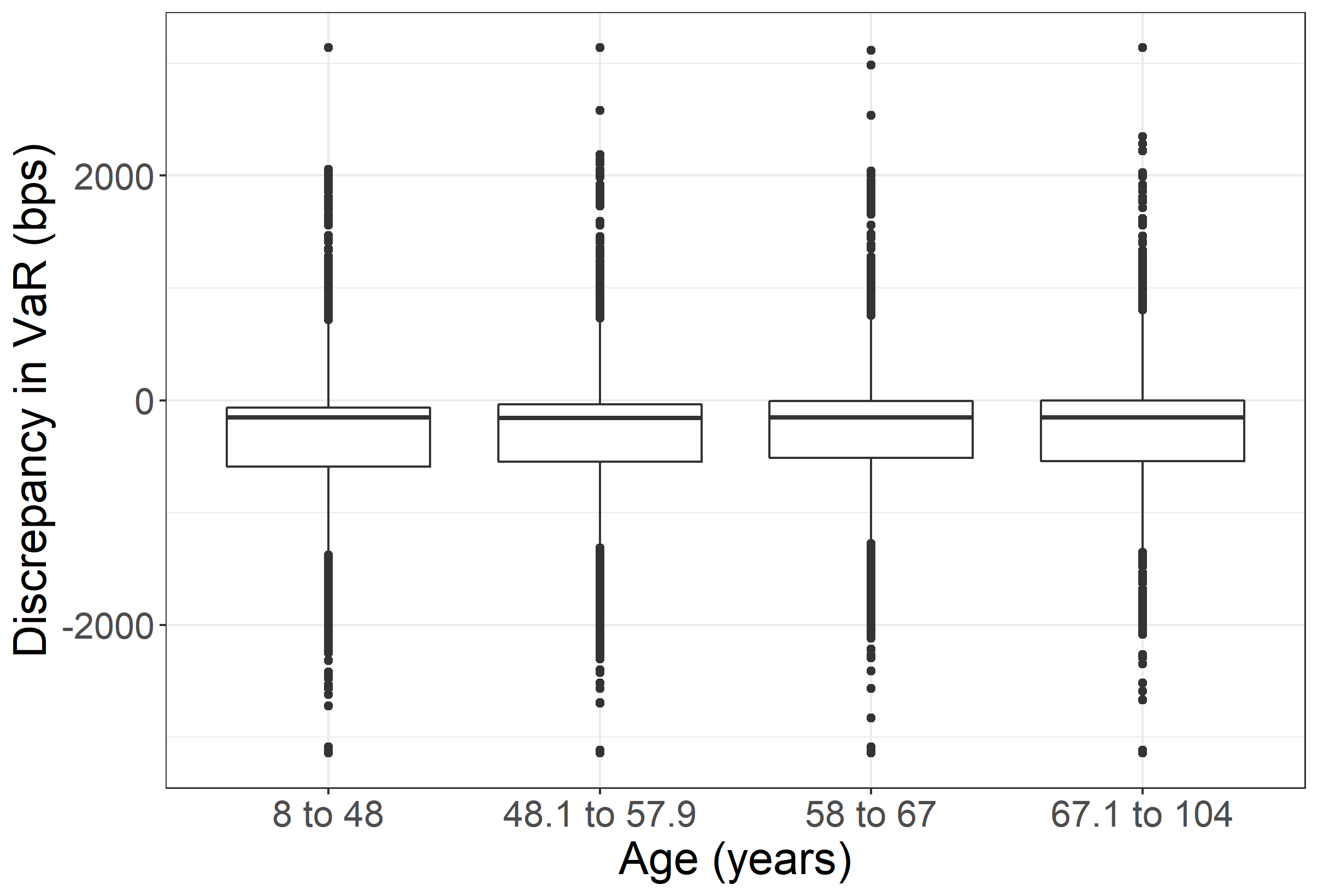} 
        \caption{Discrepancy in VaR} \label{fig:advisorDiscrepancyAge}
    \end{subfigure}
     \caption{Boxplots of profile VaR (top left panel), portfolio VaR (top right panel) and the discrepancy in VaR (lower panel) on August 12th 2019 by client ages. Each boxplot represents 25\% of the data, with the boxplots in each panel from left to right represent the first, second, third, and fourth quartiles of client ages.}
     \label{fig:boxplotsAge}
\end{figure}

Figure \ref{fig:boxplotsAge} shows boxplots of VaR for each of the quartiles of client ages. We can see that as income increases, both profile (Figure \ref{fig:advisorPrescribedVaRAge}) and portfolio (Figure \ref{fig:advisorActualVaRAge}) show that as client age increases, VaR decreases. The discrepancy in Figure \ref{fig:advisorDiscrepancyAge} shows that regardless of client age, most clients are similarly under-risked.

\begin{figure}
    \centering
    \begin{subfigure}[t]{0.45\textwidth}
        \centering
        \includegraphics[width=\linewidth]{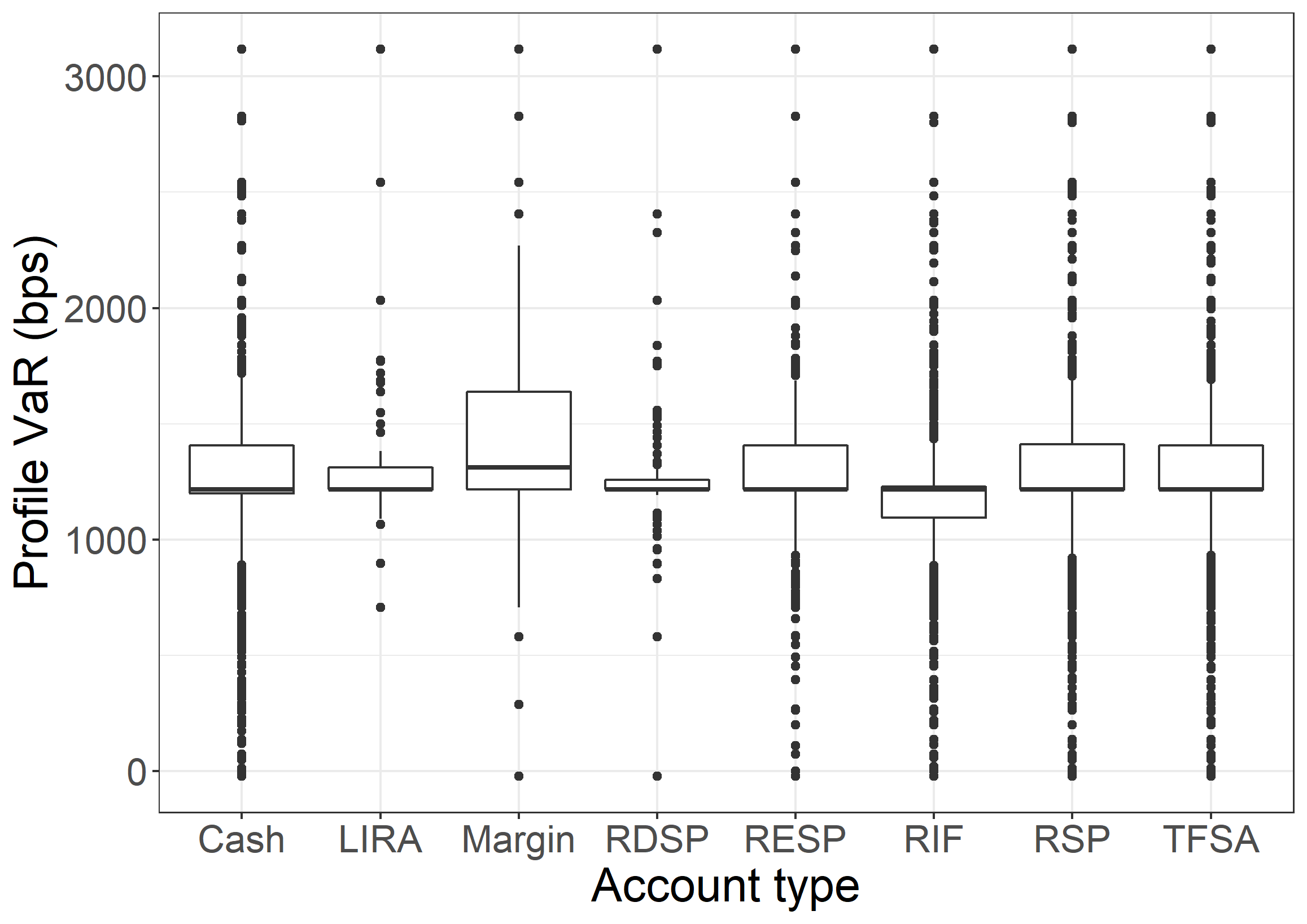}
        \caption{Profile VaR} \label{fig:advisorPrescribedVaRAcctType}
    \end{subfigure}
    \hfill
    \begin{subfigure}[t]{0.45\textwidth}
        \centering
        \includegraphics[width=\linewidth]{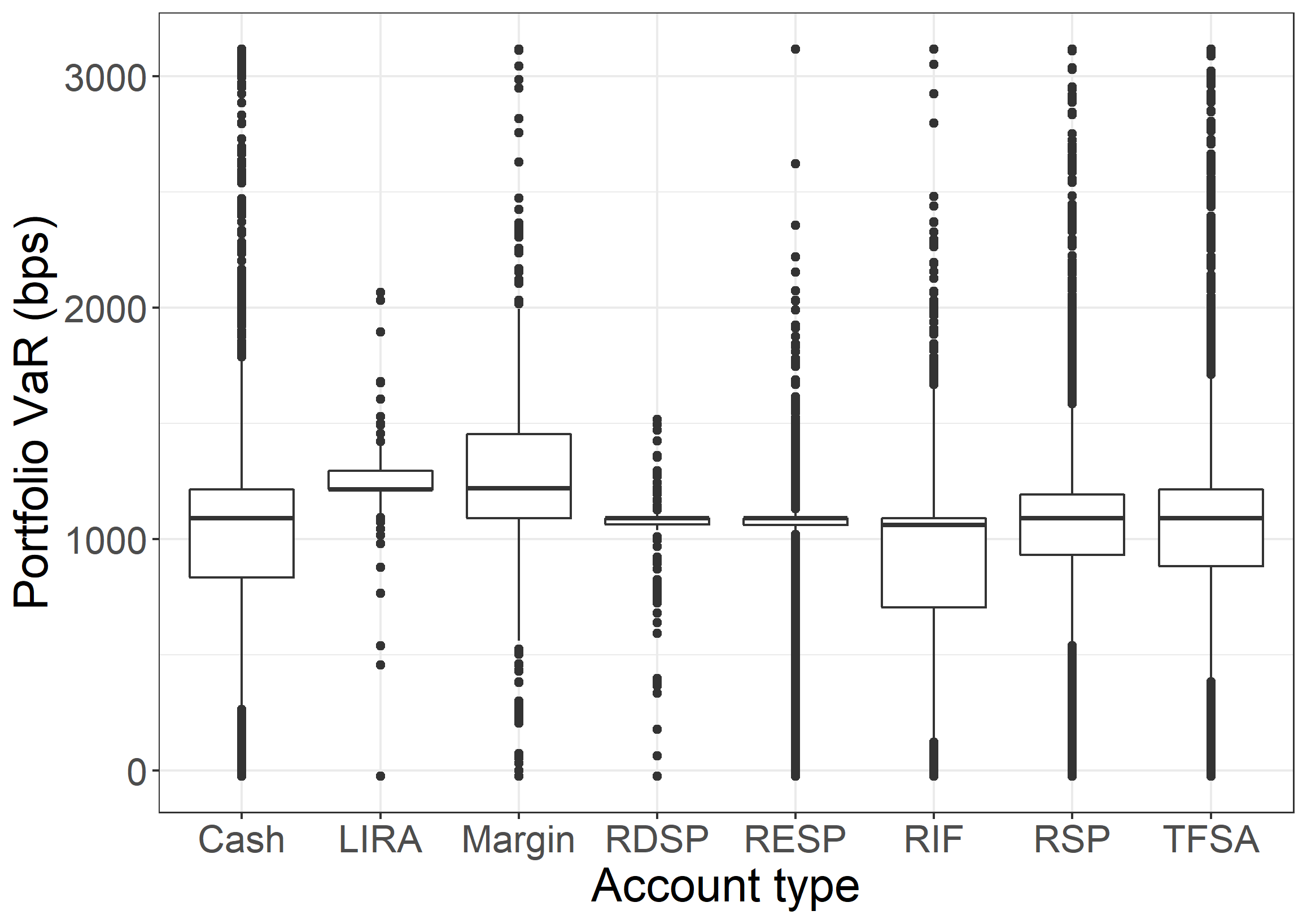} 
        \caption{Portfolio VaR} \label{fig:advisorActualVaRAcctType}
    \end{subfigure}
    \vspace{1cm}
    \begin{subfigure}[t]{0.45\textwidth}
        \centering
        \includegraphics[width=\linewidth]{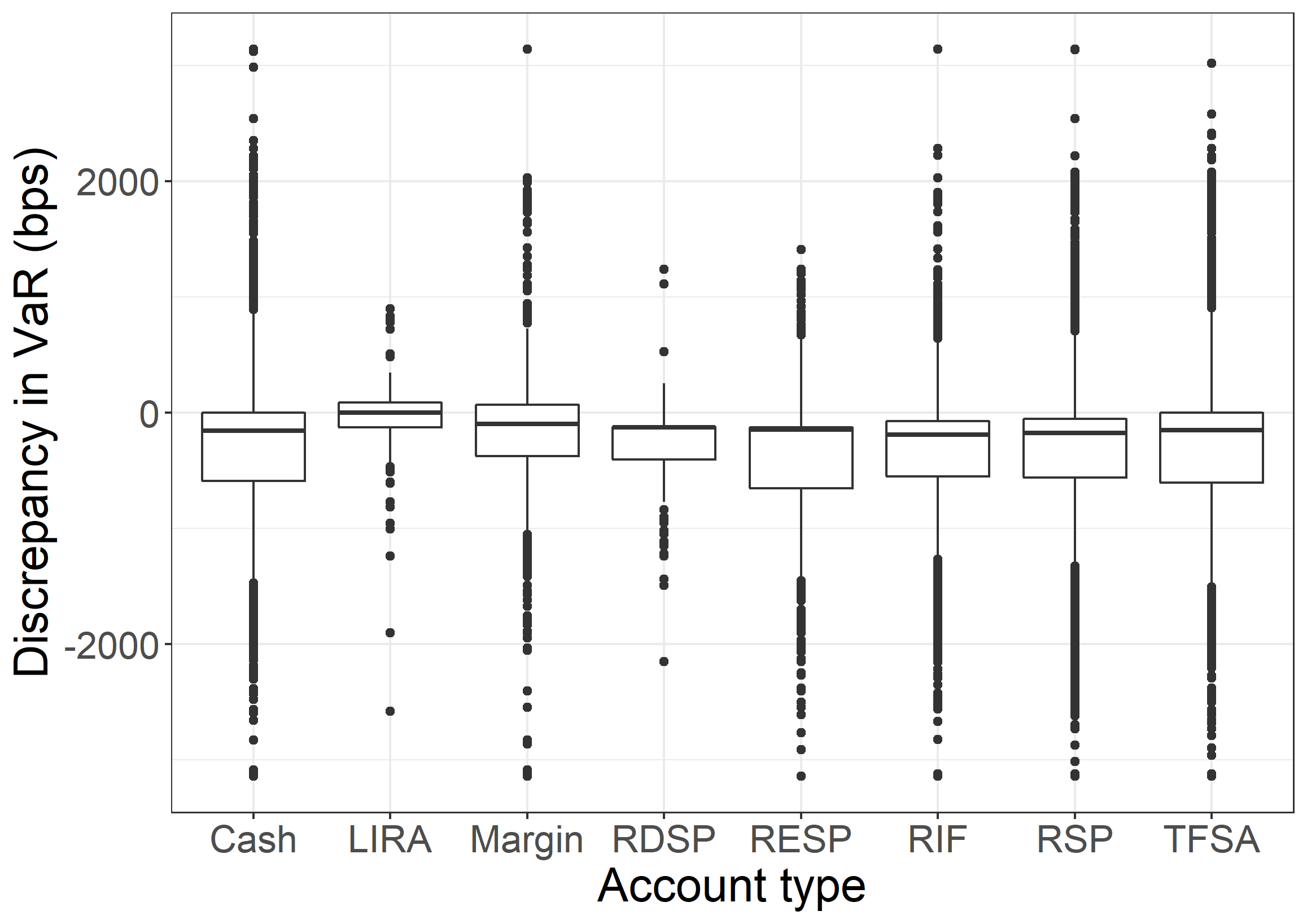} 
        \caption{Discrepancy in VaR} \label{fig:advisorDiscrepancyAcctType}
    \end{subfigure}
     \caption{Boxplots of profile VaR (top left panel), portfolio VaR (top right panel) and the discrepancy in VaR (lower panel) on August 12th 2019 by account type.}
     \label{fig:boxplotsAccountType}
\end{figure}

\begin{figure}
    \centering
    \begin{subfigure}[t]{0.45\textwidth}
        \centering
        \includegraphics[width=\linewidth]{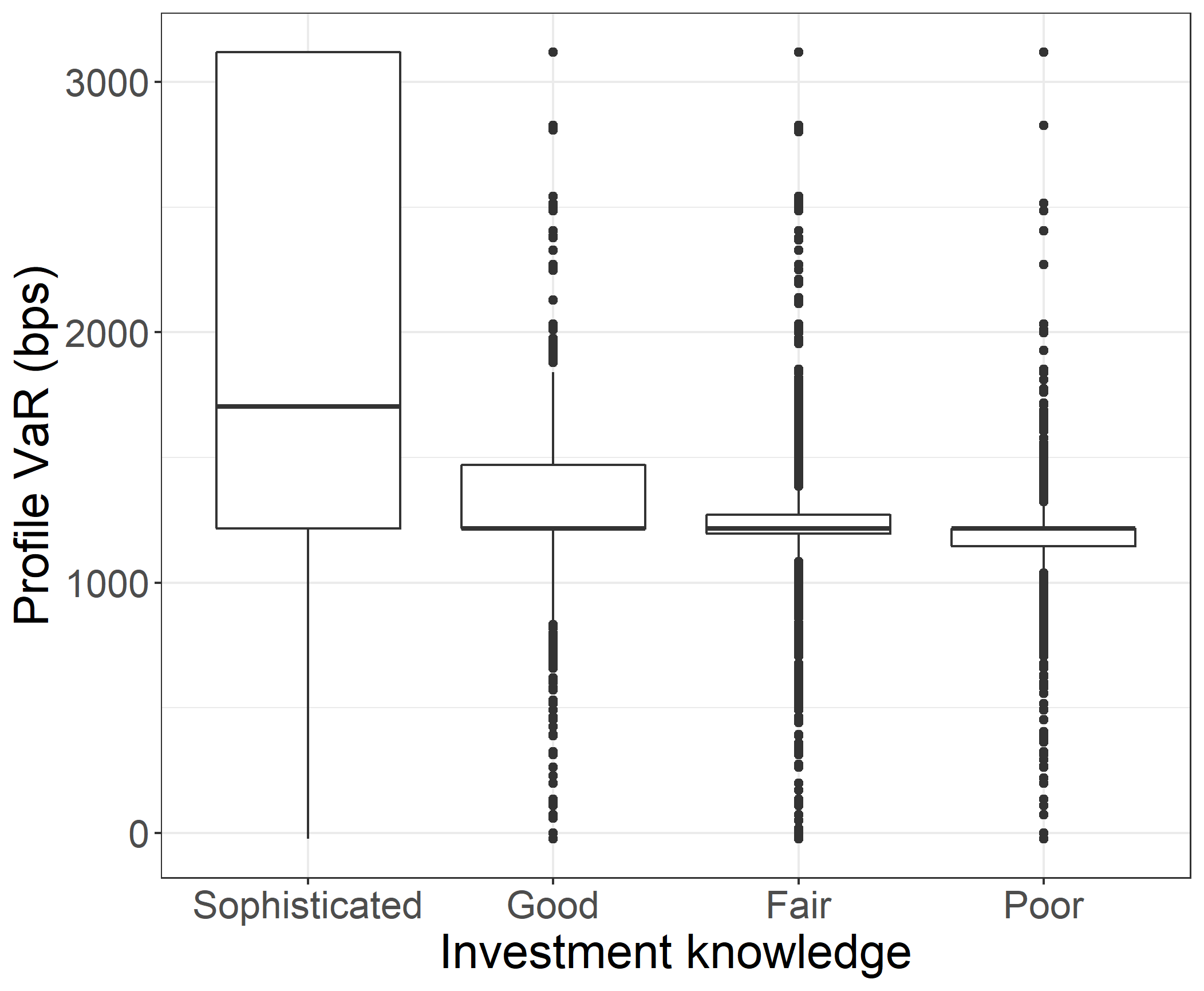}
        \caption{Profile VaR} \label{fig:advisorPrescribedVaRInvestKnow}
    \end{subfigure}
    \hfill
    \begin{subfigure}[t]{0.45\textwidth}
        \centering
        \includegraphics[width=\linewidth]{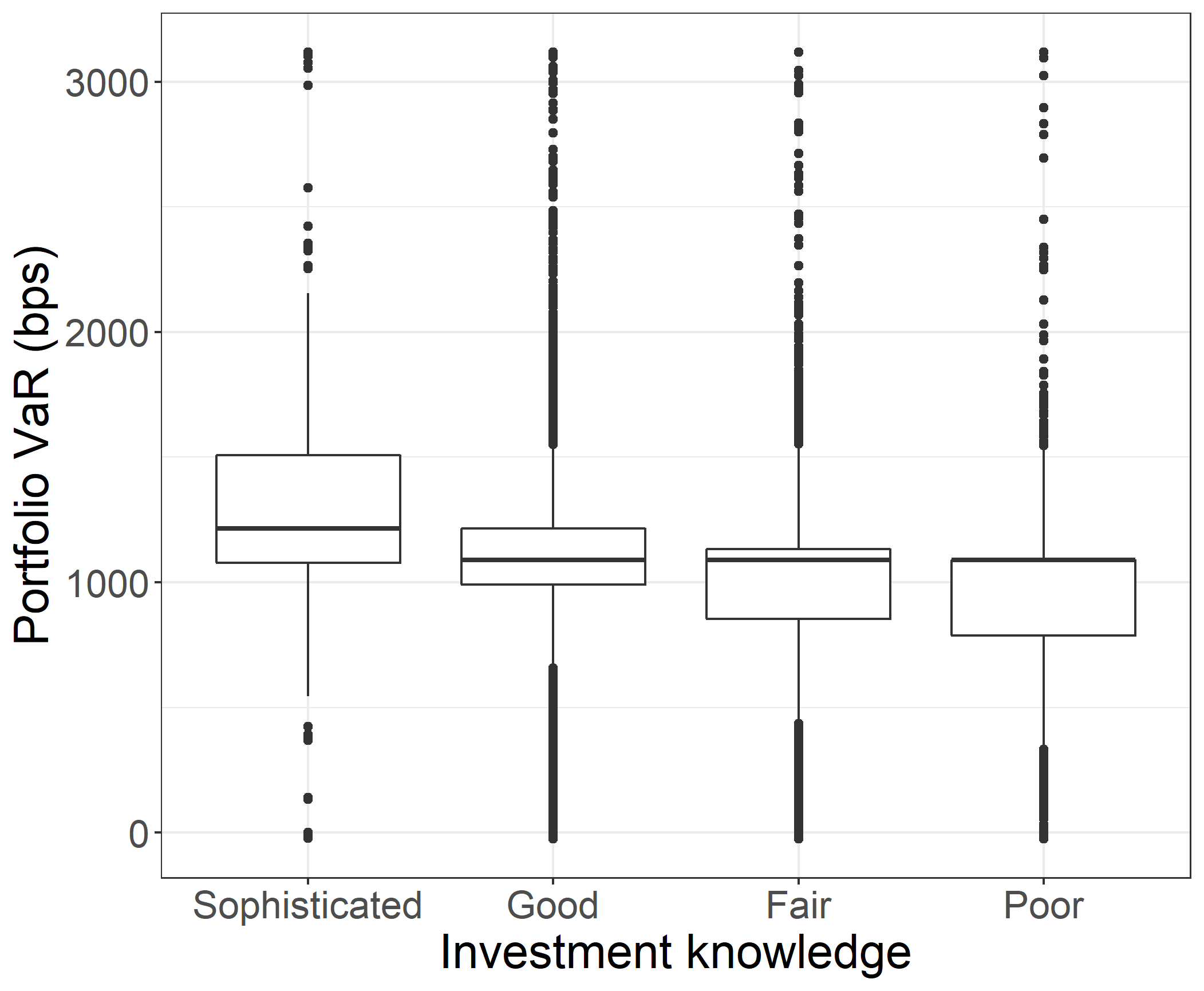} 
        \caption{Portfolio VaR} \label{fig:advisorActualVaRAInvestKnow}
    \end{subfigure}
    \vspace{1cm}
    \begin{subfigure}[t]{0.45\textwidth}
        \centering
        \includegraphics[width=\linewidth]{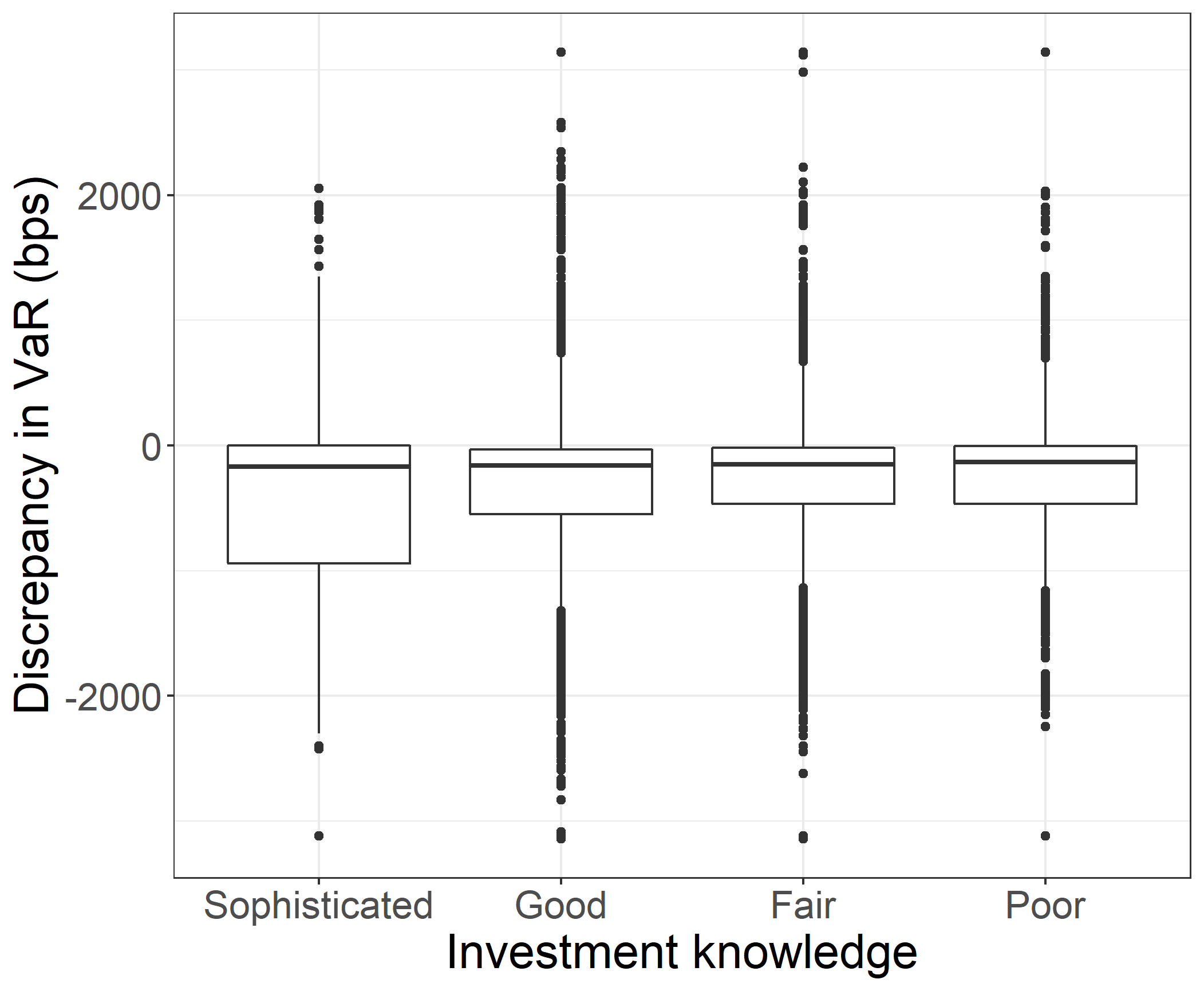} 
        \caption{Discrepancy in VaR} \label{fig:advisorDiscrepancyInvestKnow}
    \end{subfigure}
     \caption{Boxplots of profile VaR (top left panel), portfolio VaR (top right panel) and the discrepancy in VaR (lower panel) on August 12th 2019 by investment knowledge.}
     \label{fig:boxplotsInvestmentKnowledge}
\end{figure}

\begin{figure}
    \centering
    \begin{subfigure}[t]{0.45\textwidth}
        \centering
        \includegraphics[width=\linewidth]{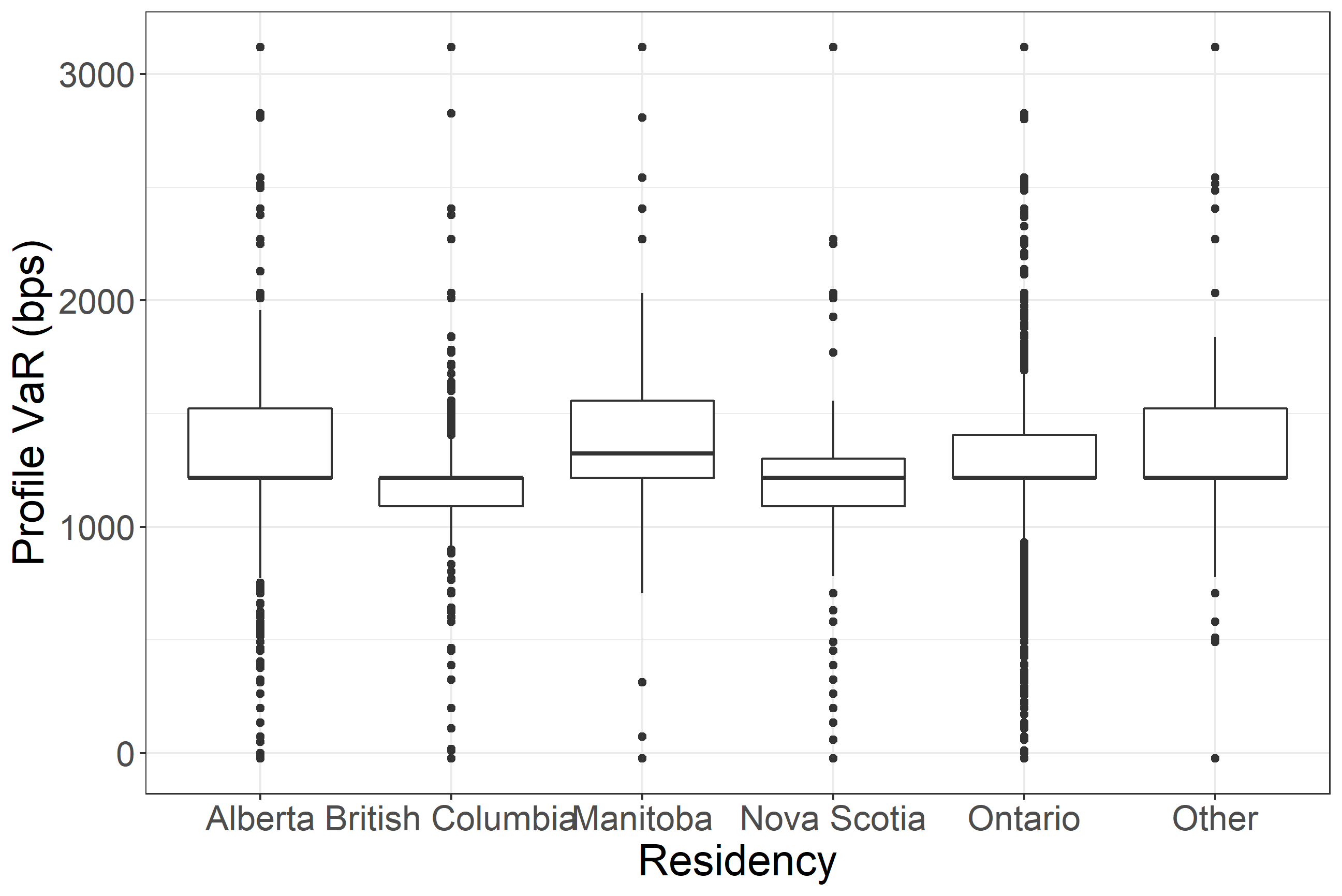}
        \caption{Profile VaR} \label{fig:advisorPrescribedVaRResidency}
    \end{subfigure}
    \hfill
    \begin{subfigure}[t]{0.45\textwidth}
        \centering
        \includegraphics[width=\linewidth]{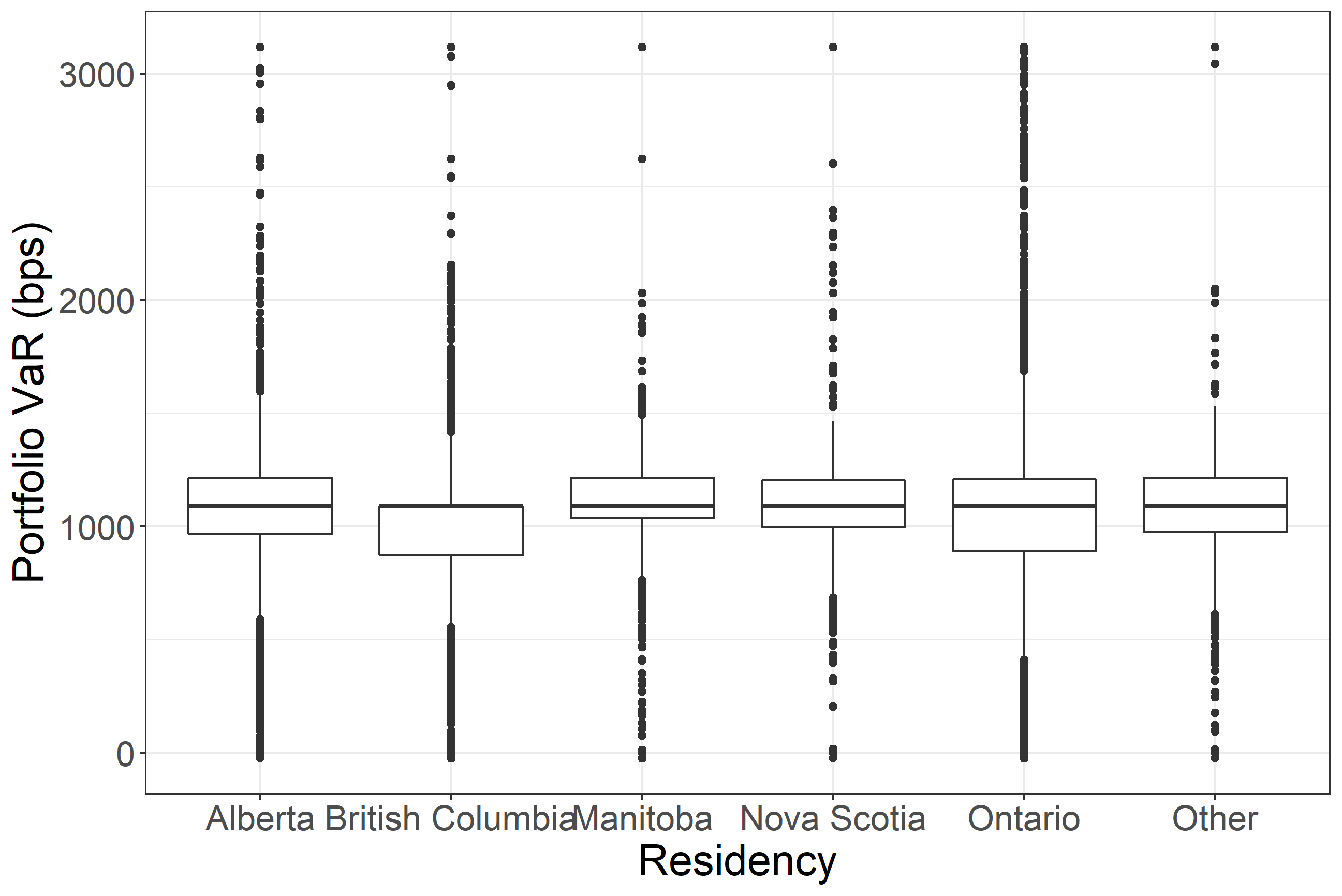} 
        \caption{Portfolio VaR} \label{fig:advisorActualVaRResidency}
    \end{subfigure}
    \vspace{1cm}
    \begin{subfigure}[t]{0.45\textwidth}
        \centering
        \includegraphics[width=\linewidth]{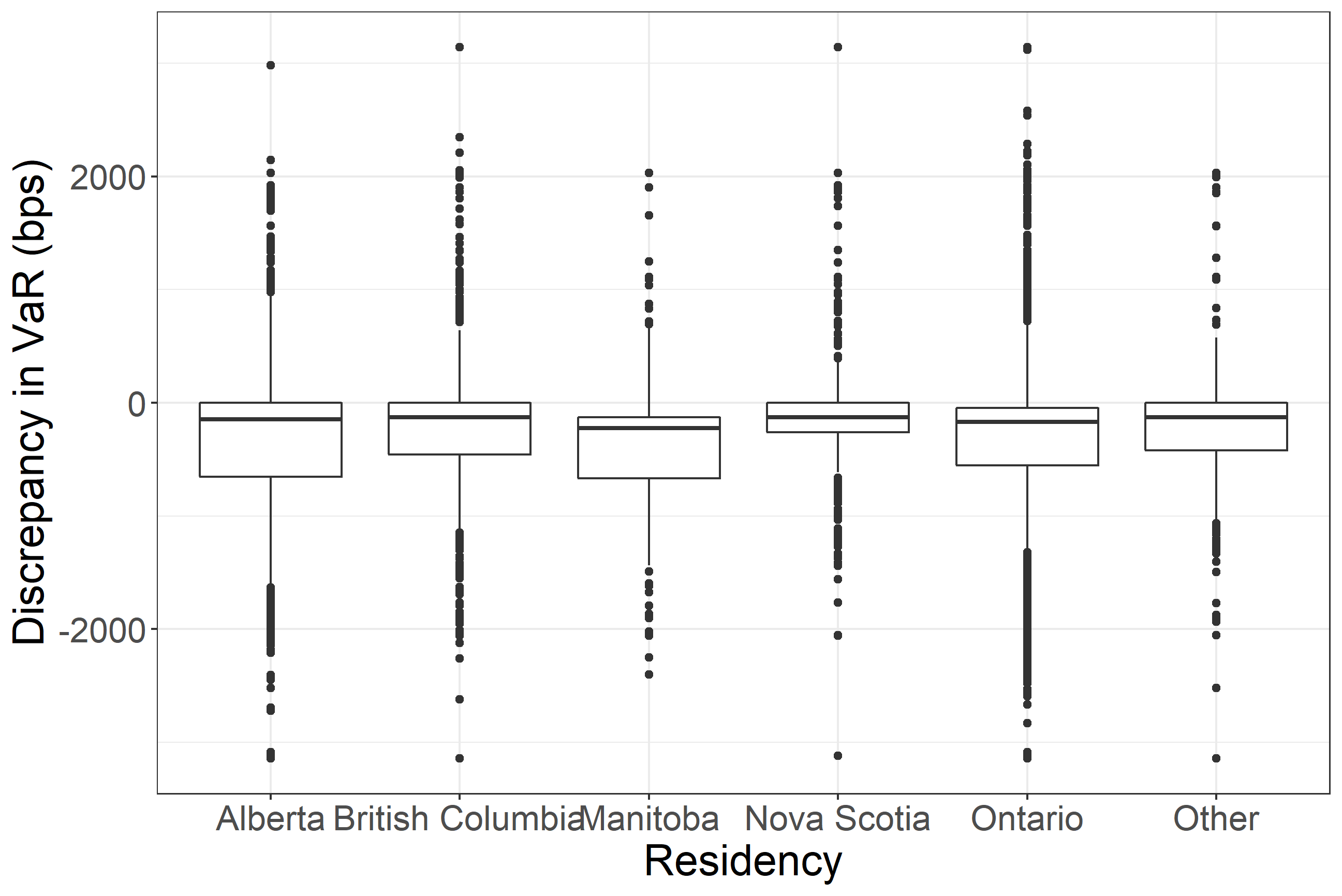} 
        \caption{Discrepancy in VaR} \label{fig:advisorDiscrepancyResidency}
    \end{subfigure}
     \caption{Boxplots of profile VaR (top left panel), portfolio VaR (top right panel) and the discrepancy in VaR (lower panel) on August 12th 2019 by residency location.}
     \label{fig:boxplotsResidency}
\end{figure}

\begin{figure}
    \centering
    \begin{subfigure}[t]{0.45\textwidth}
        \centering
        \includegraphics[width=\linewidth]{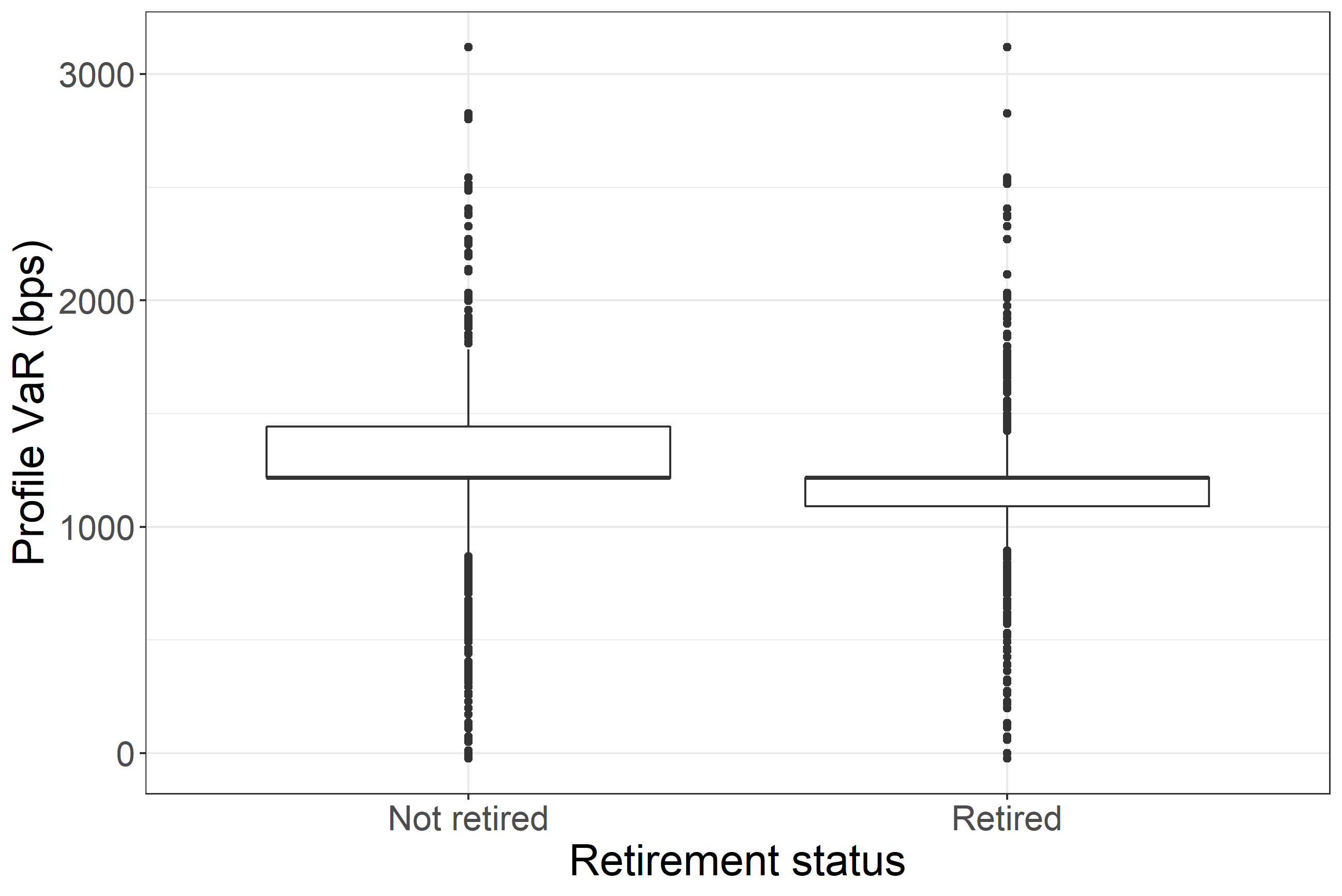}
        \caption{Profile VaR} \label{fig:advisorPrescribedVaRRetirement}
    \end{subfigure}
    \hfill
    \begin{subfigure}[t]{0.45\textwidth}
        \centering
        \includegraphics[width=\linewidth]{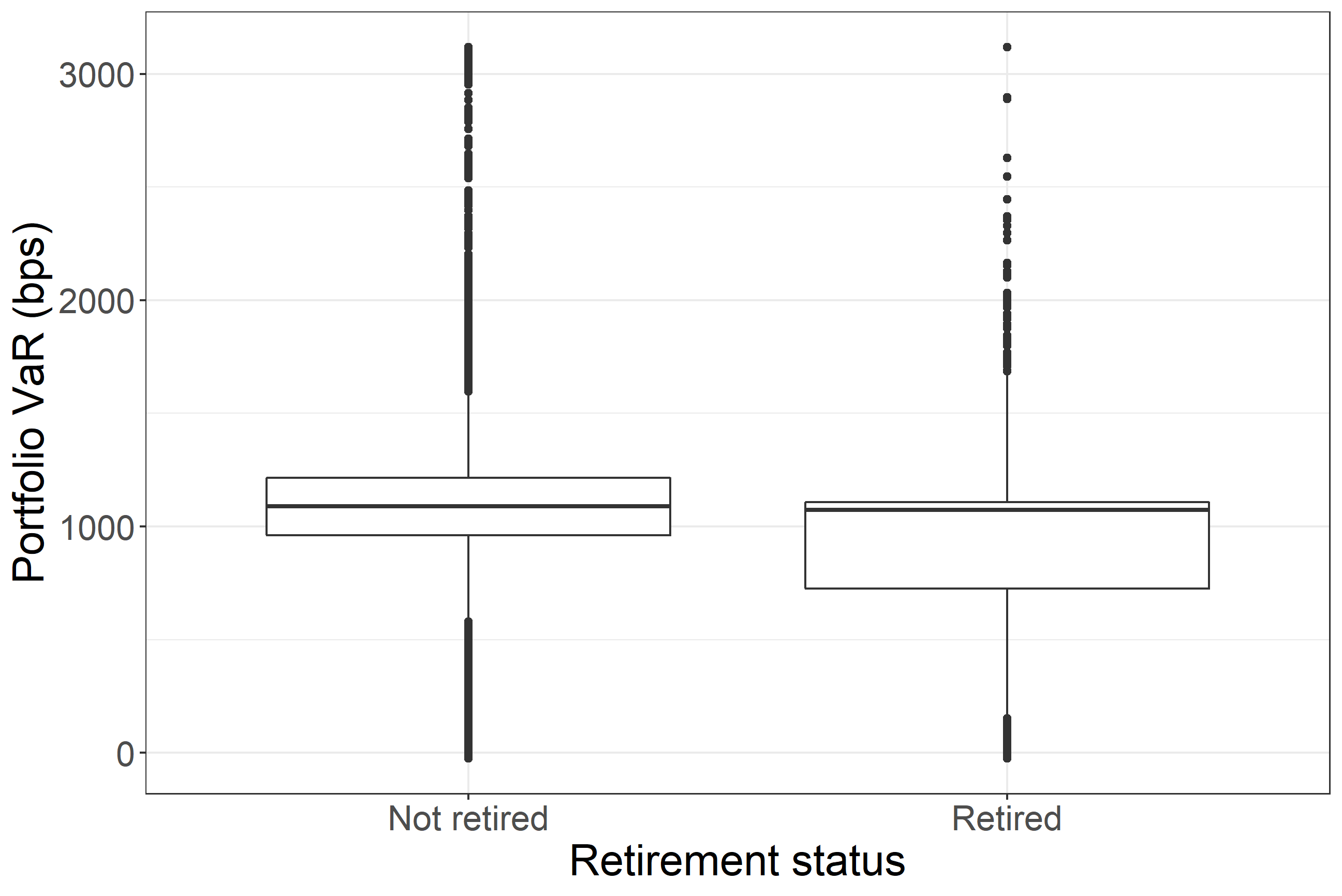} 
        \caption{Portfolio VaR} \label{fig:advisorActualVaRRetirement}
    \end{subfigure}
    \vspace{1cm}
    \begin{subfigure}[t]{0.45\textwidth}
        \centering
        \includegraphics[width=\linewidth]{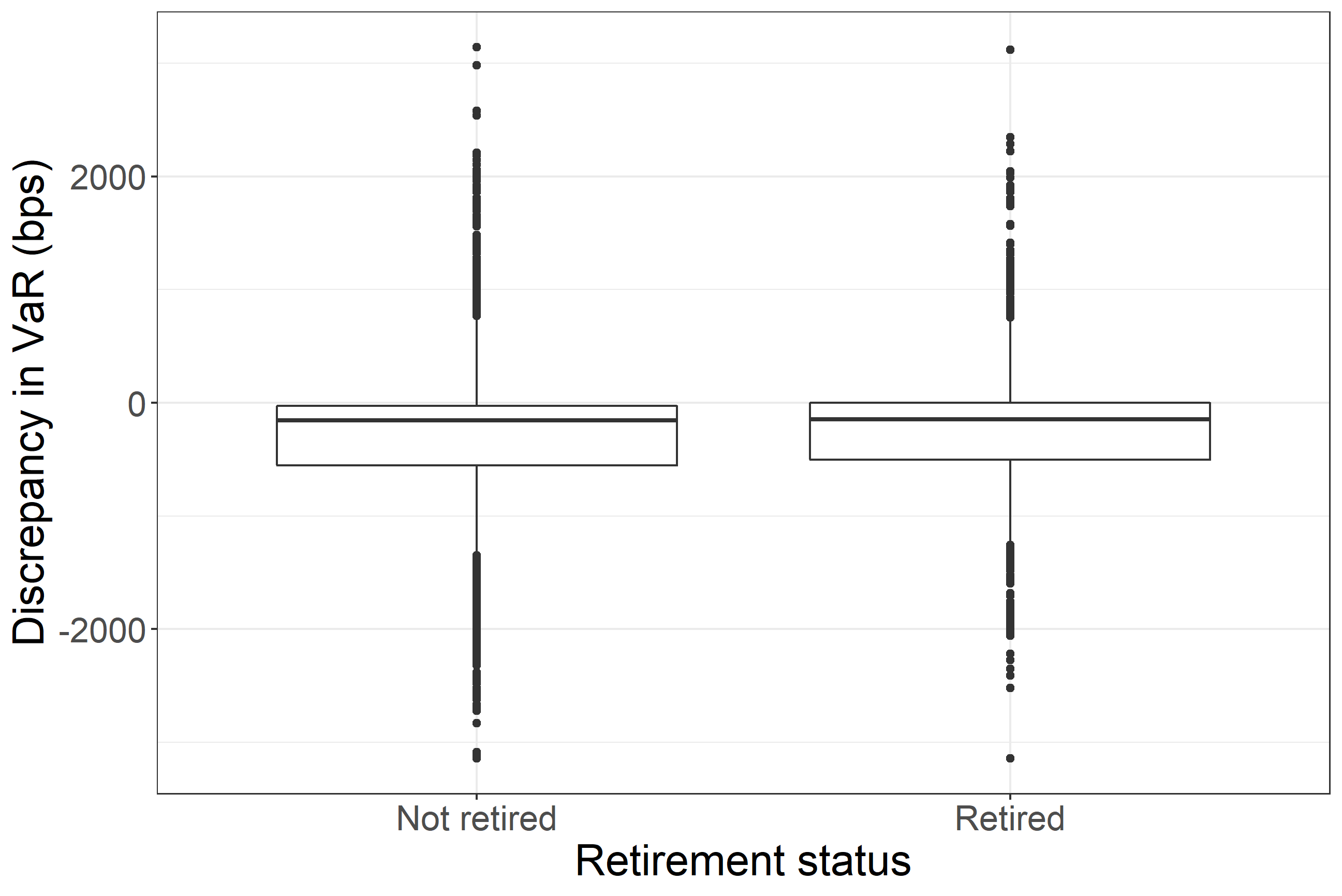} 
        \caption{Discrepancy in VaR} \label{fig:advisorDiscrepancyRetirement}
    \end{subfigure}
     \caption{Boxplots of profile VaR (top left panel), portfolio VaR (top right panel) and the discrepancy in VaR (lower panel) on August 12th 2019 by retirement status.}
     \label{fig:boxplotsRetirement}
\end{figure}

Figure \ref{fig:boxplotsAccountType} shows boxplots of VaR for each of the account types. 
Figure \ref{fig:boxplotsInvestmentKnowledge} shows boxplots of VaR for each of the investment knowledge types.
Figure \ref{fig:boxplotsResidency} shows boxplots of VaR for each of the residency locations.
Figure \ref{fig:boxplotsRetirement} shows boxplots of VaR for retirement status.
Figure \ref{fig:boxplotsRetirement} shows boxplots of VaR for gender.
\begin{figure}
    \centering
    \begin{subfigure}[t]{0.45\textwidth}
        \centering
        \includegraphics[width=\linewidth]{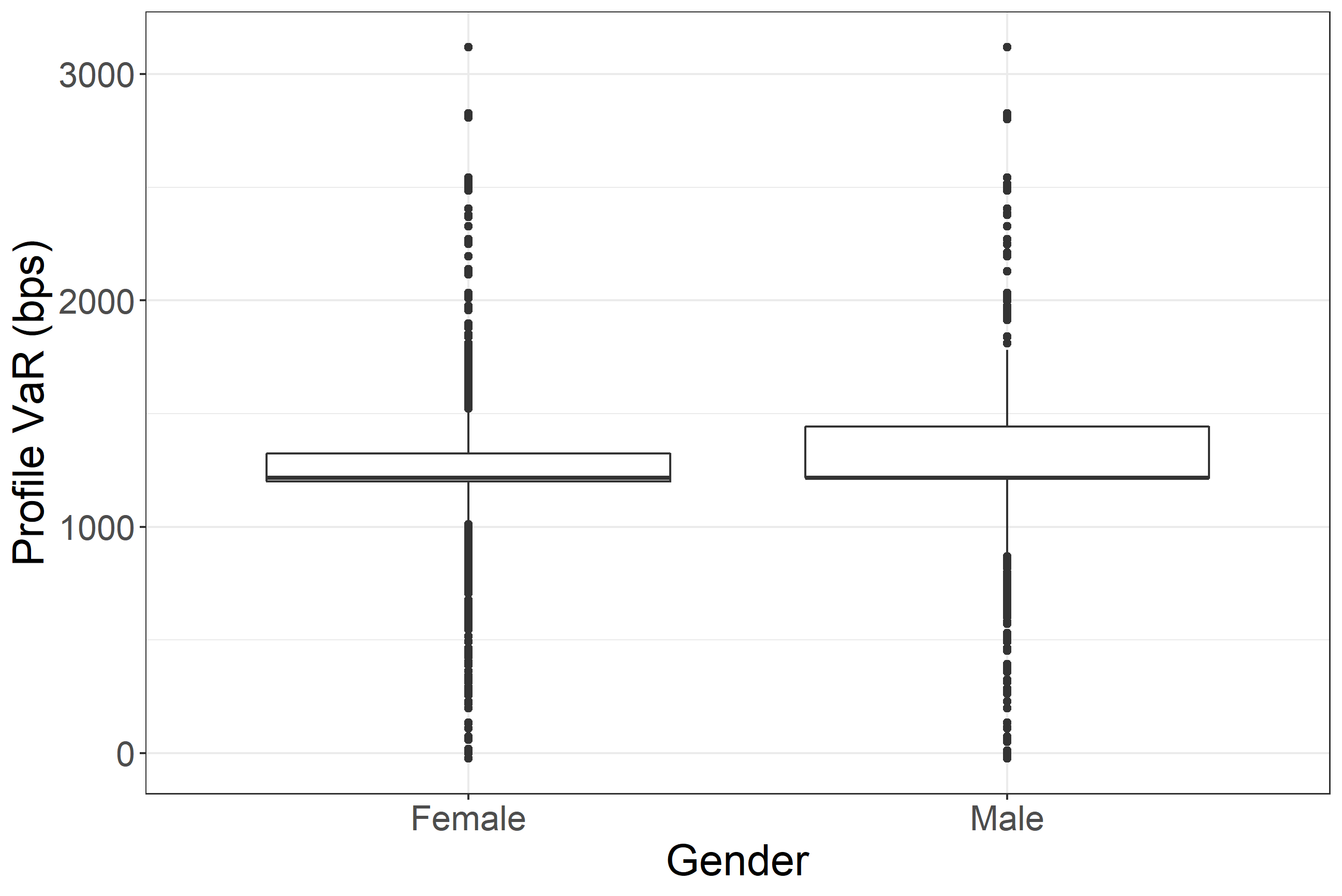}
        \caption{Profile VaR} \label{fig:advisorPrescribedVaRGender}
    \end{subfigure}
    \hfill
    \begin{subfigure}[t]{0.45\textwidth}
        \centering
        \includegraphics[width=\linewidth]{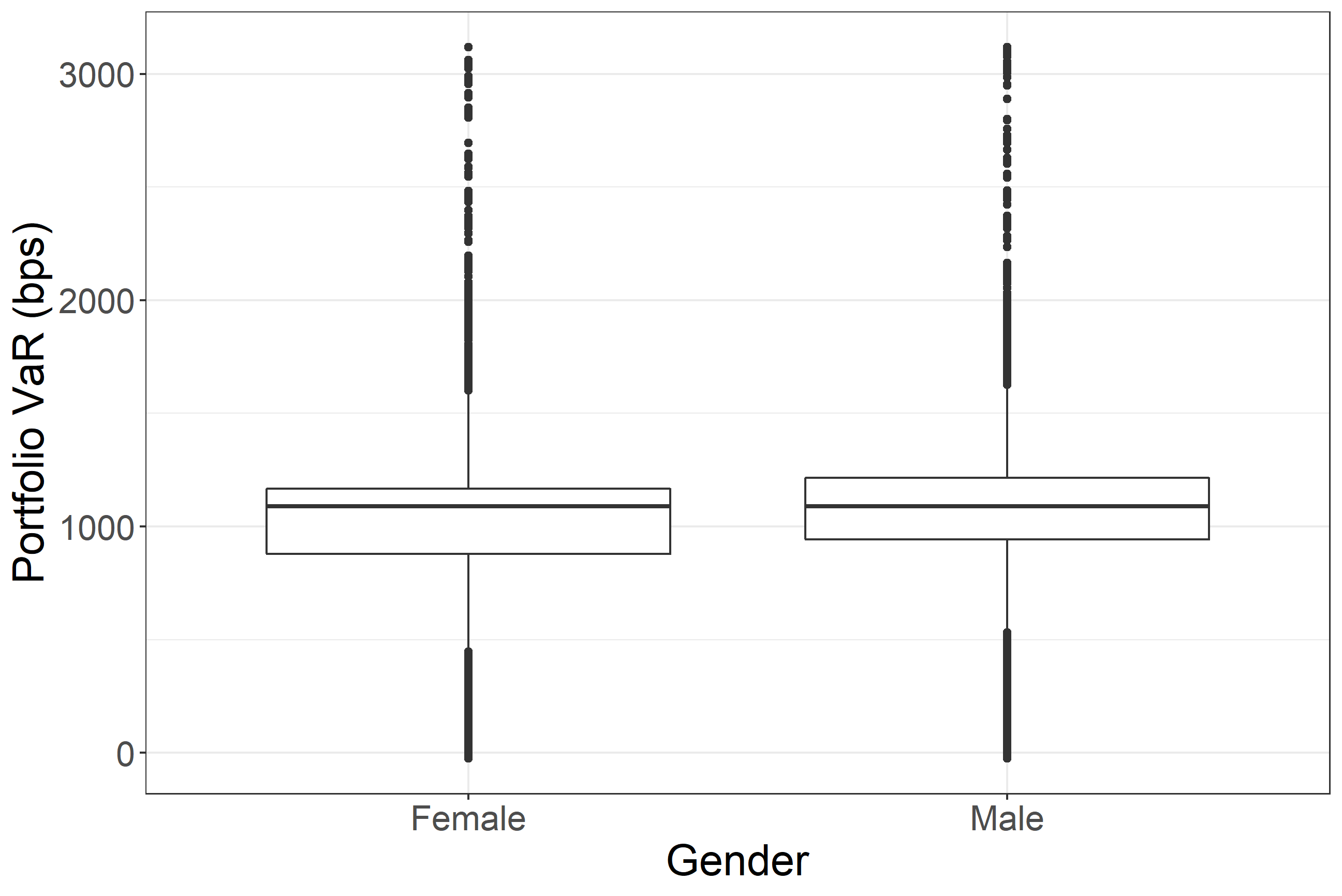} 
        \caption{Portfolio VaR} \label{fig:advisorActualVaRGender}
    \end{subfigure}
    \vspace{1cm}
    \begin{subfigure}[t]{0.45\textwidth}
        \centering
        \includegraphics[width=\linewidth]{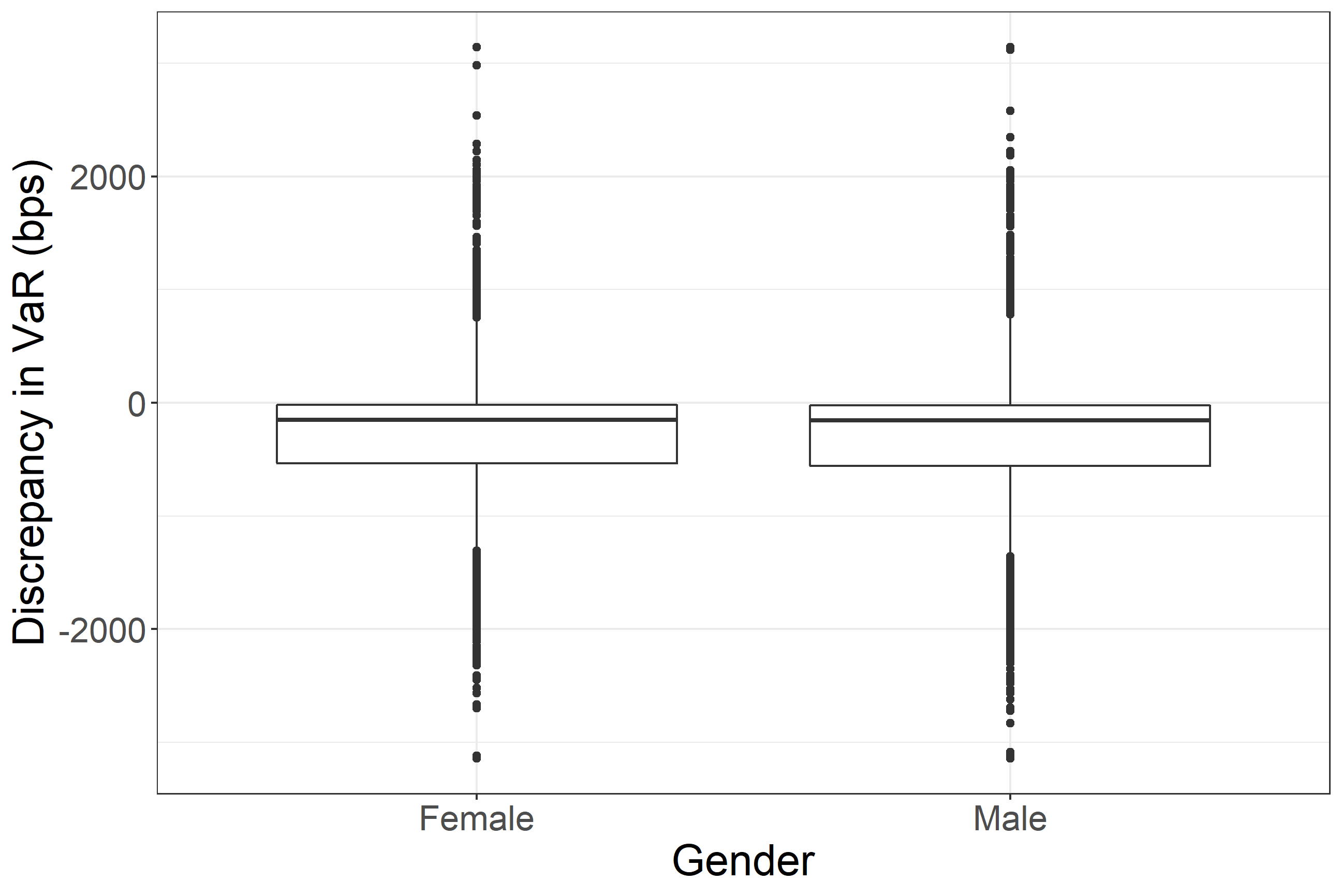} 
        \caption{Discrepancy in VaR} \label{fig:advisorDiscrepancyGender}
    \end{subfigure}
     \caption{Boxplots of profile VaR (top left panel), portfolio VaR (top right panel) and the discrepancy in VaR (lower panel) on August 12th 2019 by gender.}
     \label{fig:boxplotsGender}
\end{figure}
Figure \ref{fig:boxplotsRetirement} shows boxplots of VaR for marital status.
\begin{figure}
    \centering
    \begin{subfigure}[t]{0.45\textwidth}
        \centering
        \includegraphics[width=\linewidth]{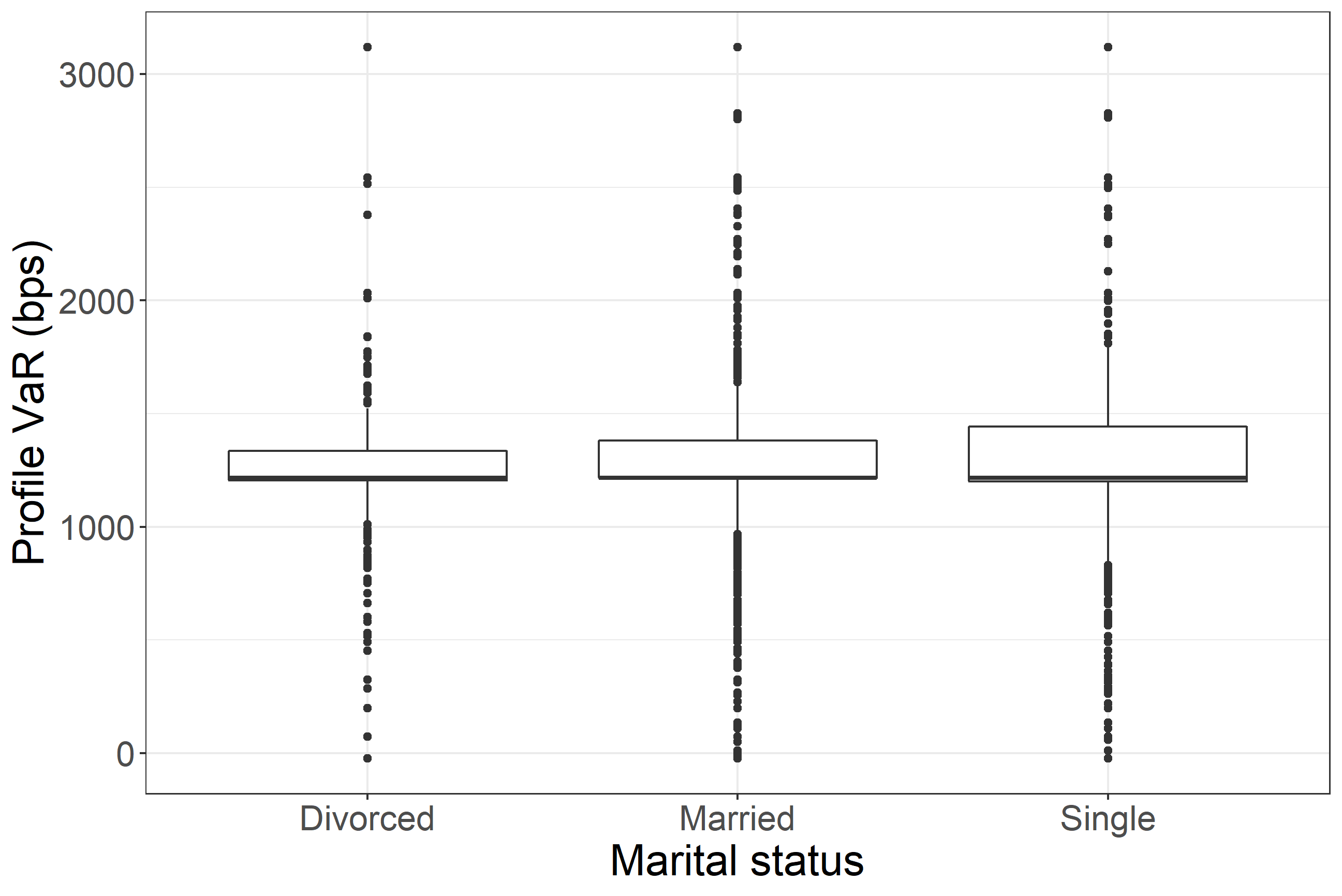}
        \caption{Profile VaR} \label{fig:advisorPrescribedVaRMarital}
    \end{subfigure}
    \hfill
    \begin{subfigure}[t]{0.45\textwidth}
        \centering
        \includegraphics[width=\linewidth]{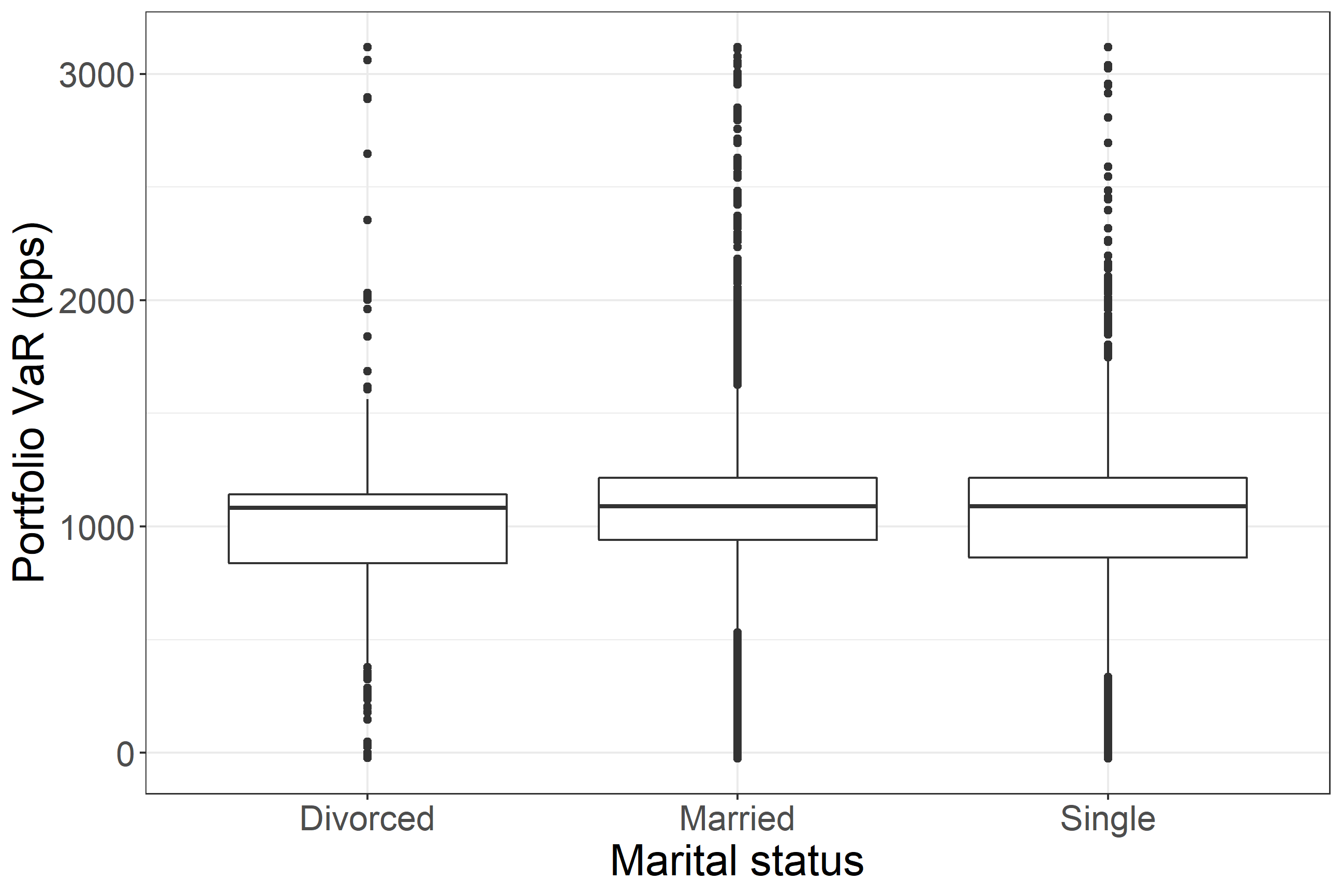} 
        \caption{Portfolio VaR} \label{fig:advisorActualVaRMarital}
    \end{subfigure}
    \vspace{1cm}
    \begin{subfigure}[t]{0.45\textwidth}
        \centering
        \includegraphics[width=\linewidth]{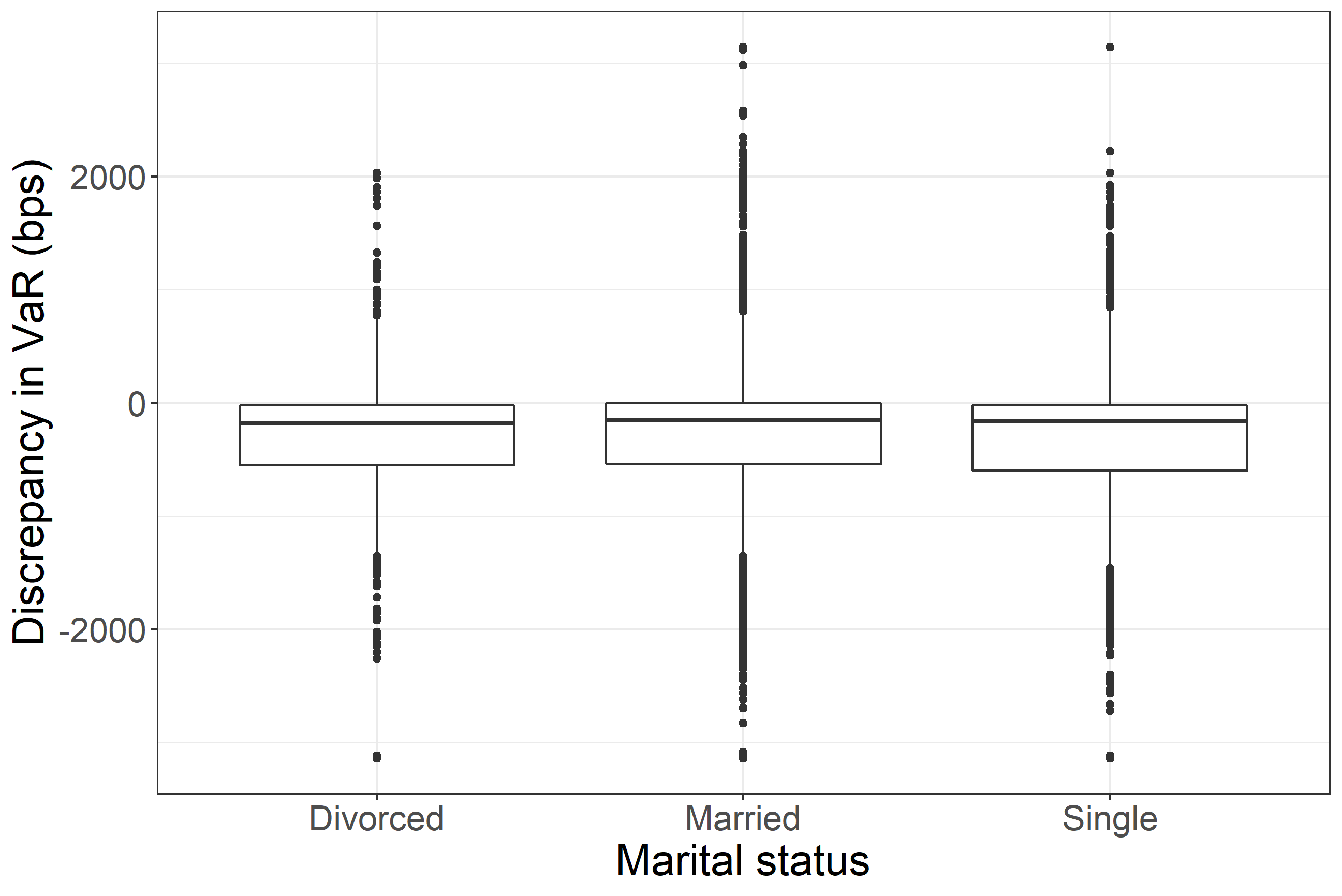} 
        \caption{Discrepancy in VaR} \label{fig:advisorDiscrepancyMarital}
    \end{subfigure}
     \caption{Boxplots of profile VaR (top left panel), portfolio VaR (top right panel) and the discrepancy in VaR (lower panel) on August 12th 2019 by marital status.}
     \label{fig:boxplotsMarital}
\end{figure}

\begin{figure}
    \centering
    \begin{subfigure}[t]{0.45\textwidth}
        \centering
        \includegraphics[width=\linewidth]{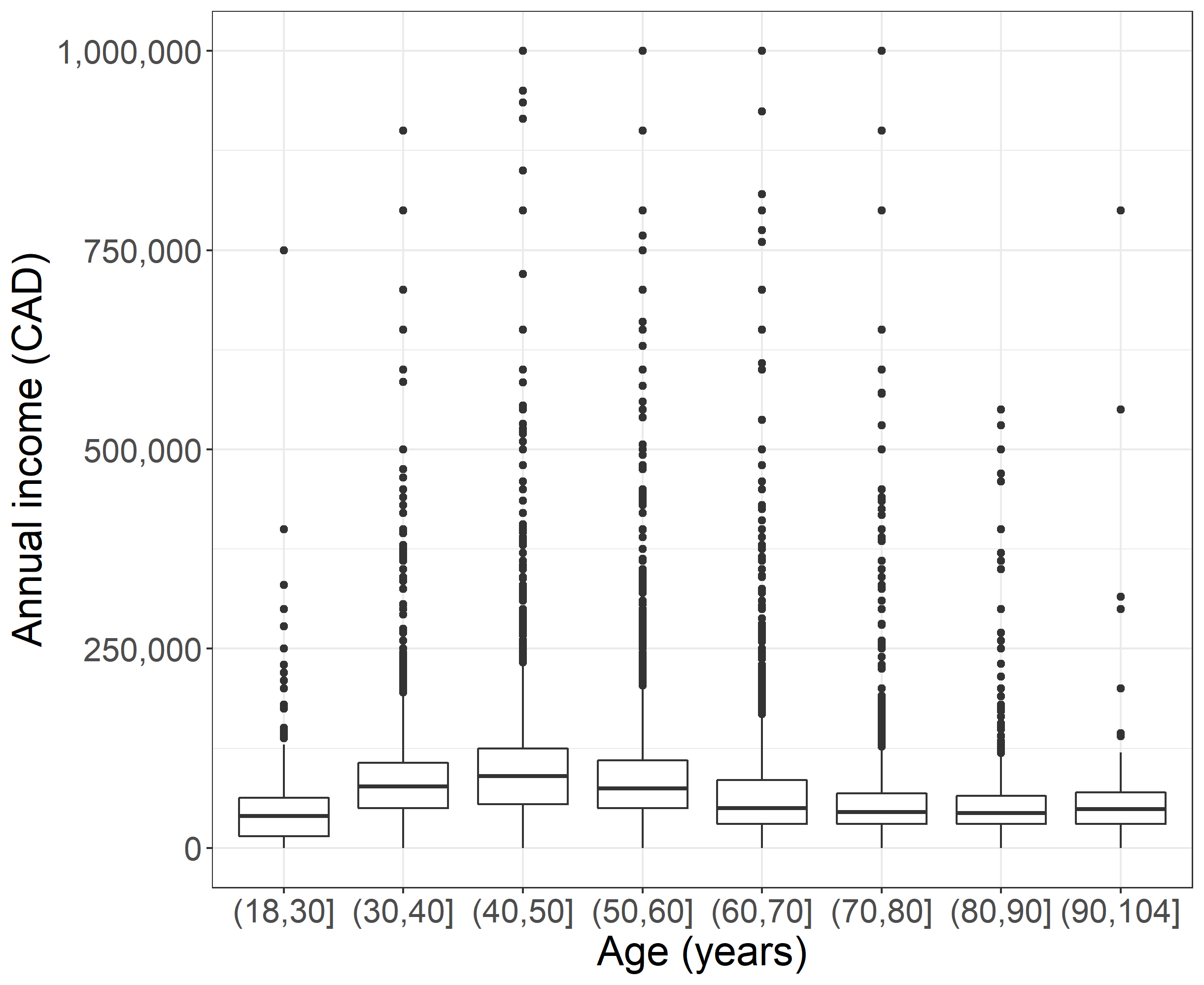}
        \caption{Annual income} \label{fig:ageAndIncome}
    \end{subfigure}
    \hfill
    \begin{subfigure}[t]{0.45\textwidth}
        \centering
        \includegraphics[width=\linewidth]{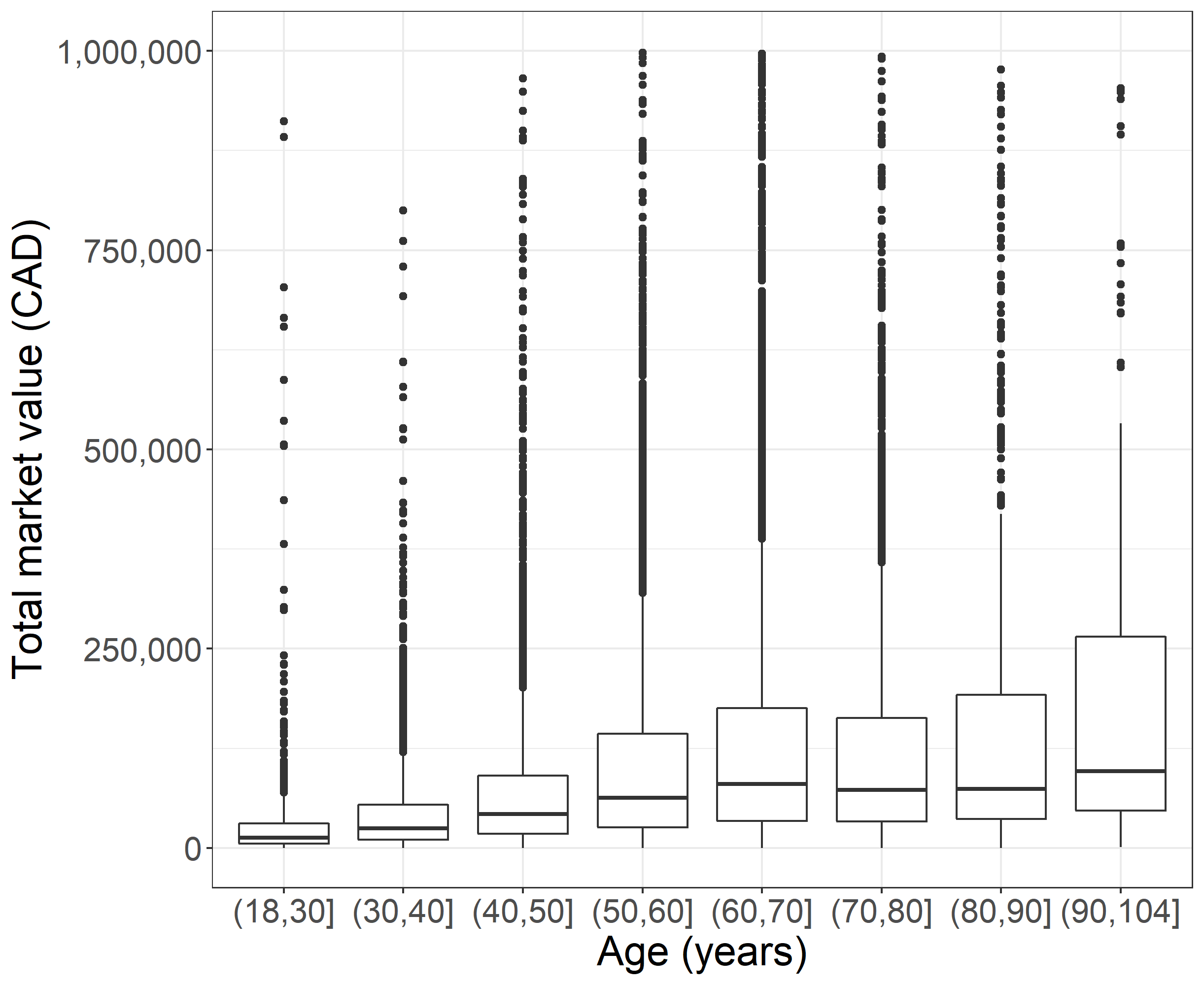} 
        \caption{Total market value} \label{fig:ageAndMktValue}
    \end{subfigure}
    \vspace{1cm}
    \begin{subfigure}[t]{0.45\textwidth}
        \centering
        \includegraphics[width=\linewidth]{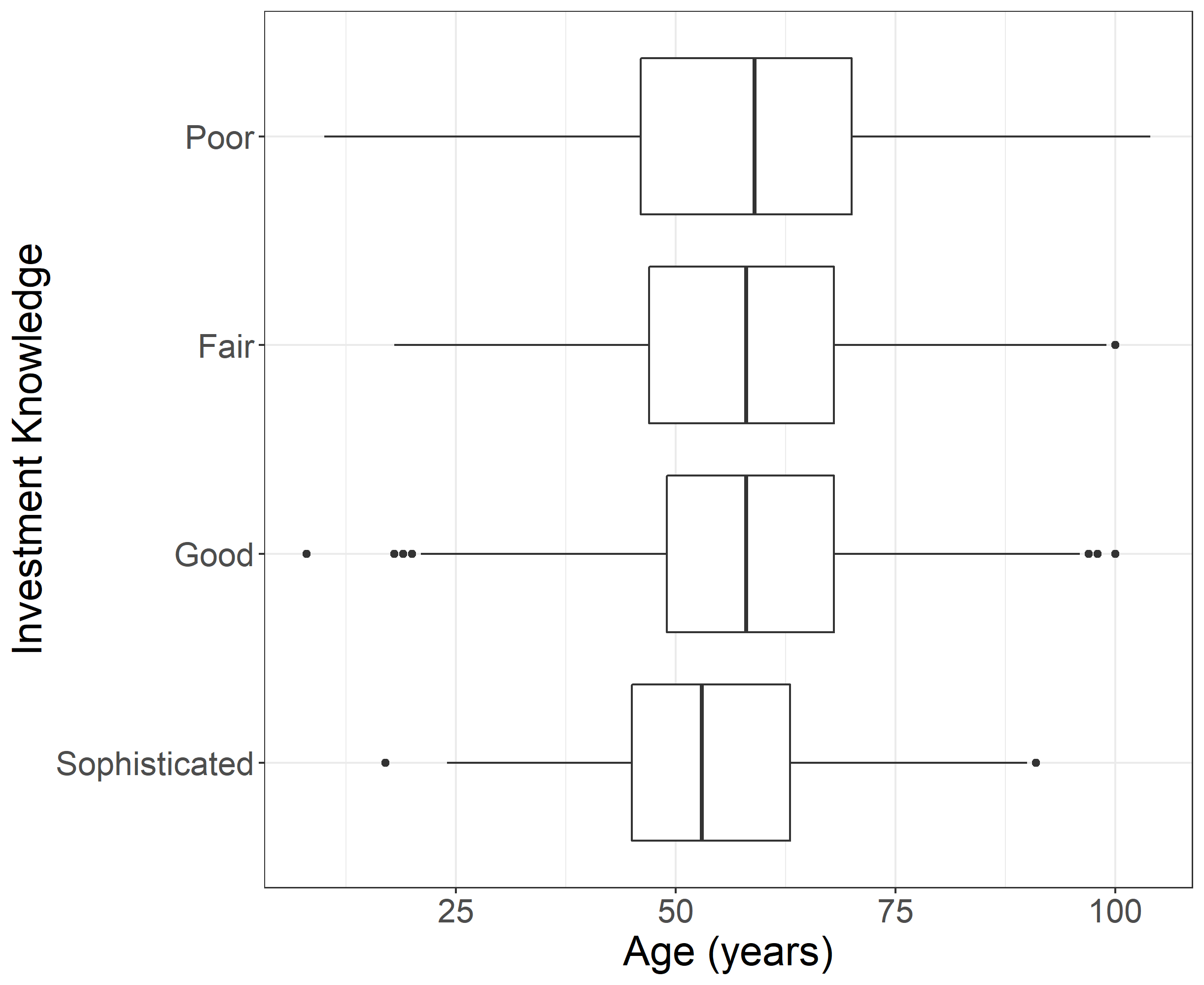} 
        \caption{Investment knowledge} \label{fig:ageAndInvKnow}
    \end{subfigure}
     \caption{Boxplots of age against annual income (top left panel), total portfolio market value (top right panel), and investment knowledge (lower panel).}
     \label{fig:ageVersus}
\end{figure}

Since age is related to the accumulation of experience and wealth, we consider the distribution of ages against annual income, market value, and investment knowledge in Figure \ref{fig:ageVersus}. Figure \ref{fig:ageAndIncome} show a comparison of age and income quartiles. Similarly, Figure \ref{fig:ageAndMktValue} shows age and total asset market value quartiles. Figure \ref{fig:ageAndInvKnow} shows age and investment knowledge quartiles.

% \section*{Appendix D - Statistical investigations (not for actual paper)}

% Figure \ref{fig:bootstrapsIncome} shows the pointwise mean over time with bootstrap confidence intervals for each annual income quartile.

% \begin{figure}
%     \centering
%     \begin{subfigure}[t]{0.45\textwidth}
%         \centering
%         \includegraphics[width=\linewidth]{diagrams/stats/incomeBootstrapsProfile.png}
%         \caption{Profile VaR} \label{fig:bootstrapsClusterProfile}
%     \end{subfigure}
%     \hfill
%     \begin{subfigure}[t]{0.45\textwidth}
%         \centering
%         \includegraphics[width=\linewidth]{diagrams/stats/incomeBootstrapsPortfolio.png} 
%         \caption{Portfolio VaR} \label{fig:bootstrapsClusterPortfolio}
%     \end{subfigure}
%     \vspace{1cm}
%     \begin{subfigure}[t]{0.45\textwidth}
%         \centering
%         \includegraphics[width=\linewidth]{diagrams/stats/incomeBootstraps.png} 
%         \caption{Discrepancy in VaR} \label{fig:bootstrapsClusterDiscrepancy}
%     \end{subfigure}
%      \caption{Mean of daily profile VaR (top left panel), portfolio VaR (top right panel) and discrepancy in VaR (lower panel) for annual income quartiles with 95\% bootstrapped confidence intervals ($B=9999$ re-samples).}
%      \label{fig:bootstrapsIncome}
% \end{figure}

\end{document}